\documentclass[aps,twocolumn,superscriptaddress,english,10pt,nofootinbib]{revtex4}

\usepackage{float}

\usepackage{amssymb, amsmath, bm, dcolumn, epsf, graphicx, latexsym, slashed, simplewick}
\usepackage[utf8]{inputenc}
\usepackage[normalem]{ulem}
\usepackage{orcidlink}

\usepackage{color}

\def\be{\begin{equation}}
\def\ee{\end{equation}}
\def\bea{\begin{eqnarray}}
\def\eea{\end{eqnarray}}

\usepackage{hyperref}
\usepackage{comment}

\usepackage[normalem]{ulem}

\begin{document}

\title{The Atacama Cosmology Telescope: Constraints on Pre-Recombination Early Dark Energy}

\author{J.~Colin Hill\,\orcidlink{0000-0002-9539-0835}}
\affiliation{Department of Physics, Columbia University, New York, NY 10027, USA}
\affiliation{Center for Computational Astrophysics, Flatiron Institute, New York, NY 10010, USA}
\author{Erminia Calabrese\,\orcidlink{0000-0003-0837-0068}}
\affiliation{School of Physics and Astronomy, Cardiff University, The Parade, Cardiff, Wales CF24 3AA, UK}
\author{Simone Aiola\,\orcidlink{0000-0002-1035-1854}}
\affiliation{Center for Computational Astrophysics, Flatiron Institute, New York, NY 10010, USA}
\author{Nicholas Battaglia}
\affiliation{Department of Astronomy, Cornell University, Ithaca, NY 14853, USA}
\author{Boris Bolliet}
\affiliation{Department of Physics, Columbia University, New York, NY 10027, USA}
\author{Steve~K.~Choi\,\orcidlink{0000-0002-9113-7058}}
\affiliation{Department of Physics, Cornell University, Ithaca, NY 14853, USA}
\affiliation{Department of Astronomy, Cornell University, Ithaca, NY 14853, USA}
\author{Mark J.~Devlin\,\orcidlink{0000-0002-3169-9761}}
\affiliation{Department of Physics and Astronomy, University of Pennsylvania, 209 South 33rd Street, Philadelphia, PA 19104, USA}
\author{Adriaan J.~Duivenvoorden\,\orcidlink{0000-0003-2856-2382}}
\affiliation{Department of Physics, Jadwin Hall, Princeton University, Princeton, NJ 08544, USA}
\author{Jo Dunkley\,\orcidlink{0000-0002-7450-2586}}
\affiliation{Department of Astrophysical Sciences, Princeton University, Peyton Hall, Princeton, NJ 08544, USA}
\affiliation{Department of Physics, Jadwin Hall, Princeton University, Princeton, NJ 08544, USA}
\author{Simone Ferraro\,\orcidlink{0000-0003-4992-7854}}
\affiliation{Lawrence Berkeley National Laboratory, One Cyclotron Road, Berkeley, CA 94720, USA}
\affiliation{Berkeley Center for Cosmological Physics, UC Berkeley, CA 94720, USA}
\author{Patricio A.~Gallardo\,\orcidlink{0000-0001-9731-3617}}
\affiliation{Kavli Institute for Cosmological Physics, University of Chicago, Chicago, IL 60637, USA}
\author{Vera Gluscevic}
\affiliation{Department of Physics and Astronomy, University of Southern California, Los Angeles, CA, 90007, USA}
\author{Matthew Hasselfield\,\orcidlink{0000-0002-2408-9201}}
\affiliation{Center for Computational Astrophysics, Flatiron Institute, New York, NY 10010, USA}
\author{Matt Hilton,\orcidlink{0000-0002-8490-8117}}
\affiliation{Astrophysics Research Centre, University of KwaZulu-Natal, Westville Campus, Durban 4041, South Africa}
\affiliation{School of Mathematics, Statistics \& Computer Science, University of KwaZulu-Natal, Westville Campus, Durban 4041, South Africa}
\author{Adam D.~Hincks\,\orcidlink{0000-0003-1690-6678}}
\affiliation{David A. Dunlap Department of Astronomy \& Astrophysics, University of Toronto, 50 St. George St., Toronto, ON M5S 3H4, Canada}
\author{Ren\'ee~Hlo\v zek\,\orcidlink{0000-0002-0965-7864}}
\affiliation{David A. Dunlap Department of Astronomy \& Astrophysics, University of Toronto, 50 St. George St., Toronto, ON M5S 3H4, Canada}
\affiliation{Dunlap Institute of Astronomy \& Astrophysics, 50 St. George St., Toronto, ON M5S 3H4, Canada}
\author{Brian~J.~Koopman\,\orcidlink{0000-0003-0744-2808}}
\affiliation{Department of Physics, Yale University, New Haven, CT 06520, USA}
\author{Arthur Kosowsky\,\orcidlink{0000-0002-3734-331X}}
\affiliation{Department of Physics and Astronomy, University of Pittsburgh, Pittsburgh, PA 15260, USA}
\author{Adrien La Posta}
\affiliation{Universit\'{e} Paris-Saclay, CNRS/IN2P3, IJCLab, 91405 Orsay, France}
\author{Thibaut Louis}
\affiliation{Universit\'{e} Paris-Saclay, CNRS/IN2P3, IJCLab, 91405 Orsay, France}
\author{Mathew S.~Madhavacheril\,\orcidlink{0000-0001-6740-5350}}
\affiliation{Perimeter Institute for Theoretical Physics, 31 Caroline Street N, Waterloo ON N2L 2Y5, Canada}
\affiliation{Department of Physics and Astronomy, University of Southern California, Los Angeles, CA, 90007, USA}
\author{Jeff McMahon}
\affiliation{Department of Astronomy and Astrophysics, University of Chicago, Chicago, IL 60637, USA}
\affiliation{Kavli Institute for Cosmological Physics, University of Chicago, Chicago, IL 60637, USA}
\affiliation{Department of Physics, University of Chicago, Chicago, IL 60637, USA}
\affiliation{Enrico Fermi Institute, University of Chicago, Chicago, IL 60637, USA}
\author{Kavilan Moodley}
\affiliation{Astrophysics Research Centre, University of KwaZulu-Natal, Westville Campus, Durban 4041, South Africa}
\affiliation{School of Mathematics, Statistics \& Computer Science, University of KwaZulu-Natal, Westville Campus, Durban 4041, South Africa}
\author{Sigurd Naess\,\orcidlink{0000-0002-4478-7111}}
\affiliation{Center for Computational Astrophysics, Flatiron Institute, New York, NY 10010, USA}
\author{Umberto Natale}
\affiliation{School of Physics and Astronomy, Cardiff University, The Parade, Cardiff, Wales CF24 3AA, UK}
\author{Federico Nati\,\orcidlink{0000-0002-8307-5088}}
\affiliation{Department of Physics, University of Milano-Bicocca, Piazza della Scienza 3, 20126 Milano (MI), Italy}
\author{Laura Newburgh}
\affiliation{Department of Physics, Yale University, New Haven, CT 06520, USA}
\author{Michael D.~Niemack\,\orcidlink{0000-0001-7125-3580}}
\affiliation{Department of Physics, Cornell University, Ithaca, NY 14853, USA}
\affiliation{Department of Astronomy, Cornell University, Ithaca, NY 14853, USA}
\affiliation{Kavli Institute at Cornell for Nanoscale Science, Cornell University, Ithaca, NY 14853, USA}
\author{Lyman A.~Page\,\orcidlink{0000-0002-9828-3525}}
\affiliation{Department of Physics, Jadwin Hall, Princeton University, Princeton, NJ 08544, USA}
\author{Bruce Partridge}
\affiliation{Department of Astronomy, Haverford College, Haverford, PA 19041, USA}
\author{Frank J.~Qu}
\affiliation{DAMTP, Centre for Mathematical Sciences, University of Cambridge, Wilberforce Road, Cambridge CB3 OWA, UK}
\author{Maria Salatino}
\affiliation{Stanford University, Stanford, CA 94305, USA}
\affiliation{Kavli Institute for Particle Astrophysics and Cosmology, Stanford, CA 94305, USA}
\author{Alessandro Schillaci}
\affiliation{Department of Physics, California Institute of Technology, Pasadena, CA 91125, USA}
\author{Neelima Sehgal\,\orcidlink{0000-0002-9674-4527}}
\affiliation{Physics and Astronomy Department, Stony Brook University, Stony Brook, NY 11794, USA}
\author{Blake D.~Sherwin}
\affiliation{DAMTP, Centre for Mathematical Sciences, University of Cambridge, Wilberforce Road, Cambridge CB3 OWA, UK}
\affiliation{Kavli Institute for Cosmology, University of Cambridge, Madingley Road, Cambridge CB3 OHA, UK}
\author{Crist\'obal Sif\'on\,\orcidlink{0000-0002-8149-1352}}
\affiliation{Instituto de F\'isica, Pontificia Universidad Cat\'olica de Valpara\'iso, Casilla 4059, Valpara\'iso, Chile}
\author{David N.~Spergel}
\affiliation{Center for Computational Astrophysics, Flatiron Institute, New York, NY 10010, USA}
\affiliation{Department of Astrophysical Sciences, Princeton University, Peyton Hall, Princeton, NJ 08544, USA}
\author{Suzanne T.~Staggs\,\orcidlink{0000-0002-7020-7301}}
\affiliation{Department of Physics, Jadwin Hall, Princeton University, Princeton, NJ 08544, USA}
\author{Emilie R.~Storer}
\affiliation{Department of Physics, Jadwin Hall, Princeton University, Princeton, NJ 08544, USA}
\author{Alexander van Engelen}
\affiliation{School of Earth and Space Exploration, Arizona State University, Tempe, AZ 85287, USA}
\author{Eve M.~Vavagiakis}
\affiliation{Department of Physics, Cornell University, Ithaca, NY 14853, USA}
\author{Edward J.~Wollack\,\orcidlink{0000-0002-7567-4451}}
\affiliation{NASA Goddard Space Flight Center, 8800 Greenbelt Rd, Greenbelt, MD 20771, USA}
\author{Zhilei Xu\,\orcidlink{0000-0001-5112-2567}}
\affiliation{Kavli Institute for Astrophysics and Space Research, MIT, 77 Massachusetts Avenue, Cambridge, MA 02139, USA}

\date{\today}


\begin{abstract}
The early dark energy (EDE) scenario aims to increase the value of the Hubble constant ($H_0$) inferred from cosmic microwave background (CMB) data 
over that found in the standard cosmological model ($\Lambda$CDM), via the introduction of a new form of energy density in the early universe.  The EDE component briefly accelerates cosmic expansion just prior to recombination, which reduces the physical size of the sound horizon imprinted in the CMB.  Previous work has found that non-zero EDE is not preferred by \emph{Planck} CMB power spectrum data alone, which yield a 95\% confidence level (CL) upper limit $f_{\rm EDE} < 0.087$ on the maximal fractional contribution of the EDE field to the cosmic energy budget.  In this paper, we fit the EDE model to CMB data from the Atacama Cosmology Telescope (ACT) Data Release 4.  We find that a combination of ACT, large-scale \emph{Planck} TT (similar to \emph{WMAP}), \emph{Planck} CMB lensing, and BAO data prefers the existence of EDE at $>99.7$\% CL: $f_{\rm EDE} = 0.091^{+0.020}_{-0.036}$, with $H_0 = 70.9^{+1.0}_{-2.0} \,\, {\rm km/s/Mpc}$ (both 68\% CL).  From a model-selection standpoint, we find that EDE is favored over $\Lambda$CDM by these data at roughly $3\sigma$ significance.  In contrast, a joint analysis of the full \emph{Planck} and ACT data yields no evidence for EDE, as previously found for \emph{Planck} alone.  We show that the preference for EDE in ACT alone is driven by its TE and EE power spectrum data.  The tight constraint on EDE from \emph{Planck} alone is driven by its high-$\ell$ TT power spectrum data.  Understanding whether these differing constraints are physical in nature, due to systematics, or simply a rare statistical fluctuation is of high priority.  The best-fit EDE models to ACT and \emph{Planck} exhibit coherent differences across a wide range of multipoles in TE and EE, indicating that a powerful test of this scenario is anticipated with near-future data from ACT and other ground-based experiments.
\end{abstract}


\maketitle


\section{Introduction}
\label{sec:intro}

The Hubble constant, $H_0$, is a fundamental quantity in cosmology, which parameterizes the current expansion rate and hence sets the overall scale of the universe.  Its value can be determined using multiple observational probes, including both ``indirect'' probes that depend on the assumption of a cosmological model and local, ``direct'' probes that do not.  Probes in the former category include the cosmic microwave background (CMB) temperature and polarization anisotropy power spectra, as measured by \emph{Planck}~\cite{Planck2018overview}, \emph{WMAP}~\cite{Hinshaw2013}, the Atacama Cosmology Telescope (ACT)~\cite{Aiola2020}, the South Pole Telescope (SPT)~\cite{SPT-3G:2021eoc}, and other experiments, as well as various large-scale structure (LSS) data sets (e.g.,~\cite{Ivanov:2019hqk,Philcox:2020vvt,eBOSS:2020yzd,Philcox:2020xbv}).  Probes in the latter category include the classical distance ladder using Type Ia supernovae (SNIa) calibrated by various means (e.g.,~Cepheids~\cite{Riess:2019cxk,Riess:2020fzl} or the tip of the red giant branch (TRGB)~\cite{Freedman:2020dne,Freedman:2021ahq}) or strong gravitational lensing time delay distances~\cite{Wong:2019kwg,Birrer:2020tax}.

Some of the direct probes have inferred values of $H_0$ that are higher than the value predicted by the best-fit $\Lambda$ cold dark matter ($\Lambda$CDM) model to CMB data~\cite{Planck2018parameters,Aiola2020} or to other indirect cosmological data (e.g., Big Bang nucleosynthesis in combination with baryon acoustic oscillation and gravitational lensing data~\cite{Abbott:2017smn}).  Perhaps most well-known is the discrepancy between the most statistically precise probes in each category, the \emph{Planck} CMB data~\cite{Planck2018parameters} (indirect) and Cepheid-calibrated SNIa distances from SH0ES~\cite{Riess:2020fzl} (direct), which is significant at $\approx 4\sigma$.  However, other direct probes have inferred values of $H_0$ that agree with the $\Lambda$CDM-predicted value from the CMB and LSS, including TRGB-calibrated SNIa~\cite{Freedman:2021ahq} and the latest strong lensing time delay data~\cite{Birrer:2020tax}.  Nevertheless, the error bars are sufficiently large that these measurements are also consistent with the higher $H_0$ value from SH0ES.  We refer the reader to Refs.~\cite{Verde:2019ivm,Knox:2019rjx,Freedman:2021ahq,DiValentino:2021izs} for a selection of reviews with various perspectives on the observational situation.  In this work, we focus entirely on indirect, cosmological probes of $H_0$, with the goal of assessing the extent to which our inference of this parameter from these data can be changed by the assumption of a different cosmological model.  We do not try to assess the global concordance of any particular model.

Attempts to increase the value of $H_0$ inferred from indirect probes have led to the development of numerous new theoretical scenarios beyond $\Lambda$CDM~\cite{Knox:2019rjx,DiValentino:2021izs}.  Amongst the hypotheses to date are strongly interacting neutrinos~\cite{Kreisch:2019yzn,Brinckmann:2020bcn}, primordial magnetic fields~\cite{Jedamzik:2020krr,Thiele:2021okz}, and varying fundamental constants~\cite{Hart:2019dxi,Sekiguchi2021}.  A thread unifying many of these approaches is a decrease in the sound horizon at recombination as compared to its $\Lambda$CDM-inferred value; to maintain agreement with the observed angular size of the sound horizon, a higher $H_0$ value is subsequently inferred when fitting CMB data to such models~\cite{Aylor:2018drw,Knox:2019rjx}.  While this approach may not suffice to dramatically increase the inferred value of $H_0$ when other observational constraints are folded into the analysis (particularly on the matter density at low redshifts)~\cite{Jedamzik:2020zmd}, it has nevertheless spurred much of the theoretical exploration in this area.

In this paper, we focus specifically on the ``early dark energy'' (EDE) proposal for increasing the CMB-inferred value of $H_0$~\cite{Poulin:2018cxd,Smith:2019ihp,Agrawal:2019lmo,Lin:2019qug}, which falls into the general class of sound-horizon-decreasing scenarios.\footnote{Note that other classes of early dark energy scenarios have been investigated over the past two decades~(e.g.,~\cite{Zlatev:1998tr,PhysRevD.58.023503,Doran:2006kp,PhysRevD.83.023011,PhysRevD.83.123504} and references therein).}  In the EDE scenario, a new field is introduced that acts to briefly accelerate cosmic expansion (relative to its $\Lambda$CDM behavior) just prior to recombination, e.g., around matter-radiation equality.  This increase in $H(z)$ leads to a decrease in the sound horizon at recombination, $r_s^*$, which subsequently yields a higher $H_0$ in fits to CMB data.

This qualitative picture suffices to explain the EDE scenario at the background level.  However, the detailed predictions of the scenario depend on the behavior of perturbations, which are significantly more model-dependent.  A variety of detailed mechanisms have been proposed to implement the basic EDE idea~(e.g.,~\cite{Poulin:2018cxd,Smith:2019ihp,Agrawal:2019lmo,Lin:2019qug,Alexander:2019rsc,Niedermann:2019olb,Sakstein:2019fmf,Lin:2020jcb,Ye:2020btb,Braglia:2020bym,Gogoi:2020qif,Allali:2021azp,Karwal:2021vpk,McDonough:2021pdg}), with varying levels of phenomenological success in matching the full range of high-precision cosmological data available today.  Here we focus on the model studied in Refs.~\cite{Poulin:2018cxd,Smith:2019ihp,Hill:2020osr,Ivanov:2019hqk,DAmico:2020ods,Vagnozzi:2021gjh}, which is amongst the more successful in fitting data, although its theoretical construction is somewhat {\it ad hoc}, requiring non-negligible fine-tuning.  In this model, the EDE is the potential energy of a new (pseudo)-scalar field, $\phi$, which is extremely light ($m \sim 10^{-27}$ eV) and thus initially frozen on its potential due to Hubble friction (i.e., frozen at a single value of $\phi$), with an effective equation of state $w_{\phi} = -1$.  When $H \sim m$, which occurs at $z \sim z_{\rm eq}$ for such light masses, the field begins to roll and eventually oscillate around the minimum of its potential, at which point its contribution to the cosmic energy budget redshifts away.  The rate at which its energy density redshifts while the field oscillates around its minimum is crucial to the success of the model: it must redshift away faster than matter (i.e., faster than $a^{-3}$) so as to avoid spoiling late-time cosmology.  This puts constraints on the form of the potential, as discussed in Sec.~\ref{sec:theory}; satisfying these requires some level of fine-tuning.  In addition, accounting properly for the behavior of perturbations in the EDE field $\phi$, which are generated as soon as the field starts to roll and $w_{\phi}$ deviates from $-1$, is essential for the model to fit CMB data~\cite{Poulin:2018cxd}.

It was first shown in Ref.~\cite{Poulin:2018cxd} that this EDE model could potentially increase the value of $H_0$ inferred in fits to cosmological data, to values as high as $H_0 \approx 71$ km/s/Mpc.  However, Ref.~\cite{Hill:2020osr} pointed out that \emph{Planck} data alone do not prefer the existence of an EDE component, and that the inferred value of $H_0$ when fitting the model to \emph{Planck} still disagrees with that measured by SH0ES at $>3\sigma$.  In addition, Refs.~\cite{Hill:2020osr,Ivanov:2020ril,DAmico:2020ods} demonstrated that as a consequence of raising $H_0$ via the EDE model fit to CMB data, the value of the physical cold dark matter (CDM) density ($\Omega_c h^2$) also increases substantially, which leads to an increased difference between CMB and large-scale structure (LSS) measurements of the low-redshift amplitude of fluctuations, $\sigma_8$ (and its closely related counterpart $S_8 \equiv \sigma_8 (\Omega_m / 0.3)^{0.5}$).  The reason for the increase in $\Omega_c h^2$ in the CMB fit is straightforward, as elucidated in detail in Ref.~\cite{Vagnozzi:2021gjh}: the presence of the EDE leads to an enhanced early integrated Sachs-Wolfe (ISW) effect, which suppresses the growth of perturbations; this suppression is counteracted by an increase in the CDM density and a somewhat smaller increase in the scalar spectral index $n_s$.  The outcome is that the combination of \emph{Planck} CMB data and a variety of LSS data put tight constraints on the EDE model, and effectively restrict its ability to significantly increase the inferred value of $H_0$ (see, however, Refs.~\cite{Murgia:2020ryi,Smith:2020rxx}).  This result was recently further verified in the context of a broader class of EDE scenarios in Ref.~\cite{Gomez-Valent:2021cbe}.  Ref.~\cite{Boylan-Kolchin:2021fvy} also pointed out that EDE models yielding a significant increase in $H_0$ predict a best-fit age of the Universe that is in tension with (i.e., younger than) the ages of individual stars, stellar clusters, and ultra-faint dwarf galaxies.

In nearly all EDE analyses thus far, the only CMB data set considered has been that from \emph{Planck}.  It is worthwhile to investigate whether current conclusions regarding the EDE model are robust to the use of alternative CMB data.  This question motivates the analysis undertaken in this paper.  In particular, we consider the CMB power spectrum measurements in ACT Data Release 4 (DR4)~\cite{Choi2020,Aiola2020} as an alternative to \emph{Planck}.\footnote{While this paper was in collaboration review, Ref.~\cite{Schoneberg:2021qvd} presented EDE constraints derived from ACT DR4 + \emph{WMAP} + BAO + Pantheon supernova data (the latter in ``uncalibrated'' form, i.e., without the use of a SH0ES or TRGB anchor), which appear to agree with our findings.}  While not quite as statistically powerful as the full \emph{Planck} data set, the ACT DR4 data are nevertheless able to tightly constrain all $\Lambda$CDM parameters (apart from the optical depth $\tau$, which requires very large-scale EE power spectrum measurements).  The combination of ACT DR4 with \emph{WMAP} data, the latter of which primarily probe $\ell \lesssim 650$ (and primarily TT), forms a joint data set with statistical power approaching that of \emph{Planck}, with completely independent systematics.  A key result of Ref.~\cite{Aiola2020} was the demonstration that a blind analysis of ACT DR4 and \emph{WMAP} data yields $\Lambda$CDM cosmological parameters in excellent agreement with those of \emph{Planck} (see Appendix~\ref{app:ha} below for a minor update to the ACT DR4 $\Lambda$CDM results).  For example, within the $\Lambda$CDM model, the combination of ACT DR4 and \emph{WMAP} (TT,TE,EE + $\tau$ prior) yields $H_0 = 67.9 \pm 1.1$ km/s/Mpc (see Table~\ref{tab:comparison} in Appendix~\ref{app:ha}), while \emph{Planck} yields $H_0 = 67.27 \pm 0.60$ km/s/Mpc (TT,TE,EE + lowE)~\cite{Planck2018parameters}.

However, consistency of $\Lambda$CDM parameters does not necessarily guarantee consistency of parameters in the context of other cosmological models.  Indeed, we find in this work that ACT DR4 prefers moderately different parameters in the EDE model than those preferred by \emph{Planck}.  This preference strengthens with the inclusion of large-scale ($\ell_{\rm max} = 650$) \emph{Planck} TT data, which are used here as a proxy for \emph{WMAP} (for reasons of convenience described in Sec.~\ref{sec:data}).  Further including BAO data and \emph{Planck} CMB lensing data weakens the preference somewhat, in line with general arguments regarding sound-horizon-reduction scenarios~\cite{Jedamzik:2020zmd}, but nevertheless a non-zero amount of EDE with an accordingly higher value of $H_0$ is preferred.  This preference stands in contrast to the EDE results from \emph{Planck} data~\cite{Hill:2020osr}.  As we show below, the different EDE parameter preferences from ACT and \emph{Planck} are driven by the ACT DR4 TE and EE power spectrum data.\footnote{Some evidence of TE data driving ACT parameters toward slightly discordant values in $\Lambda$CDM was presented in Ref.~\cite{Aiola2020} (see their Fig.~14), while similar results in the context of a different EDE model were presented in Ref.~\cite{Lin:2020jcb}.}  Understanding whether these differences are physical or driven by uncharacterized systematic effects in the ACT data is of crucial importance.

We emphasize that our goal in this paper is \emph{not} to assess the global concordance of the EDE model with respect to all current cosmological data sets.  Instead, we focus solely on assessing the robustness of current EDE constraints to our choice of CMB data.  We use only a limited set of external data with very well-understood systematics in order to break parameter degeneracies and tighten constraints.  In particular, we do not include local measurements of $H_0$ or low-redshift LSS measurements of $S_8$, e.g., from galaxy weak lensing or full-shape galaxy power spectrum data.  An important reason to perform such an analysis without using local $H_0$ or low-redshift $S_8$ data is that one should first assess whether a new cosmological model fit to CMB data naturally yields significant parameter shifts compared to $\Lambda$CDM, or at least much broader parameter error bars.  For example, such a model should naturally yield a higher value of $H_0$ than that obtained when fitting $\Lambda$CDM to CMB data, without requiring a high $H_0$ from a local distance ladder probe to artificially increase the inferred $H_0$ in a joint analysis (see Refs.~\cite{Hill:2020osr,Vagnozzi:2021gjh} for discussions of this point).  Put simply, if one uses a local probe with a ``high'' value of $H_0$ in a joint data analysis, then one cannot subsequently compare the jointly-inferred $H_0$ to that probe on its own, as the inferences are clearly not independent.  Similar statements apply to $S_8$.

The remainder of this paper is organized as follows.  Sec.~\ref{sec:theory} briefly reviews the theory underlying the EDE scenario.  Sec.~\ref{sec:data} describes the data sets that are used in this analysis.  Sec.~\ref{sec:analysis} presents constraints on cosmological parameters in the EDE and $\Lambda$CDM models derived from these data sets.  Sec.~\ref{sec:discussion} summarizes our findings and prospects for upcoming data.  Appendix~\ref{app:ha} collects results related to the numerical accuracy of our theoretical calculations, including a minor update to the ACT DR4 $\Lambda$CDM parameter constraints, while Appendix~\ref{app:plots} contains additional posterior and residual plots.  For the busy reader, a concise summary of our main results is given in Sec.~\ref{subsec:summary}.


\section{Theory}
\label{sec:theory}

The basic physics of the EDE scenario has been described in detail in several previous works~(e.g.,~\cite{Poulin:2018cxd,Lin:2019qug,Smith:2019ihp,Agrawal:2019lmo,Alexander:2019rsc,Hill:2020osr,Ivanov:2020ril}), and thus we present only a brief overview in the following.  This scenario posits the existence of a new fundamental field that slightly accelerates the cosmic expansion rate just prior to recombination, contributing $\approx 10\%$ of the total energy density near matter-radiation equality.  This burst of acceleration acts to decrease the physical size of the sound horizon imprinted in the CMB (compared to its value in $\Lambda$CDM), thereby increasing the inferred value of $H_0$ from CMB power spectrum data.  In order to avoid phenomenological disruption to the late-time universe, the EDE field's energy density must rapidly redshift away after recombination.  This scenario can be realized by taking the EDE field to be a light scalar, which is initially frozen on its potential while $H \gg m$, where $m$ is the mass of the field.\footnote{We adopt natural units with $c=\hbar=k_B=1$ when discussing the theoretical background here.}  When $H \approx m$ (see, e.g., Eq.~(7) of Ref.~\cite{Smith:2019ihp} for a more precise criterion), the field begins to roll and eventually it oscillates around the minimum of its potential, the shape of which determines how rapidly the field's energy density redshifts away.

For such a field to significantly decrease the physical size of the sound horizon, we must have $H \approx m$ in the decade of scale factor evolution just prior to $z_* = 1100$ (the redshift of recombination), which implies $m \approx 10^{-27} - 10^{-28}$ eV.  Thus the field must be many orders of magnitude lighter than any in the Standard Model.  From a particle physics perspective, the only known example of such an extremely light field is the axion~\cite{Peccei:1977hh,Wilczek:1977pj,Weinberg:1977ma}.  However, the standard axion potential does not suffice for the EDE scenario, as its energy density redshifts as matter, and thus its presence would lead to conflicts with late-time cosmology.  This situation thus motivates the consideration of axion-like potentials of the form (e.g.,~\cite{Kamionkowski:2014zda}),
\be
V(\phi) = m^2 f^2 \left( 1 - \cos (\phi/f)\right)^n \, ,
\label{eq.V}
\ee
where $f$ is the so-called axion decay constant and $n$ is a power-law index.  The standard axion potential corresponds to $n=1$; in the EDE scenario the value of $n$ is not specified {\it a priori}, but phenomenological considerations require $n \geq 2$ (see discussion below).  The field evolves according to the Klein-Gordon equation:
\be
\ddot{\phi} + 3H\dot{\phi} + \frac{dV}{d\phi} = 0 \,,
\label{eq.KG}
\ee
where dots denote derivatives with respect to cosmic time.  From Eq.~\eqref{eq.KG}, it is clear that when $H \gg m$ (i.e., at early times), the Hubble friction term dominates and $\phi$ is not dynamical, thus rendering $V(\phi)$ an effective contribution to dark energy.  At late times, when $\phi$ is near the minimum of the potential, we have $V \sim \phi^{2n}$ and the oscillations of $\phi$ yield an effective equation of state $w_{\phi} = (n-1)/(n+1)$~\cite{PhysRevD.28.1243}.  Thus the EDE energy density decays away like radiation ($\propto a^{-4}$) for $n=2$, while for $n \rightarrow \infty$ it decays away like kinetic energy ($\propto a^{-6}$).  In contrast, if $n=1$ (the usual axion potential), then $w_\phi = 0$ near the potential minimum, i.e., the EDE energy density redshifts like that of matter; clearly this is not phenomenologically viable, and thus sets the restriction that $n \geq 2$ (for integer $n$).  We compute the evolution of the field $\phi$ at the background level by solving the Klein-Gordon equation.  We compute the evolution of perturbations $\delta \phi$ by solving the perturbed Klein-Gordon equation, without adopting an effective fluid approximation (such approximations have been argued to impact the numerical accuracy of such calculations~\cite{Agrawal:2019lmo}).

The potential in Eq.~\eqref{eq.V} involves significant fine-tuning: for integer $n$, one must fine-tune $n$ terms in the instanton expansion that generates the EDE potential to be very small (see discussion in Ref.~\cite{Hill:2020osr}).  For arbitrary real-valued $n$, an infinite number of terms must be fine-tuned.  Motivated by this, and by previous data analyses, we restrict our analysis to integer values of $n$ --- primarily $n=3$, which has been shown to match current data~\cite{Smith:2019ihp}.  Although constraints on this parameter are weak, values $n>5$ are disfavored~\cite{Smith:2019ihp}, thereby leaving $n = 2, 3$, or $4$ as reasonable integer values to analyze.  Thus, we also consider $n=2$ rather than $n=3$ as a test, and verify that our results are not sensitive to this choice.  

We choose to parametrize the model in terms of ``effective'' parameters, rather than the physical parameters appearing in the scalar field potential $(m, f)$.  These effective parameters are defined by the redshift $z_c$ at which the EDE makes its largest fractional contribution $f_{\rm EDE}$ to the total cosmic energy budget, $f_{\rm EDE}(z_c) = 8 \pi G \rho_{\rm EDE}(z_c)/(3 H^2(z_c))$.  For brevity, we will generally denote $f_{\rm EDE}(z_c) \equiv f_{\rm EDE}$.  As shown in Ref.~\cite{Hill:2020osr}, due to the highly nonlinear relation between $(m,f)$ and $(z_c, f_{\rm EDE})$, sampling the model using uniform priors on the physical scalar field parameters (or their logarithms) can yield noticeably different results than using uniform priors on the effective parameters.  In fact, a uniform prior on $f_{\rm EDE}$ corresponds to a prior on the decay constant $f$ that peaks at $f \approx 0.6 M_{\rm Pl}$ and asserts the existence of super-Planckian decay constants $f > M_{\rm Pl}$~\cite{Hill:2020osr}.  Nevertheless, as the closer connection of $(z_c, f_{\rm EDE})$ to observables leads to more efficient convergence of the posterior sampling, and for consistency with previous works, we adopt the effective parametrization.  The final parameter that completes the EDE scenario is the initial field displacement $\theta_i \equiv \phi_i/f$.  Thus the model contains three additional free parameters beyond those in $\Lambda$CDM.

We do not consider any extensions of $\Lambda$CDM beyond the EDE parameters, although it is worth keeping in mind that some well-motivated extensions (e.g., massive neutrinos) lead to physical effects that are partially degenerate with those of EDE in the CMB and LSS observables.  Following the \emph{Planck} convention~\cite{Planck2018parameters}, we set the sum of the neutrino massses $\sum m_{\nu} = 0.06$ eV, with one massive eigenstate and two massless eigenstates.  We set the effective number of relativistic species $N_{\rm eff} = 3.046$.  The primordial helium fraction is determined via Big Bang nucleosynthesis.  Nonlinear corrections to the linear matter power spectrum are computed via the {\tt Halofit} prescription~\cite{Smith:2002dz,Takahashi2012}, although this has very little impact on our results, as the observables that we consider are all dominated by linear modes.


\section{Data Sets}
\label{sec:data}

We primarily focus on CMB data in this analysis, but also consider a limited collection of additional cosmological data sets, as detailed below.

\textbf{ACT DR4 CMB}:
Our primary data set, used in all analyses presented here, is comprised of the ACT DR4 CMB power spectra~\cite{Choi2020}.  We consider the multifrequency TT, TE, and EE power spectra from both the ``wide'' and ``deep'' patches analyzed in ACT DR4, which are derived from ACT data collected through 2016.  We use the {\tt actpollite\_dr4} likelihood implemented in {\tt pyactlike},\footnote{\url{https://github.com/ACTCollaboration/pyactlike/}} which also uses data from the ACT MBAC DR2 data set~\cite{Das:2013zf}.  In this likelihood, the contributions to the ACT power spectra from non-CMB foregrounds have already been marginalized over following the procedure described in~\cite{Choi2020,Aiola2020}, yielding a set of CMB-only bandpowers whose covariance includes the effects of noise, foregrounds, beam uncertainty, and calibration uncertainty.  The likelihood for the ACT DR4 data thus depends on only the cosmological parameters (six for $\Lambda$CDM or nine for EDE) and one nuisance parameter, the overall polarization efficiency, $y_p$.  Unless stated otherwise, we use the full multipole range for ACT DR4 given in~\cite{Choi2020,Aiola2020}: $600 \leq \ell \leq 4125$ (TT) and $350 \leq \ell \leq 4125$ (TE/EE).

Due to atmospheric and other $1/f$ noise, the ACT data do not probe the largest angular scales on the sky; thus, the optical depth $\tau$ is not independently constrained well by the ACT data, as it is primarily determined by the low-$\ell$ EE power spectrum.  In all analyses, we thus impose a Gaussian prior $\tau = 0.065 \pm 0.015$, identical to that used in Ref.~\cite{Aiola2020}.

\textbf{\emph{Planck} 2018 CMB}:
To complement the ACT power spectra, we consider the \emph{Planck} PR3 (2018) multifrequency TT, TE, and EE power spectra~\cite{Planck2018likelihood,Planck2018overview,Planck2018parameters}.  The \emph{Planck} data cover a wider sky area than ACT, while also possessing lower large-scale noise due to observing from space rather than through Earth's atmosphere, both of which yield more precise low-$\ell$ data for \emph{Planck} compared to ACT.  However, the white noise levels and angular resolution of \emph{Planck} are worse than those of ACT.\footnote{\emph{Planck} is cosmic-variance (CV)-limited in the TT power spectrum at roughly $\ell \lesssim 1600$.}  Thus, the ACT data can probe the TT power spectrum at higher $\ell$ than \emph{Planck}, and also probe the TE and EE power spectra with better precision over a wide range of scales owing to ACT's lower noise level.  The two data sets are thus highly complementary.  When combining ACT DR4 with the full \emph{Planck} data set, we restrict the $\ell$ range of the ACT data so as to ensure that information is not double-counted, following the method developed in Ref.~\cite{Aiola2020}.  In particular, for this combination, we set $\ell_{\rm min}^{\rm ACT,TT} = 1800$ for the minimum multipole used in the ACT TT likelihood; no minimum multipole cut is imposed in the ACT TE or EE likelihoods.  These multipole cuts were explicitly validated in Ref.~\cite{Aiola2020}.

Note that we do not include low-$\ell$ EE data from \emph{Planck}, choosing instead to constrain the optical depth $\tau$ via a Gaussian prior following~\cite{Aiola2020}, as described above.  The \emph{Planck} likelihoods used here thus comprise the {\tt Plik\_HM} high-$\ell$ likelihood, which spans $30 \leq \ell \leq 2508$ (TT) and $30 \leq \ell \leq 1996$ (TE/EE), and the {\tt Commander} low-$\ell$ TT likelihood, which covers $2 \leq \ell \leq 29$.  To allow straightforward extraction of various $\ell$-range subsets of the data (see below), we use the full {\tt Plik} likelihood, rather than the ``{\tt Plik\_lite}'' version.

We extract a \emph{WMAP}-like subset of the \emph{Planck} data for use in some combined analyses with ACT.\footnote{We use this \emph{WMAP}-like subset of the \emph{Planck} data rather than \emph{WMAP} data itself because the {\tt Cobaya} software package does not include a \emph{WMAP} likelihood.}  This combination allows the use of the full ACT DR4 data set, with no minimum multipole cuts in TT, TE, or EE (analogous to the combined ACT+\emph{WMAP} analyses in~\cite{Aiola2020}).  We define this subset of the \emph{Planck} data by considering only the TT power spectrum, and including data up to a maximum multipole such that the derived $\Lambda$CDM parameters have error bars matching those for the \emph{WMAP} likelihood used in~\cite{Aiola2020}.  (The $\tau$ prior mentioned above is used in all analyses.)  We find that $\ell_{\rm max} = 650$ satisfies this requirement, and also yields central values of the $\Lambda$CDM parameters very close to those from \emph{WMAP}, as previously found elsewhere~\cite{Planck2016shifts}.  Indeed, it has been shown that the \emph{Planck} and \emph{WMAP} data agree very closely over this multipole range at the level of the CMB power spectrum data points themselves (see Fig.~48 of Ref.~\cite{Planck2015likelihood}).  For brevity, we will sometimes refer to this \emph{Planck} data subset as ``\emph{PlanckTT650}''.

\textbf{\emph{Planck} 2018 CMB Lensing}:
In addition to the \emph{Planck} primary CMB power spectra, we consider the \emph{Planck} reconstructed CMB lensing potential power spectrum~\cite{Planck2018lensing}.  The CMB lensing power spectrum probes structure over a broad range of redshifts, peaking at $z \approx 1-2$.  The scale cuts used in the \emph{Planck} lensing power spectrum likelihood include modes with $8 \leq L \leq 400$, for which non-linear corrections are negligible~\cite{Lewis:2006fu}.  The lensed CMB TT/TE/EE power spectra also carry some lensing-related information, but the reconstructed lensing potential power spectrum is a more sensitive probe, with the \emph{Planck} detection significance reaching $40\sigma$~\cite{Planck2018lensing}.

\textbf{BAO}:
As a probe of the (relative) cosmic expansion history at low redshifts, we consider baryon acoustic oscillation (BAO) data from the SDSS DR7 main galaxy sample~\cite{Ross:2014qpa} at $z = 0.15$, the 6dF galaxy redshift survey~\cite{2011MNRAS.416.3017B} at $z = 0.106$, and from the SDSS BOSS DR12~\cite{Alam:2016hwk} LOWZ and CMASS galaxy samples at $z = 0.38$, $0.51$, and $0.61$.  These BAO data do not provide absolute distances to these redshifts, but rather relative distances normalized to the sound horizon at the end of the baryon drag epoch.  To be conservative, we do not consider redshift-space distortion or full-shape galaxy power spectrum data in this work.

\textbf{Local $H_0$ Data:}
We do not include any local measurements of $H_0$ in the data sets analyzed here, as our goal is to focus on the inference of $H_0$ from indirect probes within non-standard cosmological models.  For visualization purposes only, we display in some plots the most recent constraint on $H_0$ inferred via the TRGB-calibrated SNIa distance ladder, $H_0 = 69.8 \pm 0.6 \, ({\rm stat.}) \pm 1.6 \, ({\rm syst.})$ km/s/Mpc~\cite{Freedman:2021ahq}, and the most recent constraint inferred via the Cepheid-calibrated SNIa distance ladder from SH0ES, $H_0 = 73.2 \pm 1.3$ km/s/Mpc~\cite{Riess:2020fzl}.

\textbf{$S_8$ Data:}
To be conservative and concentrate our focus on the CMB, we do not consider low-redshift galaxy lensing or full-shape galaxy power spectrum data in this work; the CMB lensing power spectrum is the only direct probe of the low-redshift amplitude of structure ($\sigma_8$) considered here.  For visualization purposes only, we display in some plots the most recent constraint on $S_8 \equiv \sigma_8 (\Omega_m/0.3)^{0.5}$ inferred from the DES-Y3 ``3$\times$2pt'' analysis (galaxy clustering + galaxy-galaxy lensing + cosmic shear), $S_8 = 0.776 \pm 0.017$~\cite{DES:2021wwk}.

We consider four primary data set combinations: (i) ACT DR4 alone; (ii) ACT DR4 + \emph{Planck} 2018 TT ($\ell_{\rm max} = 650$); (iii) ACT DR4 + \emph{Planck} 2018 TT ($\ell_{\rm max} = 650$) + \emph{Planck} 2018 CMB lensing + BAO; (iv) ACT DR4 + \emph{Planck} 2018 TT+TE+EE.  The motivation for considering ACT DR4 alone is to provide an independent test of the EDE constraints derived from \emph{Planck} primary CMB power spectra in Ref.~\cite{Hill:2020osr}.  However, since ACT does not measure large-scale TT modes due to atmospheric noise, and since the constraining power of ACT is not as strong as that of \emph{Planck}, it is useful to add a complementary data set.  Thus we combine ACT with the large-scale \emph{Planck} TT data ($\ell < 650$), with the latter acting as a \emph{WMAP}-like data set that is independent of ACT and fills in missing modes in the power spectra.  The same strategy was used in Ref.~\cite{Aiola2020} to motivate the combination of ACT and \emph{WMAP} as a joint data set with constraining power similar to that of \emph{Planck}.  To further break parameter degeneracies while sticking to well-understood, linear probes, we then add \emph{Planck} CMB lensing data and BAO measurements to the joint data set.  Finally, we consider the full \emph{Planck} primary CMB data alone (using results from Ref.~\cite{Hill:2020osr}) and the combination of these data with ACT DR4, with associated multipole cuts to ensure that no power spectrum modes are double-counted.  This combination is largely driven by, and consistent with, \emph{Planck} alone.


\section{Analysis}
\label{sec:analysis}

We perform cosmological parameter inference by sampling from the parameter posterior distributions using Markov chain Monte Carlo (MCMC) methods.  We use the publicly available MCMC code {\tt Cobaya}~\cite{Torrado:2020dgo},\footnote{\url{https://github.com/CobayaSampler/cobaya}} which implements the Metropolis-Hastings algorithm~\cite{LewisBridle2002,Lewis2013,Neal2005}.  We assess convergence of the MCMC chains using the Gelman-Rubin~\cite{Gelman:1992zz} criterion, requiring $R-1 < 0.03$.  To obtain the best-fit parameter values, we apply the ``BOBYQA'' likelihood maximization method using the maximum a posteriori point of the converged MCMC chains as a starting point, as implemented in {\tt Cobaya}~\cite{Powell2009,Cartis2018a,Cartis2018b}.  We obtain parameter confidence intervals from the MCMC chains using {\tt GetDist}~\cite{Lewis:2019xzd}.\footnote{\url{https://github.com/cmbant/getdist}}  Our MCMC chains and {\tt Cobaya} input files are publicly available.\footnote{The chains are available at \url{https://lambda.gsfc.nasa.gov/product/act/actpol_prod_table.html} or \url{https://users.flatironinstitute.org/~chill/H21_data/}.}

The parameter posteriors in most EDE model fits are non-Gaussian.  To determine 68\% marginalized confidence intervals from the asymmetric posteriors, we compute the interval between two points with the highest equal marginalized probability density (the ``credible interval''); this is the default method in {\tt GetDist}.  One can also compute an interval such that each tail of the associated two-tail limit contains 16\% of the samples (the ``equal-tail interval'').  For Gaussian posteriors, these methods produce identical results, but for non-Gaussian (namely, skewed) posteriors the results can differ.  Previous EDE analyses have considered both approaches~\cite{Hill:2020osr,Ivanov:2019hqk}, due to the skewed EDE parameter posteriors.  Here, we adopt solely the first approach.  We quote the posterior mean as the central value of credible intervals.  For upper limits, we quote the one-sided 95\% confidence level (CL), the point where the cumulative probability distribution function reaches 0.95.

We compute theoretical predictions for the $\Lambda$CDM and EDE models using {\tt CLASS\_EDE}~\cite{Hill:2020osr},\footnote{\url{https://github.com/mwt5345/class_ede}} a modified version of the Einstein-Boltzmann code {\tt CLASS}~\cite{CLASS2011Overview,Blas2011}.\footnote{\url{http://class-code.net}}  We find that the standard precision settings for {\tt CLASS} do not yield sufficiently accurate theoretical predictions for the CMB TT, TE, and EE power spectra in the ACT DR4 likelihood.  This is to be expected, as the default {\tt CLASS} precision settings are calibrated to yield sufficient accuracy for the analysis of \emph{Planck} data, which extends only to $\ell = 2500$, as compared to $\ell = 4325$ for ACT.\footnote{Note that ACT measures secondary anisotropies out to much higher multipoles, but the CMB power spectrum information rapidly falls off above $\ell \approx 4500$.}  Thus, we run {\tt CLASS} with all CMB-relevant precision parameters set to increased levels, and evaluate the theoretical spectra to multipoles well beyond our actual $\ell_{\rm max}$ cut, a choice that is necessary to obtain precise results for the effect of lensing on the CMB power spectra.\footnote{In detail, we set all CMB-relevant {\tt CLASS} precision parameters to their values in the file {\tt cl\_ref.pre} that is distributed with {\tt CLASS}, and also set {\tt l\_max\_scalars} = 11000.}  We verify the accuracy of our $\Lambda$CDM TT/TE/EE power spectrum calculations by comparing to results from {\tt CAMB}~\cite{Lewis:1999bs}\footnote{\url{http://camb.info}} obtained when running this code with all CMB- and lensing-related precision settings set to substantially increased values.  We find that the two codes agree extremely well ($\ll 1$\% difference) at all $\ell < 4500$ when run in these high-precision configurations, which serves to validate the accuracy of our theoretical calculations.  Our $\Lambda$CDM results reproduce those of Ref.~\cite{Choi2020} to within $< 0.1\sigma$  when adopting the same Boltzmann precision settings.  Appendix~\ref{app:ha} presents further investigation related to the theoretical precision issue for ACT DR4, including updated $\Lambda$CDM results.  Ref.~\cite{McCarthy2021} investigates implications for upcoming CMB experiments if this precision issue (amongst other lensing-related effects) is not treated properly.

We sample the parameter space spanned by $\{ f_{\rm EDE}, \mathrm{log}_{10}(z_c), \theta_i, \ln(10^{10} A_\mathrm{s}), n_s, 100\theta_s, \Omega_b h^2, \Omega_c h^2, \tau, \\ y_p \}$.  We use broad, uninformative priors on the standard $\Lambda$CDM parameters: the physical baryon density ($\Omega_b h^2$), the physical CDM density ($\Omega_c h^2$), the angular size of the sound horizon ($100\theta_s$), the logarithm of the amplitude of the power spectrum of scalar perturbations ($\ln(10^{10} A_\mathrm{s})$), the spectral index of the power spectrum of scalar perturbations ($n_s$), and the Thomson-scattering optical depth ($\tau$).  For the ACT polarization efficiency, we use a prior range $y_p \in [0.9,1.1]$, which is much broader than the data-determined constraint on this parameter~\cite{Choi2020, Aiola2020}.  When including \emph{Planck} data, we also sample over the full set of nuisance parameters included in the \emph{Planck} likelihoods~\cite{Planck2018likelihood}.  For the EDE parameters, we adopt the following uniform priors: $f_{\rm EDE} \in [0.001, 0.5]$, $\mathrm{log}_{10}(z_c) \in [3, 4.3]$, $\theta_i \in [0.1, 3.1]$.  Note that the choice of the prior range for $\mathrm{log}_{10}(z_c)$ is important; if this is extended to arbitrarily high redshifts, then a vast region of parameter space is opened up in which $f_{\rm EDE}$ can take on large values while having no impact on the CMB or other observables, as the EDE energy density has already decayed away long before recombination.  In addition, the small difference between the physical range for $\theta_i$ ($0 \leq \theta_i \leq \pi$) and the prior range used here is due to numerical challenges that arise for $\theta_i$ values near the boundaries.  The small restriction has negligible impact on our results, and matches the same choice made in many previous works studying this model~\cite{Poulin:2018cxd,Smith:2019ihp,Hill:2020osr,Ivanov:2020ril}.

  \begin{table*}[ht!]
Constraints on EDE ($n=3$)  \\
  \centering
  \begin{tabular}{|c|c|c|c|c|c|}
    \hline\hline Parameter & \hspace{0cm} \begin{tabular}[t]{@{}c@{}}ACT DR4 \\ TT+TE+EE, $\tau$\end{tabular} & \hspace{0cm}\begin{tabular}[t]{@{}c@{}}ACT DR4\\ TT+TE+EE, \\  \emph{Planck} 2018 TT \\ ($\ell_{\rm max} = 650$), $\tau$\end{tabular} & \hspace{0cm}\begin{tabular}[t]{@{}c@{}}ACT DR4\\ TT+TE+EE, \\ \emph{Planck} 2018 TT \\ ($\ell_{\rm max} = 650$), \\ \emph{Planck} 2018 lensing, \\ BAO, $\tau$\end{tabular} &
    \hspace{0cm}\begin{tabular}[t]{@{}c@{}}\emph{Planck} 2018 \\ TT+TE+EE \\ (from Ref.~\cite{Hill:2020osr})\end{tabular} & 
    \hspace{0cm}\begin{tabular}[t]{@{}c@{}}ACT DR4 \\ TT+TE+EE, \\ \emph{Planck} 2018 \\ TT+TE+EE \\ (no low-$\ell$ EE), $\tau$ \end{tabular}
    \\\hline \hline

    {\boldmath$f_\mathrm{EDE} $} & $0.142^{+0.039}_{-0.072}$ & $ 0.129^{+0.028}_{-0.055} $ & $ 0.091^{+0.020}_{-0.036}$ & $ < 0.087 $ & $ < 0.124$\\

    {\boldmath$\mathrm{log}_{10}(z_c)$} & $< 3.70$ & $< 3.43$ & $< 3.36$ & $3.66^{+0.24}_{-0.28}$ & $3.54^{+0.28}_{-0.20}$ \\

    {\boldmath$\theta_i$} & $> 0.24  $ &  $ < 2.89 $ &  $ < 2.82 $ &  $ > 0.36 $&  $ > 0.51  $\\

    {\boldmath$\Omega_\mathrm{c} h^2$} & $0.1307^{+0.0054}_{-0.0120}$ & $0.1291^{+0.0051}_{-0.0098}$ & $0.1286^{+0.0027}_{-0.0063}$ & $0.1234^{+0.0019}_{-0.0038}$ & $0.1244^{+0.0025}_{-0.0051}$\\

    \hline
    
    $H_0 \, [\mathrm{km/s/Mpc}]$ & $74.5^{+2.5}_{-4.4}$ & $74.4^{+2.2}_{-3.0}$ & $70.9^{+1.0}_{-2.0}$ & $68.29^{+0.73}_{-1.20}$ & $69.17^{+0.83}_{-1.70}$\\

    $\Omega_m$ & $0.276^{+0.020}_{-0.023}$ & $0.274\pm 0.017$ & $0.3000\pm 0.0072$ & $0.3145\pm 0.0086 $&  $0.3084\pm 0.0084$\\
    
    $\sigma_8$ & $0.831^{+0.027}_{-0.043}$ & $0.827^{+0.029}_{-0.035}$ & $0.829^{+0.013}_{-0.021}$ & $0.820^{+0.009}_{-0.013}$&  $0.838^{+0.013}_{-0.015}$\\
    
    $S_8$ & $0.796 \pm 0.049$ & $0.791^{+0.040}_{-0.046}$ & $0.828^{+0.015}_{-0.018}$ & $0.839 \pm 0.018$&  $0.850 \pm 0.017 $\\
    
    \hline
  \end{tabular} 
  \caption{Marginalized constraints on key cosmological parameters in the EDE model (with power-law index $n=3$ held fixed in Eq.~\eqref{eq.V}) from ACT DR4 primary CMB data (TT+TE+EE), \emph{Planck} 2018 primary CMB data (TT+TE+EE), including a \emph{WMAP}-like subset restricted to only the large-scale ($\ell_{\rm max} = 650$) TT data; \emph{Planck} 2018 CMB lensing data; and BAO data from 6dF, SDSS DR7, and SDSS DR12 (BOSS).  Sampled parameters are shown in bold.  Upper and lower limits are given at 95\% CL.  The best-fit parameter values for these analyses are given in Sec.~\ref{sec:analysis}.  A Gaussian prior on $\tau$ is applied in all analyses, apart from the \emph{Planck}-only results, which are taken from Ref.~\cite{Hill:2020osr}.}
  \label{table:EDE-params-full}
\end{table*}

Constraints on the EDE model considered here derived from primary CMB anisotropy data alone were first presented in Ref.~\cite{Hill:2020osr}, which we briefly summarize.  Their analysis reported no evidence for the existence of EDE in the \emph{Planck} 2018 TT, TE, and EE power spectra, leading to an upper limit $f_{\rm EDE} < 0.087$ (95\% CL).  This limit is below the values generally required to raise the CMB-inferred $H_0$ to a level consistent with the direct SH0ES measurement, roughly $f_{\rm EDE} \approx 0.10-0.12$ for $z_c \sim z_{\rm eq}$~(e.g.,~\cite{Poulin:2018cxd,Smith:2019ihp,Hill:2020osr,Ivanov:2020ril}).  The central value of $H_0$ in the fit to \emph{Planck} data increased by roughly $1.5\sigma$ in EDE as compared to $\Lambda$CDM, along with a large increase in the error bar: $H_0^{\rm EDE} = 68.29^{+0.73}_{-1.20}$ km/s/Mpc vs. $H_0^{\Lambda \mathrm{CDM}} = 67.29 \pm 0.59$ km/s/Mpc.  Nevertheless, the difference between the \emph{Planck}-inferred and SH0ES-inferred $H_0$ values persisted at $3.3\sigma$ significance (using the latest SH0ES value~\cite{Riess:2020fzl}).  In addition, to counteract the EDE-induced early ISW effect in the CMB, the physical CDM density $\Omega_c h^2$ increased substantially in the EDE fit to \emph{Planck}, as compared to $\Lambda$CDM.  This led to a slight increase in $S_8$, heightening the moderate discrepancy with low-redshift structure measurements.  For a detailed discussion of the \emph{Planck}-only constraints and parameter posterior plots, see Ref.~\cite{Hill:2020osr}.  We use the MCMC chains from Ref.~\cite{Hill:2020osr} when presenting \emph{Planck}-only results in the analysis below.\footnote{The chains are publicly available at \url{http://users.flatironinstitute.org/~chill/H20_data/} . }

Despite the non-preference for EDE in \emph{Planck} primary CMB data, it is of interest to assess the robustness of this conclusion to the choice of CMB data set.  It has been shown that ACT DR4 and \emph{Planck} primary CMB constraints agree for the $\Lambda$CDM parameters, formally consistent at $2.5\sigma$ (see Appendix~\ref{app:ha} for this analysis, updated slightly from Ref.~\cite{Aiola2020}).  The ACT DR4 and \emph{WMAP} constraints also agree for $\Lambda$CDM, formally consistent at $2.4\sigma$ (see Appendix~\ref{app:ha})~\cite{Aiola2020}.  In this work, we investigate whether the same holds true for the EDE parameters.

\subsection{Summary of Main Results}
\label{subsec:summary}

We consider the following data set combinations in this work:
\begin{itemize}
    \item Sec.~\ref{subsec:ACT_alone}: ACT DR4 TT+TE+EE
    \item Sec.~\ref{subsec:ACT_P18TTlmax650}: ACT DR4 TT+TE+EE + \emph{Planck} 2018 TT ($\ell_{\rm max} = 650$)
    \item Sec.~\ref{subsec:ACT_P18TTlmax650_Lens_BAO}: ACT DR4 TT+TE+EE + \emph{Planck} 2018 TT ($\ell_{\rm max} = 650$) + \emph{Planck} 2018 CMB lensing + BAO
    \item Sec.~\ref{subsec:ACT_P18full}: ACT DR4 TT+TE+EE + \emph{Planck} 2018 TT+TE+EE (no $\ell < 30$ EE)
\end{itemize}
In all analyses, we impose a Gaussian prior on the optical depth ($\tau = 0.065 \pm 0.015$) following Ref.~\cite{Aiola2020}, as described in Sec.~\ref{sec:data}.  For comparison, we also include results obtained for \emph{Planck} 2018 TT+TE+EE data alone, which are taken from Ref.~\cite{Hill:2020osr}.\footnote{Note that the results from Ref.~\cite{Hill:2020osr} use the \emph{Planck} low-$\ell$ EE data to constrain $\tau$, rather than imposing the Gaussian prior used here.  However, this has negligible impact on the EDE parameters, as we verify in Sec.~\ref{subsec:ACT_P18full}.}

\begin{figure*}[!tp]
\includegraphics[width=\textwidth]{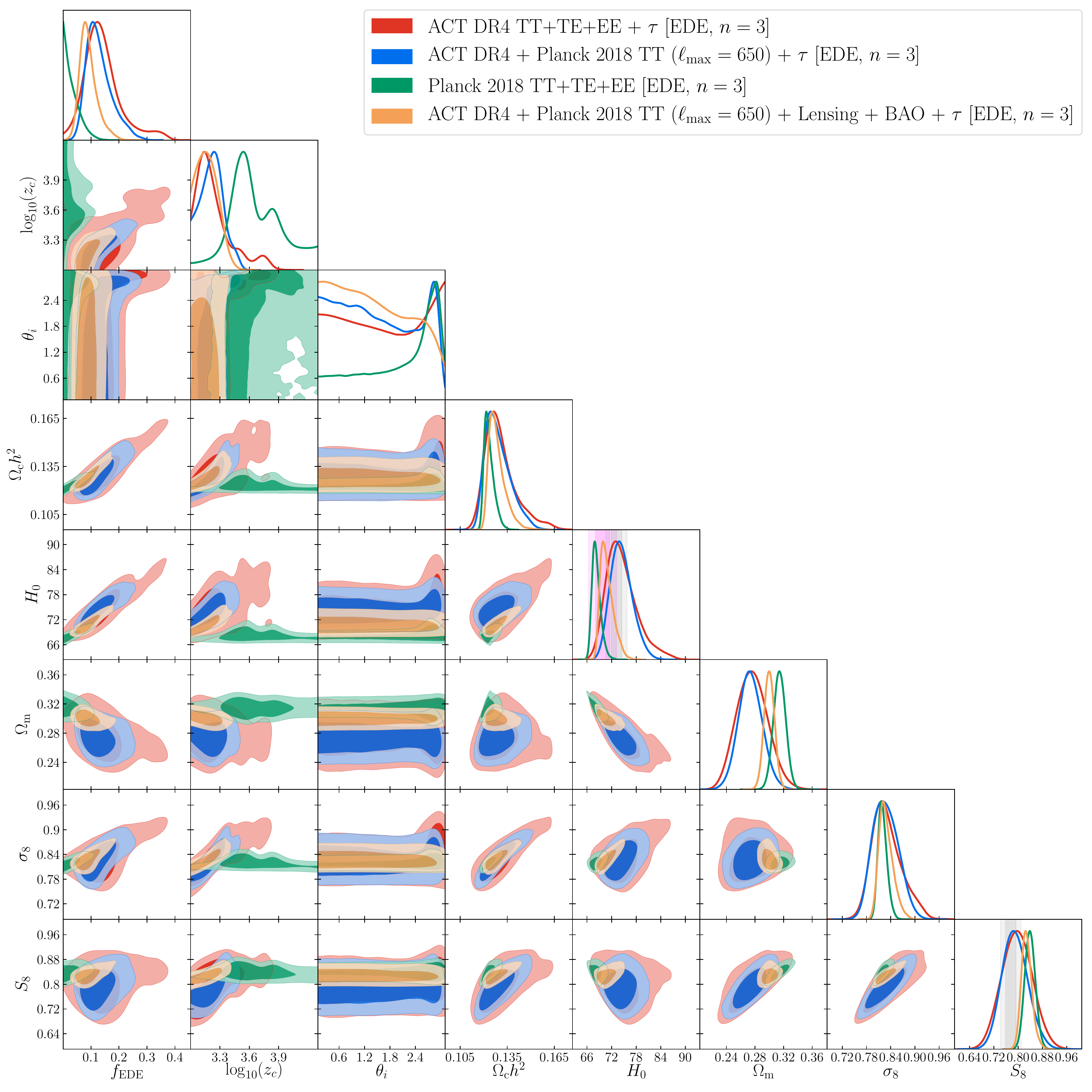}
\caption{Marginalized posteriors for the EDE parameters and a subset of other parameters of interest in fits to ACT DR4 TT+TE+EE data (red), ACT DR4 combined with large-scale ($\ell_{\rm max}=650$) \emph{Planck} 2018 TT data (blue), these same data with \emph{Planck} 2018 CMB lensing and BAO data included (orange), and the full \emph{Planck} 2018 TT+TE+EE data on their own (green, from Ref.~\cite{Hill:2020osr}).  All analyses, apart from \emph{Planck}-alone in green, impose a Gaussian prior on $\tau$, as discussed in Sec.~\ref{sec:data}.  The EDE potential power-law index $n=3$ in Eq.~\eqref{eq.V} is held fixed.  The vertical grey and magenta bands in the $H_0$ panel show the latest SH0ES~\cite{Riess:2020fzl} and TRGB~\cite{Freedman:2021ahq} constraints, respectively.  The vertical grey band in the $S_8$ panel shows the DES-Y3 constraint~\cite{DES:2021wwk}.}
\label{fig:EDE_ACT_PlanckTTlmax650_BAO_Lens}
\end{figure*}

\begin{figure*}[!tp]
\includegraphics[width=\textwidth]{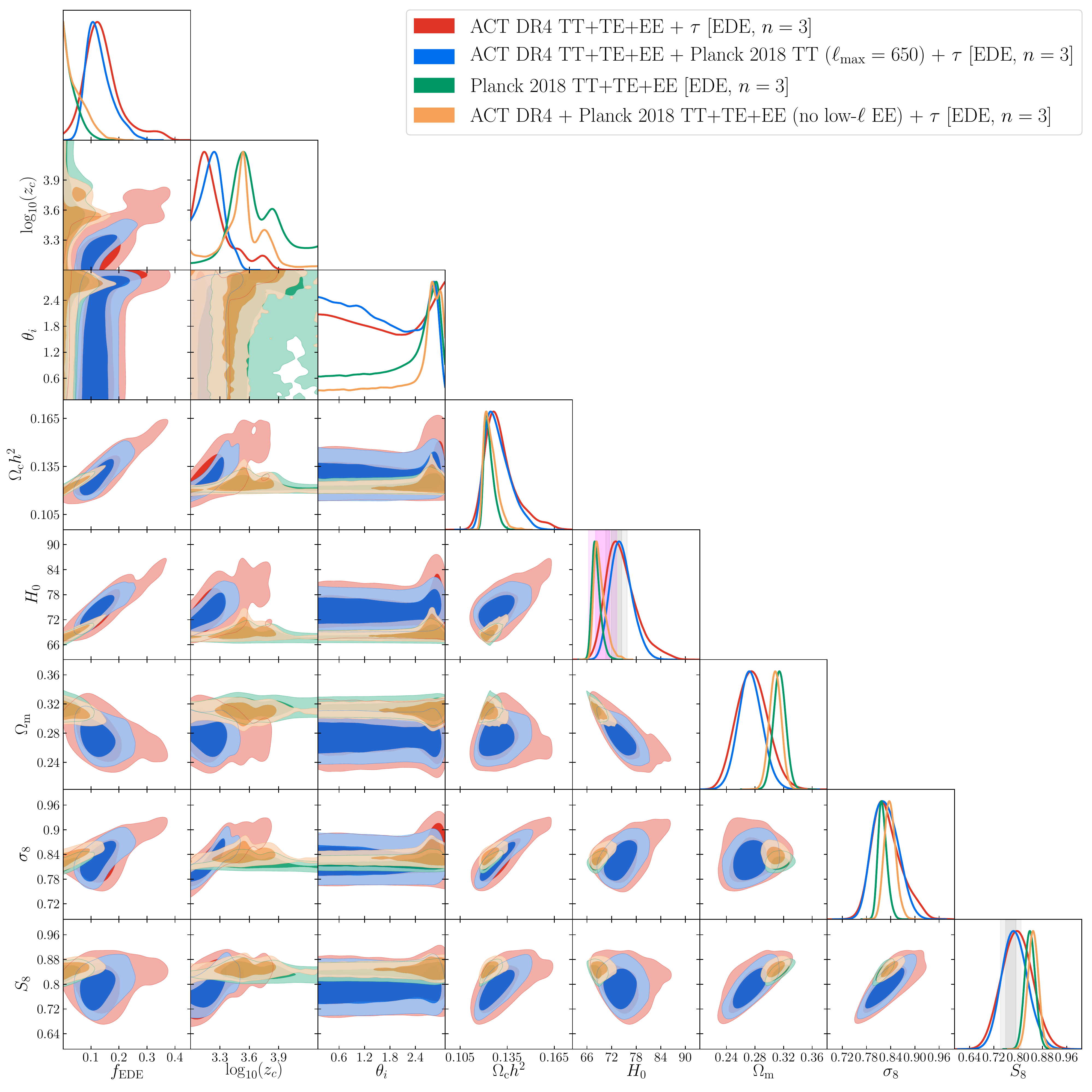}
\caption{Marginalized posteriors for the EDE parameters and a subset of other parameters of interest in fits to various data sets.  The red, blue, and green contours show the same data set combinations as in Fig.~\ref{fig:EDE_ACT_PlanckTTlmax650_BAO_Lens}, while the orange contours show the full \emph{Planck} 2018 TT+TE+EE data combined with ACT DR4 TT+TE+EE.  The low-$\ell$ \emph{Planck} EE data are excluded in lieu of the $\tau$ prior used elsewhere, for consistency.  The EDE potential power-law index $n=3$ in Eq.~\eqref{eq.V} is held fixed.  The vertical grey and magenta bands in the $H_0$ panel show the latest SH0ES~\cite{Riess:2020fzl} and TRGB~\cite{Freedman:2021ahq} constraints, respectively.  The vertical grey band in the $S_8$ panel shows the DES-Y3 constraint~\cite{DES:2021wwk}.}
\label{fig:EDE_ACT_Planckfull}
\end{figure*}

The main results of this work are summarized in Table~\ref{table:EDE-params-full} and Figs.~\ref{fig:EDE_ACT_PlanckTTlmax650_BAO_Lens} and~\ref{fig:EDE_ACT_Planckfull}, which present the marginalized posteriors and best-fit values for the EDE parameters and a subset of other parameters of interest, including $H_0$ and $S_8$.  Numerical results for other parameters in these analyses are presented in Tables~\ref{table:params-ACT-DR4},~\ref{table:params-ACT-DR4-P18TTlmax650},~\ref{table:params-ACT-DR4-P18TTlmax650-Lens-BAO}, and~\ref{table:params-ACT-DR4-P18full} in the subsections below, while additional posterior plots are presented in Appendix~\ref{app:plots}.  The $\chi^2$ values for the best-fit EDE and $\Lambda$CDM models to each combination of data sets are collected in Tables~\ref{table:chi2_ACT_alone},~\ref{table:chi2_ACT_P18TTlmax650},~\ref{table:chi2_ACT_P18TTlmax650_Lens_BAO}, and~\ref{table:chi2_ACT_P18full}, which we use to assess which model the data prefer.

We summarize the main takeaways from our analysis in the following.  The ACT DR4 data alone (red contours in Figs.~\ref{fig:EDE_ACT_PlanckTTlmax650_BAO_Lens} and~\ref{fig:EDE_ACT_Planckfull}) allow a non-negligible amount of EDE, but the uncertainties are too large to yield a clear preference:
\be
\label{eq:ACT_fEDE}
f_{\rm EDE} = 0.142^{+0.039\,+0.15}_{-0.072\,-0.13} \,\,\, (68\%/95\% \,\, {\rm CL}) \nonumber
\ee
The inferred value of $H_0$ increases compared to that in $\Lambda$CDM (see Table~\ref{table:EDE-params-full}), but the error bars increase by a factor of $2-3$ as well, thus rendering comparisons to other data sets generally uninformative: $H_0 = 74.5^{+2.5}_{-4.4}$ km/s/Mpc.  The EDE model yields a slightly better fit to the ACT DR4 data than $\Lambda$CDM, with $\Delta \chi^2 = -8.7$.  Accounting for the additional parameters in the EDE scenario, this corresponds to a $2.1\sigma$ preference, which is not significant.  This preference, albeit weak, is driven entirely by residuals in the lowest seven multipole bins of the ACT wide-patch EE data (see Figs.~\ref{fig:ACT_alone_residuals} and~\ref{fig:ACT_alone_wide_residuals}).

Including large-scale ($\ell_{\rm max} = 650$) \emph{Planck} TT data in the analysis (blue contours in Figs.~\ref{fig:EDE_ACT_PlanckTTlmax650_BAO_Lens} and~\ref{fig:EDE_ACT_Planckfull}) does not significantly change the central values of the ACT-only results, but tightens error bars by breaking some parameter degeneracies, particularly involving $n_s$ and $\Omega_b h^2$.  In contrast to the EDE analysis of the full \emph{Planck} data alone~\cite{Hill:2020osr}, a non-zero amount of EDE is preferred in this fit:
\be
\label{eq:ACT_P18TTlmax650_fEDE}
f_{\rm EDE} = 0.129^{+0.028\, +0.099\, +0.14}_{-0.055\, -0.076\, -0.084} \,\,\, (68\%/95\%/99.7\% \,\, {\rm CL}) \nonumber
\ee
The inferred $H_0$ and $S_8$ values in the EDE fit to ACT DR4 + large-scale \emph{Planck} TT data are consistent with measurements from TRGB or SH0ES and DES-Y3, respectively, albeit with large error bars: $H_0 = 74.4^{+2.2}_{-3.0}$ km/s/Mpc and $S_8 = 0.791^{+0.040}_{-0.046}$.  In contrast to the EDE fit to \emph{Planck}~\cite{Hill:2020osr}, the fit here does not yield a large value of $S_8$ despite the anticipated increase in $\Omega_c h^2$ to counteract the EDE-induced early ISW effect, because the shift in $H_0$ is large enough that $\Omega_m$ does not have to increase (in fact it decreases compared to its value in the $\Lambda$CDM fit).  The EDE model yields a better fit to the ACT DR4 + large-scale \emph{Planck} TT data than $\Lambda$CDM, with $\Delta \chi^2 = -15.4$.  Accounting for the additional parameters in the EDE scenario, this corresponds to a $3.2\sigma$ preference, which is non-negligible.  The preference is driven by a residual difference in the ACT TE power spectrum that is coherent across a wide multipole range, as well as the same residuals in the lowest several EE multipole bins seen in the ACT-only analysis, both of which the EDE model can accommodate (see Fig.~\ref{fig:ACT_P18TTlmax650_residuals}).

Including \emph{Planck} CMB lensing and BAO data has a noticeable impact (orange contours in Fig.~\ref{fig:EDE_ACT_PlanckTTlmax650_BAO_Lens}).  These data require higher $\Omega_m$ than the value preferred in the EDE fits described thus far~\cite{Planck2018lensing}, which forces $H_0$ to decrease, $f_{\rm EDE}$ to decrease, and $S_8$ to increase.  However, the parameter uncertainties also decrease significantly, and non-zero EDE remains preferred:
\be
\label{eq:ACT_P18TTlmax650_BAO_Lens_fEDE}
f_{\rm EDE} = 0.091^{+0.020\, +0.069\, +0.11}_{-0.036\, -0.056\, -0.063} \,\,\, (68\%/95\%/99.7\% \,\, {\rm CL}) \nonumber
\ee
The inferred $H_0$ value in the EDE fit decreases due to the inclusion of the CMB lensing and BAO data as described above, but is nevertheless still higher than found in $\Lambda$CDM:
\be
\label{eq:ACT_P18TTlmax650_BAO_Lens_H0}
H_0 = 70.9^{+1.0}_{-2.0} \,\, {\rm km/s/Mpc} \,\,\, (68\% \,\, {\rm CL}) \nonumber
\ee
The inferred $S_8$ value in the EDE fit is forced higher by the CMB lensing and BAO data, returning near its $\Lambda$CDM-inferred value, which is roughly $2\sigma$ higher than the DES-Y3 result.  Even with the inclusion of the CMB lensing and BAO data, the EDE model remains a better fit to the combined data set than $\Lambda$CDM, with $\Delta \chi^2 = -12.7$.  Accounting for the additional parameters in the EDE scenario, this corresponds to a $2.8\sigma$ preference, which is not significant, although not completely negligible.  The preference is primarily driven by the residual difference in the ACT TE data across a wide range of multipoles, with the EE residuals now playing a much smaller role than in the previous analyses.

Perhaps intriguingly, this data set combination, which is the most constraining we consider without including the full \emph{Planck} data set, prefers values of $z_c < z_{\rm eq}$, close to the recombination epoch.\footnote{Recall that the \emph{Planck} fit to $\Lambda$CDM yields $z_{\rm eq} = 3387 \pm 21$ (TT,TE,EE+lowE+lensing+BAO)~\cite{Planck2018parameters}, i.e., $\log_{10}(z_{\rm eq}) \approx 3.53$.}  The same preference is seen in the ACT + large-scale \emph{Planck} TT fit, but at weaker significance.  The initial field value $\theta_i$ is essentially unconstrained in these fits, but interestingly it is consistent with low values, in contrast to results seen with \emph{Planck}~\cite{Smith:2019ihp,Hill:2020osr} (see Sec.~\ref{sec:discussion} for further discussion of this point).  

Finally, we consider a joint analysis of the full \emph{Planck} 2018 TT+TE+EE data with the ACT DR4 data (orange contours in Fig.~\ref{fig:EDE_ACT_Planckfull}).  The statistical weight of \emph{Planck} is sufficiently large compared to that of ACT that the joint results are qualitatively similar to those found for \emph{Planck} alone~\cite{Hill:2020osr} (green contours in Figs.~\ref{fig:EDE_ACT_PlanckTTlmax650_BAO_Lens} and~\ref{fig:EDE_ACT_Planckfull}).  However, the ACT preference for EDE leads to a weaker upper bound on $f_{\rm EDE}$ (and a slight upward shift and larger error bar on $H_0$) from the joint fit than that found from \emph{Planck} alone.  Notably, the overall fit improvement in this case is not significant, with $\Delta \chi^2 = -6.8$, corresponding to no model-selection preference for EDE over $\Lambda$CDM (1.8$\sigma$).

We conclude that the full \emph{Planck} data set provides a different outlook on the EDE scenario than that obtained in the ACT-based fits here, apart from the results when ACT is combined with the full \emph{Planck} data.  However, we cannot easily compute overall consistency in the full EDE parameter space with a simple metric (as done in Ref.~\cite{Aiola2020} and in Appendix~\ref{app:ha} for $\Lambda$CDM) because of the strong non-Gaussianity of the posteriors.  It may be possible to use a different consistency statistic (e.g., the ``suspiciousness''~\cite{Handley:2019wlz,Lemos:2019txn}), but this would incur significant computational expense due to the much slower Boltzmann code evaluation for EDE compared to $\Lambda$CDM.  We thus primarily rely on $\chi^2$-based criteria to assess consistency.

The significant constraining power of the $\ell > 650$ \emph{Planck} TT data plays a crucial role in the \emph{Planck} vs. ACT differences seen in the EDE fits here.  In particular, we observe that the best-fit EDE model to the full \emph{Planck} data predicts TE and EE power spectra that are not in perfect agreement with the ACT DR4 data; conversely, the best-fit EDE model to the ACT data set combinations considered in Sec.~\ref{subsec:ACT_alone}-\ref{subsec:ACT_P18TTlmax650_Lens_BAO} predicts a TT power spectrum that does not precisely match the high-$\ell$ \emph{Planck} data (see Sec.~\ref{sec:discussion} for a quantification of the latter statement).  Uncovering the origin of these differences -- and whether they may be related to systematic effects in the ACT data -- is clearly of high priority.  We briefly discuss a few potentially relevant effects in the next subsection.


\subsection{Constraints from ACT DR4 Alone}
\label{subsec:ACT_alone}

\begin{table*}[htb!]
Constraints on $\Lambda$CDM and EDE ($n=3$) from ACT DR4 TT+TE+EE + $\tau$ prior \vspace{2pt} \\
  \centering
  \begin{tabular}{|c|c|c|c|c|}
    \hline\hline Parameter &$\Lambda$CDM Best-Fit~~&$\Lambda$CDM Marg.~~&~~~EDE ($n=3$) Best-Fit~~~&~~~EDE ($n=3$) Marg.\\ \hline \hline

{\boldmath$\log(10^{10} A_\mathrm{s})$} & $3.043 $ & $3.046\pm 0.030$ & $3.083                    $ & $3.067\pm 0.034            $ \\

{\boldmath$n_\mathrm{s}   $} & $1.013                  $ & $1.011\pm 0.015            $ & $1.064                    $ & $0.987^{+0.027}_{-0.047}   $ \\

{\boldmath$100\theta_\mathrm{s}$} & $1.04356                   $ & $1.04345\pm 0.00070        $ & $1.04279                   $ & $1.04247\pm 0.00079        $ \\

{\boldmath$\Omega_\mathrm{b} h^2$} & $0.02149                 $ & $0.02152\pm 0.00031        $ & $0.02214                   $ & $0.02141^{+0.00044}_{-0.00065}$ \\

{\boldmath$\Omega_\mathrm{c} h^2$} & $0.1170                   $ & $0.1167\pm 0.0037          $ & $0.1425                    $ & $0.1307^{+0.0054}_{-0.0120} $ \\

{\boldmath$\tau_\mathrm{reio}$} & $0.063                  $ & $0.064\pm 0.014            $ & $0.061                    $ & $0.065\pm 0.015            $ \\

{\boldmath$y_p            $} & $1.0009                   $ & $1.0007\pm 0.0048          $ & $0.9951                    $ & $1.0037\pm 0.0070          $ \\

{\boldmath$f_\mathrm{EDE} $} & $-$ & $-$ & $0.241                     $ & $0.142^{+0.039}_{-0.072}   $\\

{\boldmath$\mathrm{log}_{10}(z_c)$} & $-$ & $-$ & $3.72     $ & $< 3.70    $\\

{\boldmath$\theta_i$} & $-$ & $-$ & $2.97                      $ & $>0.24$        \\

    \hline

$H_0  \, [\mathrm{km/s/Mpc}]         $ & $68.2                   $ & $68.4\pm 1.5               $ & $77.6                     $ & $74.5^{+2.5}_{-4.4}        $\\

$\Omega_\mathrm{m}         $ & $0.299                    $ & $0.298\pm 0.021            $ & $0.274                    $ & $0.276^{+0.020}_{-0.023}   $\\

$\sigma_8                  $ & $0.820                    $ & $0.819\pm 0.016            $ & $0.883                    $ & $0.831^{+0.027}_{-0.043}   $\\

$S_8$ & $0.818                    $ & $0.816\pm 0.042            $& $0.844                   $ & $0.796\pm 0.049            $\\

$\mathrm{log}_{10}(f/{\mathrm{eV}})$ & $-$ & $-$ & $26.65                   $ & $27.17^{+0.34}_{-0.55}     $ \\

$\mathrm{log}_{10}(m/{\mathrm{eV}})$ & $-$ & $-$ & $-26.90                 $ & $-27.52^{+0.26}_{-0.72}    $\\

    \hline
  \end{tabular} 
  \caption{Best-fit and marginalized 68\% CL constraints on cosmological parameters in the $\Lambda$CDM and EDE ($n=3$) models, inferred from ACT DR4 primary CMB data (TT+TE+EE) in combination with a Gaussian prior on the optical depth $\tau$.  Upper and lower bounds are quoted at 95\% CL.  Sampled parameters are shown in bold (and in subsequent tables).  The associated posteriors are shown in Fig.~\ref{fig:EDE_ACT_PlanckTTlmax650_BAO_Lens} (EDE parameters) and in Fig.~\ref{fig:ACT_alone} in Appendix~\ref{app:plots} (standard $\Lambda$CDM parameters).}
  \label{table:params-ACT-DR4}
\end{table*}

\begin{table}[t!]
\centering
  \begin{tabular}{|c|c|c|}
    \hline
    Data set &~~$\Lambda$CDM~~&~~~EDE ($n=3$) ~~~\\ \hline \hline
    ACT DR4 TT &  98.8   &  98.9\\
    ACT DR4 TE &  72.8   &  76.3\\
    ACT DR4 EE &  100.9  &  89.7 \\
    \hline
    ACT DR4 TT+TE+EE &  282.5  &  273.8 \\
     $\Delta \chi^2 $ &  & $-8.7$ \\ 
    \hline
  \end{tabular}
  \caption{$\chi^2$ values for the best-fit $\Lambda$CDM and EDE models to the ACT DR4 TT+TE+EE data.  Note that the joint $\chi^2$ for TT+TE+EE is not equal to the sum of the individual $\chi^2$ due to non-negligible off-diagonal blocks in the covariance matrix.  The decrease in $\chi^2$ is 8.7 for the three-parameter EDE extension of $\Lambda$CDM.}
  \label{table:chi2_ACT_alone}
\end{table}

We begin our analysis by fitting the $\Lambda$CDM and EDE ($n=3$) models to the ACT DR4 TT+TE+EE data, in combination with the $\tau$ prior discussed in Sec.~\ref{sec:data}.  The marginalized constraints are presented in Table~\ref{table:params-ACT-DR4}, while the posteriors are shown in Fig.~\ref{fig:EDE_ACT_PlanckTTlmax650_BAO_Lens} (EDE parameters) and in Fig.~\ref{fig:ACT_alone} in Appendix~\ref{app:plots} (standard $\Lambda$CDM parameters).  Note that the $\Lambda$CDM parameter results differ slightly from those in Ref.~\cite{Aiola2020} due to our use of higher numerical precision in the Boltzmann code -- see Appendix~\ref{app:ha} for detailed discussion of this point, and validation with independent Boltzmann and MCMC implementations.  

The best-fit parameters (e.g.,~$H_0$) shift noticeably between $\Lambda$CDM and EDE when fit to ACT data, in contrast to what is observed with \emph{Planck} data, for which the parameter shifts are small~\cite{Hill:2020osr}.  However, the statistical constraining power of ACT DR4 is weaker than that of \emph{Planck}; thus ACT cannot tightly constrain the EDE parameters on its own.  The relatively large uncertainties in the ACT EDE fit render the evidence for non-zero $f_{\rm EDE}$ marginal: the 95\% CL constraint on the EDE fraction is $f_{\rm EDE} = 0.142^{+0.15}_{-0.13}$, nearly consistent with zero.

Nevertheless, it is clear that the EDE model is able to accommodate residuals in the ACT data that $\Lambda$CDM cannot.  The best-fit value for the maximal EDE fraction is $f_{\rm EDE} = 0.241$ with $\log_{10}(z_c) = 3.72$, yielding $H_0 = 77.6$ km/s/Mpc.  Note that the posteriors are highly non-Gaussian, and thus these best-fit values differ strongly from the posterior means, as seen in Table~\ref{table:params-ACT-DR4}: e.g., the marginalized constraints on $f_{\rm EDE}$ and $H_0$ are $f_{\rm EDE} = 0.142^{+0.039}_{-0.072}$ and $H_0 = 74.5^{+2.5}_{-4.4}$ km/s/Mpc, respectively.

As expected, the physical CDM density increases substantially in the EDE fit as compared to $\Lambda$CDM, in order to counteract the early ISW effect associated with the EDE field~\cite{Hill:2020osr,Ivanov:2020ril}: $\Omega_c h^2 = 0.1307^{+0.0054}_{-0.0120}$ (EDE) vs. $\Omega_c h^2 = 0.1167\pm 0.0037$ ($\Lambda$CDM).  The posterior for this parameter is very non-Gaussian in the EDE case, which is reflected by the significant difference between the posterior mean and the best-fit value of $\Omega_c h^2 = 0.1425$.  This increase in the physical CDM density produces an increase in the low-redshift amplitude of structure $\sigma_8$, along with a significant broadening of the error bars on this parameter.  Interestingly, the parameter $S_8$ actually decreases in the EDE fit as compared to $\Lambda$CDM, as the increase in $H_0$ is so large that $\Omega_m$ decreases despite the increase in $\Omega_c h^2$.  Thus, in contrast to what is seen in EDE fits to \emph{Planck} and other cosmological data, the ``$S_8$ tension'' is not worsened in the EDE fit to ACT data -- in fact it is reduced.  Nevertheless, the error bars on parameters in the EDE fit to ACT alone are sufficiently large that it is difficult to draw strong conclusions about agreement or disagreement with other data sets.

Given the parameter shifts observed between $\Lambda$CDM and EDE fits to ACT data, it is interesting to assess the associated difference in $\chi^2$ and determine the origin of any improvement that is seen.  Table~\ref{table:chi2_ACT_alone} presents $\chi^2$ values for the best-fit $\Lambda$CDM and EDE models to ACT DR4.  The $\chi^2$ are further broken down into contributions from the TT, TE, and EE power spectra, but the overall TT+TE+EE $\chi^2$ is not equal to the sum of these contributions due to the non-negligible off-diagonal blocks in the joint covariance matrix.

Both the $\Lambda$CDM and EDE models provide a good fit to the ACT power spectra.  The best-fit $\Lambda$CDM model has $\chi^2 = 282.5$ for 254 degrees of freedom (260 bandpowers minus six free parameters), corresponding to a probability-to-exceed (PTE) $= 0.106$.\footnote{Note that this value differs slightly from that found in Ref.~\cite{Aiola2020} due to the use of higher-precision Boltzmann calculations here; see Appendix~\ref{app:ha} for details.}  The best-fit EDE model has $\chi^2 = 273.8$ for 251 degrees of freedom, corresponding to PTE $= 0.154$.  Thus, in terms of the overall goodness-of-fit, both models are acceptable.

\begin{figure*}[!tp]
\includegraphics[width=0.87\textwidth]{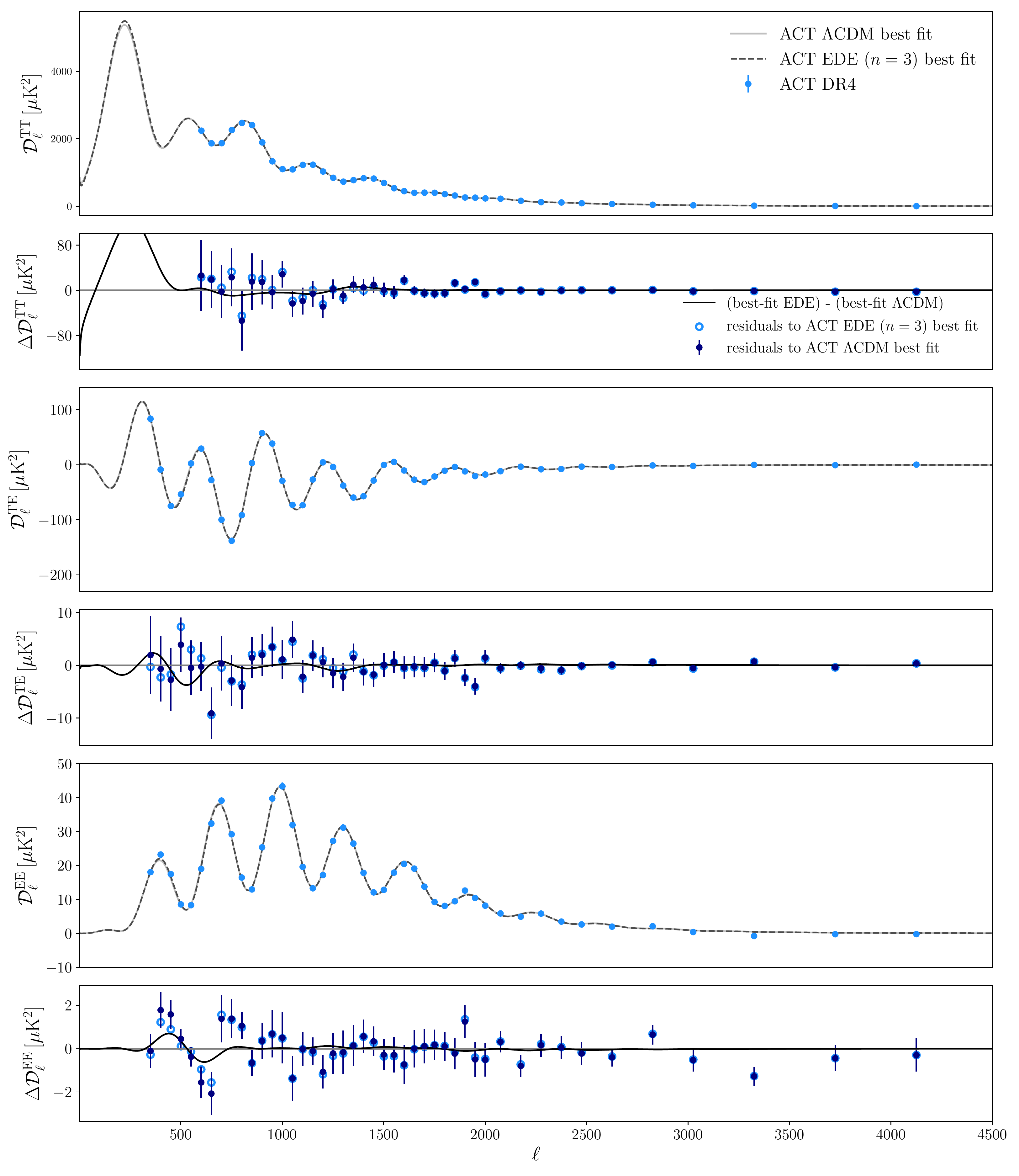}
\caption{Best-fit $\Lambda$CDM and EDE ($n=3$) models to the ACT DR4 TT (top), TE (middle), and EE (bottom) power spectrum data.  The smaller panels show the residuals of the best-fit models with respect to the data, as well as the difference between the best-fit EDE and $\Lambda$CDM models.  The EDE $\chi^2$ improvement over $\Lambda$CDM (c.f. Table~\ref{table:chi2_ACT_alone}) is driven entirely by the seven lowest multipole bins in the EE power spectrum.  Figs.~\ref{fig:ACT_alone_wide_residuals} and~\ref{fig:ACT_alone_deep_residuals} in Appendix~\ref{app:plots} show a further breakdown into the residuals for the wide- and deep-patch ACT data, respectively.}
\label{fig:ACT_alone_residuals}
\end{figure*}

We find that the best-fit EDE model yields an improvement of $\Delta \chi^2 = -8.7$ over the best-fit $\Lambda$CDM model.  Interestingly, the improvement is entirely driven by the EE power spectrum ($\Delta \chi^2_{\rm EE} = -11.2$); the EDE model actually fits the TT and TE power spectra worse than $\Lambda$CDM does, despite having three additional free parameters.  We further investigate the origin of the improvement in the EE fit below.

As the value of $\Delta \chi^2$ is expected to follow a $\chi^2$ distribution with three degrees of freedom here (because the EDE model contains three additional parameters beyond the $\Lambda$CDM model), we can compute the associated CL at which the EDE model is preferred over $\Lambda$CDM.  We find that $\Delta \chi^2 = -8.7$ corresponds to a preference for EDE over $\Lambda$CDM at the 96.6\% CL, or $2.1\sigma$, which is not significant.

To further assess the robustness of the overall $\chi^2$ improvement seen for the ACT data, given the three additional free parameters in the EDE model, we use the Akaike information criterion (AIC)~\cite{Akaike1998}: ${\rm AIC} \equiv -2 \ln \mathcal{L}_{\rm max} + 2 N_{\rm param}$, where $\mathcal{L}_{\rm max}$ is the maximum likelihood value and $N_{\rm param}$ is the number of free parameters in the model.  We thus find $\Delta {\rm AIC} = -2.7$ for the improvement of the EDE fit to the ACT DR4 data over $\Lambda$CDM, which is not significant.

To investigate the origin of the improved $\chi^2$ for EDE relative to $\Lambda$CDM, we plot the residuals of the ACT DR4 TT, TE, and EE power spectra with respect to the best-fit $\Lambda$CDM and EDE models in Fig.~\ref{fig:ACT_alone_residuals}.  The only noticeable improvement in the residuals for EDE compared to $\Lambda$CDM is in the lowest seven multipole bins in the EE power spectrum.  In fact, we can sub-divide the data further to see whether the ``wide'' patch or ``deep'' patch ACT DR4 data dominate these residuals (see Ref.~\cite{Choi2020} for definition of the deep and wide data subsets).  Residual plots for the wide and deep patches are shown in Figs.~\ref{fig:ACT_alone_wide_residuals} and~\ref{fig:ACT_alone_deep_residuals}, respectively, in Appendix~\ref{app:plots}.  By eye, it is clear that the EDE improvement in the residuals is dominated by the wide patch EE data, in particular the lowest seven multipole bins in that spectrum (this is also borne out in detail numerically).  This can also be seen by breaking the $\chi^2$ down further into contributions from the wide and deep patches.  For simplicity, we compute diagonal $\chi^2$ values here (i.e., the off-diagonal entries in the covariance matrix are not included -- only for this brief investigation), as these exactly match the intuition one would obtain from ``$\chi$-by-eye'' investigation of Figs.~\ref{fig:ACT_alone_wide_residuals} and~\ref{fig:ACT_alone_deep_residuals}.  We find an improvement of $\Delta \chi^2_{\rm EE, wide, diag} = -11.9$, whereas $\Delta \chi^2_{\rm EE, deep, diag} = -0.4$.

Thus, we conclude that the EDE improvement in $\chi^2$ over $\Lambda$CDM is localized to the lowest seven multipole bins in the wide patch EE data, while worsening the fit to the TT and TE data, and negligibly improving the fit to the EE-deep data.  This localization may hint at the presence of a low-significance systematic in the wide patch EE data, e.g., related to ground pickup or the transfer function calibration, both of which most strongly affect the lowest-$\ell$ data points in the power spectrum.  We emphasize that a vast number of systematics tests were performed in Ref.~\cite{Choi2020} to search for such issues, with no significant failures seen.  In particular, null tests comparing the power spectra from the wide and deep regions were passed successfully.\footnote{Note that the wide patch includes scans at lower elevation, and thus could potentially be more affected by ground pickup.}  Another issue to keep in mind at low-to-moderate $\ell$ in polarization is Galactic dust contamination, which is constrained in ACT DR4 using \emph{Planck} 353 GHz data~\cite{Choi2020}.  However, given their footprints on the sky, one would expect the wide patch to be more dust-contaminated than the deep patch, and thus the results here may suggest further robustness checks of the dust modeling.  Alternatively, it may instead be that we are first seeing evidence of the EDE model in the wide patch data because it is more constraining than the deep patch.  Indeed, the lowest several multipole bins in the EE-deep spectrum have sufficiently large error bars that they are consistent with both the best-fit $\Lambda$CDM model and with the residuals in the EE-wide spectrum that drive the EDE preference (see Figs.~\ref{fig:ACT_alone_wide_residuals} and~\ref{fig:ACT_alone_deep_residuals}).  Further investigation will be needed to solidify this interpretation.  We are carrying out complementary ongoing work to understand the ACT transfer function (particularly for TT at $\ell < 600$, though this multipole range is not used in any of our analyses presently), as well as other systematics studies on the ACT data in preparation for even more sensitive CMB maps in upcoming data releases.

As a final check on the robustness of the EDE parameter constraints from ACT DR4, we perform an exercise motivated by the observation in Ref.~\cite{Aiola2020} that dividing the ACT TE data by a factor of 1.05 moves the inferred cosmological parameters in $\Lambda$CDM into better agreement with those obtained independently from \emph{WMAP} or \emph{Planck}, particularly in the $\Omega_b h^2 - n_s$ plane.  While the TE data do not appear to drive the EDE results in the analysis here, this test is nevertheless of interest due to the non-zero correlation between $f_{\rm EDE}$ and $n_s$, as a result of the early ISW effect described earlier.  We thus divide the ACT TE data by 1.05 and rerun the EDE MCMC analysis described in this subsection.  The results are presented in Figs.~\ref{fig:EDE_ACT_TEresc} and~\ref{fig:ACT_alone_TEresc} in Appendix~\ref{app:plots}.

\begin{table*}[htb!]
Constraints on $\Lambda$CDM and EDE ($n=3$) from ACT DR4 TT+TE+EE + \emph{Planck} 2018 TT ($\ell_{\rm max} = 650$) + $\tau$ prior \vspace{2pt} \\
  \centering
  \begin{tabular}{|c|c|c|c|c|}
    \hline\hline Parameter &$\Lambda$CDM Best-Fit~~&$\Lambda$CDM Marg.~~&~~~EDE ($n=3$) Best-Fit~~~&~~~EDE ($n=3$) Marg.\\ \hline \hline

{\boldmath$\log(10^{10} A_\mathrm{s})$} & $3.064                    $ & $3.063\pm 0.026            $ & $3.048                    $ & $3.064\pm 0.032            $\\

{\boldmath$n_\mathrm{s}   $} & $0.9774                    $ & $0.9775\pm 0.0064          $ & $0.977                $ & $0.985^{+0.011}_{-0.018}   $\\

{\boldmath$100\theta_\mathrm{s}$} & $1.04318                   $ & $1.04311\pm 0.00061        $ & $1.04223                   $ & $1.04238\pm 0.00066        $\\

{\boldmath$\Omega_\mathrm{b} h^2$} & $0.02232                  $ & $0.02236\pm 0.00019        $ & $0.02124                  $ & $0.02162\pm 0.00044        $\\

{\boldmath$\Omega_\mathrm{c} h^2$} & $0.1187                   $ & $0.1188\pm 0.0027          $ & $0.1265                    $ & $0.1291^{+0.0051}_{-0.0098}$\\

{\boldmath$\tau_\mathrm{reio}$} & $0.061                    $ & $0.062\pm 0.013            $ & $0.061                    $ & $0.063\pm 0.014            $\\

{\boldmath$y_p            $} & $1.0019                   $ & $1.0027\pm 0.0040          $ & $1.0058                    $ & $1.0038\pm 0.0055          $\\

{\boldmath$f_\mathrm{EDE} $} & $-$ & $-$ & $0.113                    $ & $0.129^{+0.028}_{-0.055}   $\\

{\boldmath$\mathrm{log}_{10}(z_c)$} & $-$ & $-$ & $3.18           $ & $< 3.43      $\\

{\boldmath$\theta_i$} & $-$ & $-$ & $0.20                      $ & $< 2.89$        \\

    \hline

$H_0                       $ & $68.2                    $ & $68.2\pm 1.1               $ & $74.0                     $ & $74.4^{+2.2}_{-3.0}        $\\

$\Omega_\mathrm{m}         $ & $0.304                    $ & $0.305\pm 0.015            $ & $0.271                    $ & $0.274\pm 0.017            $\\

$\sigma_8                  $ & $0.820                    $ & $0.819\pm 0.013            $ & $0.813                    $ & $0.827^{+0.029}_{-0.035}   $\\

$S_8$ & $0.825             $ & $0.825\pm 0.031            $ & $0.773                    $ & $0.791^{+0.040}_{-0.046}   $\\

$\mathrm{log}_{10}(f/{\mathrm{eV}})$ & $-$ & $-$ & $28.09                   $ & $27.22^{+0.28}_{-0.56}     $ \\

$\mathrm{log}_{10}(m/{\mathrm{eV}})$ & $-$ & $-$ & $-26.30                  $ & $-27.54^{+0.19}_{-0.63}    $\\

    \hline
  \end{tabular} 
  \caption{Best-fit and marginalized 68\% CL constraints on cosmological parameters in the $\Lambda$CDM and EDE ($n=3$) models, inferred from ACT DR4 TT+TE+EE and \emph{Planck} 2018 TT ($\ell_{\rm max} = 650$) data, in combination with a Gaussian prior on the optical depth $\tau$.  Upper and lower bounds are quoted at 95\% CL.  The associated posteriors are shown in Fig.~\ref{fig:EDE_ACT_PlanckTTlmax650_BAO_Lens} (EDE parameters) and in Fig.~\ref{fig:ACT_PlanckTTlmax650} in Appendix~\ref{app:plots} (standard $\Lambda$CDM parameters).}
  \label{table:params-ACT-DR4-P18TTlmax650}
\end{table*}

\begin{table}[t!]
\centering
  \begin{tabular}{|c|c|c|}
    \hline
    Data set &~~$\Lambda$CDM~~&~~~EDE ($n=3$) ~~~\\ \hline \hline
    ACT DR4 TT &  100.6  &  97.2\\
    ACT DR4 TE &  80.9   &  74.1\\
    ACT DR4 EE &  99.4  &   95.2 \\
    \hline
    ACT DR4 TT+TE+EE &  291.1  &  275.0 \\
    \emph{Planck} 2018 low-$\ell$ TT & 21.7  & 21.9 \\
    \emph{Planck} 2018 TT ($\ell_{\rm max}=650$) & 250.3 & 250.8 \\
    \hline
    Total $\chi^2 $   & 563.1 & 547.7\\
     $\Delta \chi^2 $ &  & $-15.4$ \\ 
    \hline
  \end{tabular}
  \caption{$\chi^2$ values for the best-fit $\Lambda$CDM and EDE models to the ACT DR4 TT+TE+EE and \emph{Planck} 2018 TT ($\ell_{\rm max} = 650$) data.  Note that the joint $\chi^2$ for ACT DR4 TT+TE+EE is not equal to the sum of the individual $\chi^2$ due to non-negligible off-diagonal blocks in the covariance matrix.  The decrease in $\chi^2$ is 15.4 for the three-parameter EDE extension of $\Lambda$CDM.}
  \label{table:chi2_ACT_P18TTlmax650}
\end{table}

Several changes are seen in the posteriors in this exercise.  As expected, $\Omega_b h^2$ shifts upward into closer agreement with \emph{Planck}, but interestingly $n_s$ does not shift downward, as seen when performing this exercise in $\Lambda$CDM (see Fig.~14 of Ref.~\cite{Aiola2020}).  In fact, $n_s$ increases, as does $S_8$.  It thus appears that the early-ISW-compensation signatures of the EDE model are seen more strongly in this exercise (the results thus resemble \emph{Planck} more closely now in this regard).  The $f_{\rm EDE}$ posterior broadens noticeably, but the central value hardly changes: the marginalized constraint is now $f_{\rm EDE} = 0.146^{+0.064}_{-0.090}$ at 68\% CL, and at 95\% CL the posterior is consistent with zero, yielding an upper limit $f_{\rm EDE} < 0.285$.  The $H_0$ posterior is remarkably stable to this modification of the ACT TE data: we find $H_0 = 74.5^{+3.2}_{-4.6}$ km/s/Mpc, i.e., with unchanged central value and a moderate increase in the error bars.  Intriguingly, $z_c$ shifts to higher values closer to $z_{\rm eq}$, and a bimodality emerges in the posterior.  A moderate preference for large $\theta_i$ is also seen.  However, given the weaker evidence for non-zero $f_{\rm EDE}$ itself here, the $z_c$ and $\theta_i$ changes should be taken with a grain of salt.  Overall, we conclude that rescaling the ACT TE data does not fully erase the EDE-driven shifts seen in $\Lambda$CDM parameters in our analysis above, although the preference for non-zero EDE is weakened.


\subsection{Constraints from ACT DR4 + \emph{Planck} 2018 TT ($\ell_{\rm max} = 650$)}
\label{subsec:ACT_P18TTlmax650}

We next extend our analysis to include the large-scale ($\ell_{\rm max} = 650$) \emph{Planck} 2018 TT power spectrum data, which serve as a \emph{WMAP}-like complement to the ACT data, filling in large-scale modes that are not easily accessible from the ground (e.g., due to atmospheric $1/f$ noise or ground pickup).  We fit the $\Lambda$CDM and EDE ($n=3$) models to the ACT DR4 TT+TE+EE data in combination with this subset of the \emph{Planck} 2018 TT data and the $\tau$ prior discussed in Sec.~\ref{sec:data}.  The marginalized parameter constraints are presented in Table~\ref{table:params-ACT-DR4-P18TTlmax650}, while the posteriors are shown in Fig.~\ref{fig:EDE_ACT_PlanckTTlmax650_BAO_Lens} (EDE parameters) and in Fig.~\ref{fig:ACT_PlanckTTlmax650} in Appendix~\ref{app:plots} (standard $\Lambda$CDM parameters). The $\Lambda$CDM parameter results for this data combination are very similar to those found for ACT DR4 combined with \emph{WMAP} -- see Table~\ref{tab:comparison} in Appendix~\ref{app:ha}, which gives slightly updated results for ACT+\emph{WMAP} $\Lambda$CDM parameters using Boltzmann calculations with higher numerical accuracy than in Ref.~\cite{Aiola2020}.

As in the ACT-only analysis presented in Sec.~\ref{subsec:ACT_alone} (see Table~\ref{table:params-ACT-DR4}), the best-fit parameters shift noticeably between $\Lambda$CDM and EDE when fit to these data, in contrast to what is observed with the full \emph{Planck} data set on its own, for which the parameter shifts are small~\cite{Hill:2020osr}.  Moreover, with the large-scale \emph{Planck} data included here in addition to ACT, the statistical constraining power is sufficiently high to yield moderately significant evidence for non-zero $f_{\rm EDE}$: the 95\% CL constraint on the EDE fraction is $f_{\rm EDE} = 0.129^{+0.099}_{-0.076}$, while at 99.7\% CL we obtain $f_{\rm EDE} = 0.129^{+0.14}_{-0.084}$.  Note that the 68\% CL constraint is $f_{\rm EDE} = 0.129^{+0.028}_{-0.055}$; comparing this with the 95\% and 99.7\% CL bounds illustrates the skewed nature of the marginalized posterior.\footnote{If one instead adopts the equal-tail confidence interval approach, the 68\% CL constraint is $f_{\rm EDE} = 0.129^{+0.045}_{-0.043}$.}

Interestingly, even with the improved constraining power afforded by the inclusion of the large-scale \emph{Planck} TT data, the EDE model is still able to accommodate residuals in the ACT data -- while jointly fitting both data sets -- that $\Lambda$CDM cannot.  The best-fit value for the maximal EDE fraction is $f_{\rm EDE} = 0.113$ with $\log_{10}(z_c) = 3.18$, yielding $H_0 = 74.0$ km/s/Mpc.  As in the ACT-only analysis, the posteriors are somewhat non-Gaussian, albeit less so.  For example, most of the best-fit parameter values are now fairly close to the posterior means in Table~\ref{table:params-ACT-DR4-P18TTlmax650}: e.g., the marginalized constraint on the Hubble constant is $H_0 = 74.4^{+2.2}_{-3.0}$ km/s/Mpc.

As in the ACT-only analysis, the physical CDM density increases substantially in the EDE fit as compared to $\Lambda$CDM, so as to counteract the early ISW effect driven by the EDE field~\cite{Hill:2020osr,Ivanov:2020ril}: $\Omega_c h^2 = 0.1291^{+0.0051}_{-0.0098}$ (EDE) vs. $\Omega_c h^2 = 0.1188\pm 0.0027$ ($\Lambda$CDM).  Also as seen before, the posterior for this parameter is quite non-Gaussian in the EDE case (albeit slightly less than for ACT-alone).  However, in an intriguing difference, the best-fit value of $\sigma_8$ decreases in the EDE fit to these data as compared to $\Lambda$CDM, despite the increase in the CDM density.  Note, though, that the posterior mean value of $\sigma_8$ increases, and the error bars on $\sigma_8$ roughly triple in the EDE case over those in $\Lambda$CDM.   The best-fit and posterior mean $S_8$ values decrease substantially in the EDE fit as compared to $\Lambda$CDM, as the increase in $H_0$ is sufficiently large that $\Omega_m$ decreases despite the increase in $\Omega_c h^2$.  Thus, in contrast to what is seen in EDE fits to \emph{Planck} and other cosmological data~\cite{Hill:2020osr,Ivanov:2020ril,DAmico:2020ods}, the ``$S_8$ tension'' is not worsened in the EDE fit to ACT DR4 TT+TE+EE and large-scale \emph{Planck} TT data -- in fact it is essentially eliminated here.  However, even with the inclusion of the large-scale \emph{Planck} data, the parameter error bars in the EDE fit here are sufficiently large that it is difficult to draw strong conclusions about agreement or disagreement with other data sets.  Also, the exclusion of BAO and CMB lensing data impacts these conclusions, as will be seen in the next subsection.  Nevertheless, higher-precision small-scale CMB data, as well as careful understanding of systematics, will be needed to ascertain whether the shifts seen here are statistical fluctuations or preliminary indications of a preference for a different cosmological model than $\Lambda$CDM.

\begin{figure*}[!tp]
\includegraphics[width=0.87\textwidth]{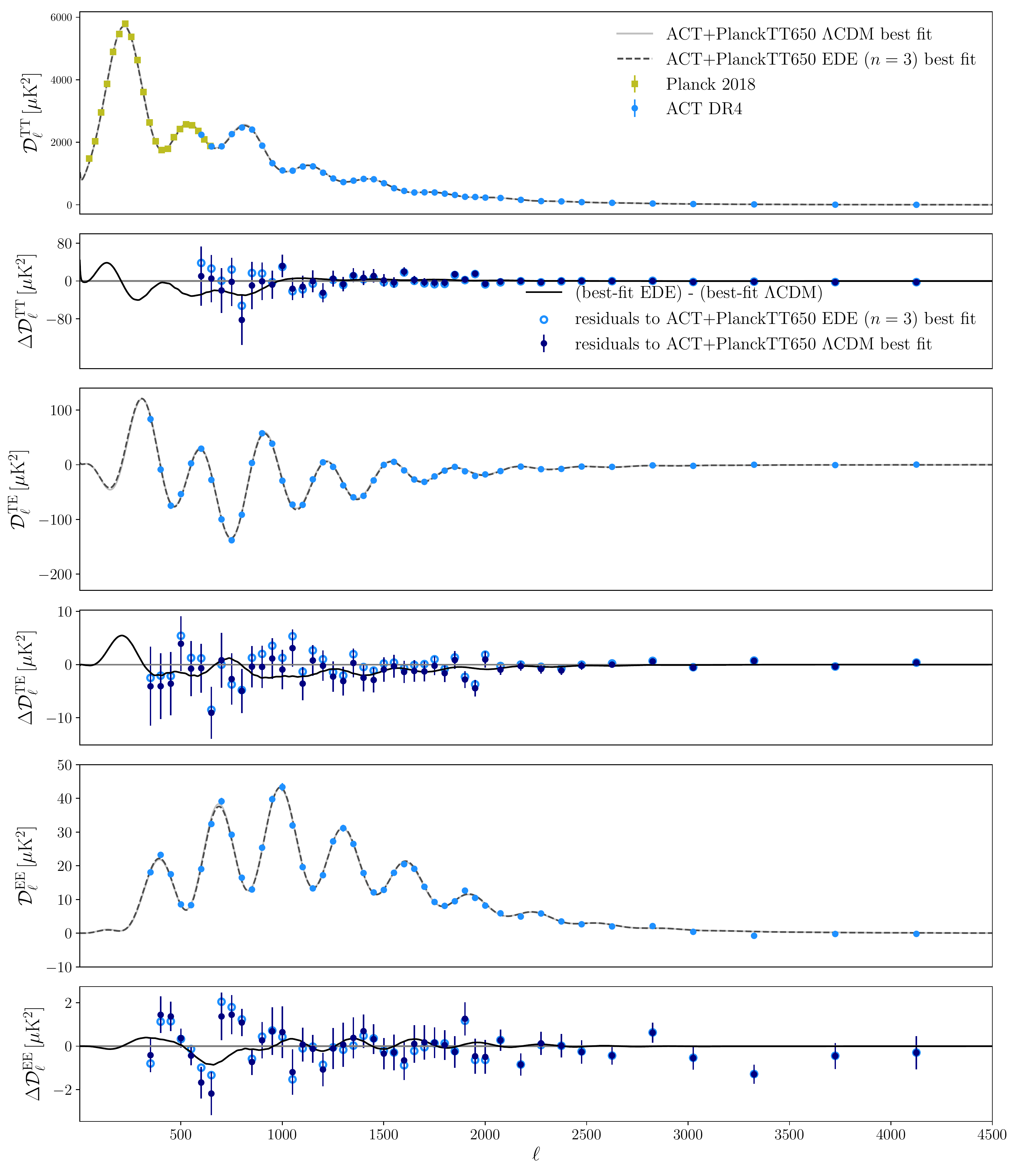}
\caption{Best-fit $\Lambda$CDM and EDE ($n=3$) models to the ACT DR4 TT (top), TE (middle), and EE (bottom) power spectrum data, fit in combination with large-scale ($\ell_{\rm max} = 650$) \emph{Planck} 2018 TT power spectrum data (yellow squares in top panel).  The smaller panels show the residuals of the best-fit models with respect to the ACT data, as well as the difference between the best-fit EDE and $\Lambda$CDM models.  The EDE $\chi^2$ improvement over $\Lambda$CDM (c.f. Table~\ref{table:chi2_ACT_P18TTlmax650}) receives its largest contribution from the ACT TE power spectrum, but moderate improvements in TT and EE are also seen.  Figs.~\ref{fig:ACT_P18TTlmax650_wide_residuals} and~\ref{fig:ACT_P18TTlmax650_deep_residuals} in Appendix~\ref{app:plots} show a further breakdown into the residuals for the wide- and deep-patch ACT data, respectively.}
\label{fig:ACT_P18TTlmax650_residuals}
\end{figure*}

The robustness of the preference for EDE over $\Lambda$CDM in this analysis can be investigated by assessing the difference in the best-fit $\chi^2$ values.  Table~\ref{table:chi2_ACT_P18TTlmax650} presents $\chi^2$ values for the best-fit $\Lambda$CDM and EDE models to ACT DR4 and \emph{Planck} TT ($\ell_{\rm max} = 650$).  The $\chi^2$ are further broken down into contributions from the TT, TE, and EE power spectra.

As in the ACT-only analysis in Sec.~\ref{subsec:ACT_alone}, both the $\Lambda$CDM and EDE models provide a good fit to the ACT power spectra within the joint ACT + \emph{Planck} TT ($\ell_{\rm max} = 650$) analysis.  The best-fit $\Lambda$CDM model has $\chi^2_{\rm ACT} = 291.1$ for 254 degrees of freedom, corresponding to PTE $= 0.055$.  The best-fit EDE model has $\chi^2_{\rm ACT} = 275.0$ for 251 degrees of freedom, corresponding to PTE $= 0.143$.  Thus, both models are acceptable fits to the ACT data here.

The best-fit EDE model yields an improvement of $\Delta \chi^2 = -15.4$ over the best-fit $\Lambda$CDM model.  The improvement is driven entirely by the ACT data; the fit to the large-scale \emph{Planck} TT data is essentially unchanged (slightly worsened).  Unlike in the ACT-only analysis in Sec.~\ref{subsec:ACT_alone}, the improvement here does not come entirely from the EE power spectrum.  Here, the largest contribution to $\Delta \chi^2$ comes from the TE power spectrum, but moderate improvements are also seen in the TT and EE power spectra.  Thus, the EDE preference cannot be as easily localized as in the ACT-only case.  Nevertheless, we further investigate the origin of the improvements in the fit below.

As before, we use the value of $\Delta \chi^2$ (assumed to be $\chi^2$-distributed with three degrees of freedom) and the AIC to assess the robustness of the overall preference for EDE over $\Lambda$CDM, given that the EDE model has three additional free parameters.  We find that $\Delta \chi^2 = -15.4$ corresponds to a preference for EDE over $\Lambda$CDM at the 99.8\% CL, or $3.2\sigma$, which is weakly significant.  We find $\Delta {\rm AIC} = -9.4$ for the improvement of the EDE fit over $\Lambda$CDM, which is moderately significant.

Where in the data does the improvement in $\chi^2$ come from?  We plot the residuals of the ACT DR4 TT, TE, and EE power spectra with respect to the best-fit $\Lambda$CDM and EDE models (to the joint data set) in Fig.~\ref{fig:ACT_P18TTlmax650_residuals}.  The main improvement that is visible by eye is in the TE power spectrum, which exhibits a difference between the best-fit models that is coherent across a wide range of multipoles (roughly $700 \lesssim \ell \lesssim 2500$).  However, the TT and EE power spectra also make non-negligible contributions to the improvement in fit.  As the $\chi^2$ improvement is distributed relatively broadly across the observable channels, it is difficult to ascertain whether a systematic effect could be responsible.  However, we do note that the EE improvement is again dominated by the lowest $\ell$ bins in the wide-patch data (see below), as in the ACT-only analysis in Sec.~\ref{subsec:ACT_alone}.  The ACT TE $\chi^2$ improvement may be associated with an overall amplitude offset in this power spectrum as compared to that predicted by the best-fit $\Lambda$CDM model to ACT + \emph{WMAP} or ACT + \emph{Planck} 2018 TT ($\ell_{\rm max} = 650$)~\cite{Aiola2020}.  While extensive investigation has not yielded evidence of a systematic effect that could cause this amplitude shift, it has been shown that dividing the ACT TE data by a factor of 1.05 moves the inferred cosmological parameters in $\Lambda$CDM into better agreement with those obtained independently from \emph{WMAP} or \emph{Planck}.  We investigate this issue in detail below.

To further examine the best-fit models in this case, we sub-divide the data again into the ``wide'' patch and ``deep'' patch ACT DR4 subsets.  Residual plots for the wide and deep patches are shown in Figs.~\ref{fig:ACT_P18TTlmax650_wide_residuals} and~\ref{fig:ACT_P18TTlmax650_deep_residuals}, respectively, in Appendix~\ref{app:plots}.  In contrast to the ACT-only analysis, the $\chi^2$ improvements are not strongly dominated by the wide or deep data individually, although this turns out to be a coincidence, as the TE-deep improvement dominates over TE-wide, while the EE-wide dominates over EE-deep.  The single largest contribution to the improvement in $\chi^2$ comes from the deep-patch TE data, while the next largest comes from the wide-patch EE data.  Computing diagonal $\chi^2$ values again for simplicity (i.e., the off-diagonal entries in the covariance matrix are not included), as these exactly match the intuition one would obtain from ``$\chi$-by-eye'' investigation of Figs.~\ref{fig:ACT_P18TTlmax650_wide_residuals} and~\ref{fig:ACT_P18TTlmax650_deep_residuals}, we find an improvement of $\Delta \chi^2_{\rm TE, wide, diag} = -1.2$, whereas $\Delta \chi^2_{\rm TE, deep, diag} = -5.8$.   Interestingly, the $\chi^2$ improvement in the TE-deep data is not localized in $\ell$, but is rather spread evenly in small amounts across a broad range of multipole bins (see Fig.~\ref{fig:ACT_P18TTlmax650_deep_residuals}).  The EE data exhibit the opposite behavior, with the wide patch contributing a larger improvement than the deep patch: $\Delta \chi^2_{\rm EE, wide, diag} = -4.2$, while $\Delta \chi^2_{\rm EE, deep, diag} = -0.7$.  As in Sec.~\ref{subsec:ACT_alone}, the lowest seven multipole bins in the wide-EE data are responsible for the improvement.

Thus, as in the ACT-only analysis, the wide-patch EE data play an important role in the EDE preference, but here the TE data (particularly from the deep patch) have an even more significant impact than EE.  The TT wide and deep data exhibit smaller (and relatively similar) improvements to one another in the EDE fit compared to $\Lambda$CDM.  We conclude that the same TE feature seen in Ref.~\cite{Aiola2020} may drive the moderate EDE preference seen in this analysis.  Further work, and higher-precision data, will be needed to ascertain whether this is a real feature of the CMB sky.

As a cross-check on the stability of the EDE results presented in this subsection, we rerun our MCMC analysis for a model with power-law index $n=2$ in Eq.~\eqref{eq.V}.  We obtain very small shifts ($\lesssim 0.5\sigma$) in $f_{\rm EDE}$, $\log_{10}(z_c)$, and the standard cosmological parameters, as compared to those found in the $n=3$ analysis.  The main change is a preference for somewhat smaller values of the initial field displacement $\theta_i$, which is nevertheless not tightly constrained in the $n=3$ analysis anyhow.  This shift arises due to the partial degeneracy between $n$ and $\theta_i$ (see Ref.~\cite{Smith:2019ihp} for a detailed discussion of the role played by $\theta_i$).\footnote{Briefly, with all other parameters held fixed, $\theta_i$ sets the frequency of oscillations of the EDE field $\phi$ at the background level, which subsequently determines the effective perturbation sound speed.}  We conclude that although current data do not constrain $n$ precisely, our results are robust to the choice of reasonable values for this parameter.

Finally, to assess the robustness of the EDE parameter constraints in this analysis, we perform the same TE-rescaling exercise done at the end of Sec.~\ref{subsec:ACT_alone}.  We divide the ACT TE data by a factor of 1.05 and rerun the EDE MCMC analysis.  The results are presented in Figs.~\ref{fig:EDE_ACT_TEresc} and~\ref{fig:ACT_PlanckTTlmax650_TEresc} in Appendix~\ref{app:plots}.

In general, the posterior changes seen in this exercise are relatively similar to those seen in the ACT-only analysis in the previous subsection: $\Omega_b h^2$ shifts upward into closer agreement with \emph{Planck}, while $n_s$ and $S_8$ increase, exhibiting more noticeable compensation effects for the EDE-induced early ISW effect.  The $f_{\rm EDE}$ posterior broadens, but the central value is very stable: the marginalized constraint is now $f_{\rm EDE} = 0.128^{+0.050\, +0.10}_{-0.058\, -0.10}$ (68\%/95\% CL), while at 99.7\% CL it is consistent with zero.  The $H_0$ constraint remains remarkably stable: we find $H_0 = 74.1^{+2.5}_{-2.8}$ km/s/Mpc, very similar to the result in Table~\ref{table:params-ACT-DR4-P18TTlmax650}.  The critical redshift $z_c$ shifts upward toward $z_{\rm eq}$, but the bimodality seen in this exercise for the ACT-only analysis is no longer present.  In addition, $\theta_i$ is essentially unconstrained.  Qualitatively, the results are more similar to those from \emph{Planck} alone~\cite{Hill:2020osr} than found without the TE rescaling, but the persistence of the high value of $H_0$ is intriguing.  As in the ACT-only case, we conclude that rescaling the ACT TE data does not fully erase the EDE-driven shifts seen in $\Lambda$CDM parameters in our analysis above, although the preference for non-zero EDE is weakened.  Given the important role played by the TE data in the EDE preference in this joint analysis, these results motivate careful scrutiny to ascertain whether a physical or systematic explanation for such a rescaling of the TE data exists.


\subsection{Constraints from ACT DR4 + \emph{Planck} 2018 TT ($\ell_{\rm max} = 650$) + \emph{Planck} 2018 CMB Lensing + BAO}
\label{subsec:ACT_P18TTlmax650_Lens_BAO}

\begin{table*}[htb!]
Constraints on $\Lambda$CDM and EDE ($n=3$) from ACT DR4 TT+TE+EE + \emph{Planck} 2018 TT ($\ell_{\rm max} = 650$) + \emph{Planck} 2018 CMB Lensing + BAO + $\tau$ prior \vspace{2pt} \\
  \centering
  \begin{tabular}{|c|c|c|c|c|}
    \hline\hline Parameter &$\Lambda$CDM Best-Fit~~&$\Lambda$CDM Marg.~~&~~~EDE ($n=3$) Best-Fit~~~&~~~EDE ($n=3$) Marg.\\ \hline \hline

{\boldmath$\log(10^{10} A_\mathrm{s})$} & $3.060                    $ & $3.058\pm 0.021            $ & $3.050                    $ & $3.060\pm 0.021            $\\

{\boldmath$n_\mathrm{s}   $} & $0.9785                   $ & $0.9775\pm 0.0046          $ & $0.9684                    $ & $0.9728^{+0.0073}_{-0.0130} $\\

{\boldmath$100\theta_\mathrm{s}$} & $1.04328                   $ & $1.04310\pm 0.00057        $ & $1.04246                   $ & $1.04228\pm 0.00062        $\\

{\boldmath$\Omega_\mathrm{b} h^2$} & $0.02239                  $ & $0.02236\pm 0.00017        $ & $0.02120                $ & $0.02154^{+0.00035}_{-0.00042}$\\

{\boldmath$\Omega_\mathrm{c} h^2$} & $0.1185                  $ & $0.1187\pm 0.0012          $ & $0.1289                   $ & $0.1286^{+0.0027}_{-0.0063}$\\

{\boldmath$\tau_\mathrm{reio}$} & $0.060                    $ & $0.059\pm 0.011            $ & $0.055                   $ & $0.059\pm 0.013            $\\

{\boldmath$y_p            $} & $1.0028                   $ & $1.0028\pm 0.0041          $ & $1.0037                    $ & $1.0051\pm 0.0051          $\\

{\boldmath$f_\mathrm{EDE} $} & $-$ & $-$ & $0.091                    $ & $0.091^{+0.020}_{-0.036}   $\\

{\boldmath$\mathrm{log}_{10}(z_c)$} & $-$ & $-$ & $3.16                    $ & $< 3.36    $ \\

{\boldmath$\theta_i$} & $-$ & $-$ & $0.72                      $ & $< 2.82                 $        \\

    \hline

$H_0                       $ & $68.40                     $ & $68.25\pm 0.51             $ & $70.9                     $ & $70.9^{+1.0}_{-2.0}      $\\

$\Omega_\mathrm{m}         $ & $0.3025                    $ & $0.3042\pm 0.0068          $ & $0.3003                    $ & $0.3000\pm 0.0072          $\\

$\sigma_8                  $ & $0.8175                    $ & $0.8171\pm 0.0080          $ & $0.825                    $ & $0.829^{+0.013}_{-0.021}   $\\

$S_8$ & $0.821                    $ & $0.823\pm 0.012         $ & $0.825                    $ & $0.828^{+0.015}_{-0.018} $\\

$\mathrm{log}_{10}(f/{\mathrm{eV}})$ & $-$ & $-$ & $27.47                   $ & $27.19^{+0.33}_{-0.52}     $\\

$\mathrm{log}_{10}(m/{\mathrm{eV}})$ & $-$ & $-$ & $-27.42                   $ & $-27.60^{+0.17}_{-0.66}    $\\

    \hline
  \end{tabular} 
  \caption{Best-fit and marginalized 68\% CL constraints on cosmological parameters in the $\Lambda$CDM and EDE ($n=3$) models, inferred from ACT DR4 TT+TE+EE, \emph{Planck} 2018 TT ($\ell_{\rm max} = 650$), \emph{Planck} 2018 CMB lensing, and BAO data, in combination with a Gaussian prior on the optical depth $\tau$.  Upper and lower bounds are quoted at 95\% CL.  The associated posteriors are shown in Fig.~\ref{fig:EDE_ACT_PlanckTTlmax650_BAO_Lens} (EDE parameters) and in Fig.~\ref{fig:ACT_PlanckTTlmax650_BAO_Lens} in Appendix~\ref{app:plots} (standard $\Lambda$CDM parameters).}
  \label{table:params-ACT-DR4-P18TTlmax650-Lens-BAO}
\end{table*}

\begin{table}[t!]
\centering
  \begin{tabular}{|c|c|c|}
    \hline
    Data set &~~$\Lambda$CDM~~&~~~EDE ($n=3$) ~~~\\ \hline \hline
    ACT DR4 TT &  100.8  &  97.3 \\
    ACT DR4 TE &  81.6   &  71.8 \\
    ACT DR4 EE &  100.0  &  96.5 \\
    \hline
    ACT DR4 TT+TE+EE &  292.1  &  276.1 \\
    \emph{Planck} 2018 low-$\ell$ TT & 21.5  & 23.5 \\
    \emph{Planck} 2018 TT ($\ell_{\rm max}=650$) & 251.1 & 251.3 \\
    \emph{Planck} 2018 CMB Lensing &  8.7  &  9.4 \\
    BAO (6dF) & 0.007 & 0.03 \\
    BAO (DR7 MGS) & 2.0 & 2.3 \\
    BAO (DR12 BOSS) & 3.3 & 3.4 \\
    \hline
    Total $\chi^2 $   & 578.7 & 566.0\\
     $\Delta \chi^2 $ &  & $-12.7$ \\ 
    \hline
  \end{tabular}
  \caption{$\chi^2$ values for the best-fit $\Lambda$CDM and EDE models to the ACT DR4 TT+TE+EE, \emph{Planck} 2018 TT ($\ell_{\rm max} = 650$), \emph{Planck} 2018 CMB lensing, and BAO data.  Note that the joint $\chi^2$ for ACT DR4 TT+TE+EE is not equal to the sum of the individual $\chi^2$ due to non-negligible off-diagonal blocks in the covariance matrix.  The decrease in $\chi^2$ is 12.7 for the three-parameter EDE extension of $\Lambda$CDM.}
  \label{table:chi2_ACT_P18TTlmax650_Lens_BAO}
\end{table}

We now extend our analysis to include the \emph{Planck} 2018 CMB lensing power spectrum data~\cite{Planck2018lensing}, which primarily constrain $\sigma_8$ and $\Omega_m$, and the compilation of BAO data described in Sec.~\ref{sec:data}, which primarily constrain $\Omega_m$ via the distance-redshift relation.  These additional data sets thus break degeneracies in the CMB-only data analyses presented thus far, particularly the so-called ``geometric degeneracy'' between $\Omega_m$ and $H_0$~(e.g.,~\cite{Sherwin2011}).  We fit the $\Lambda$CDM and EDE ($n=3$) models to the ACT DR4 TT+TE+EE data, the large-scale ($\ell_{\rm max} = 650$) \emph{Planck} TT data, the \emph{Planck} 2018 CMB lensing data, and the BAO data in combination with the $\tau$ prior discussed in Sec.~\ref{sec:data}.  The marginalized parameter constraints are presented in Table~\ref{table:params-ACT-DR4-P18TTlmax650-Lens-BAO}, while the posteriors are shown in Fig.~\ref{fig:EDE_ACT_PlanckTTlmax650_BAO_Lens} (EDE parameters) and in Fig.~\ref{fig:ACT_PlanckTTlmax650_BAO_Lens} in Appendix~\ref{app:plots} (standard $\Lambda$CDM parameters).

As in the ACT-only and ACT + \emph{Planck} 2018 TT ($\ell_{\rm max} = 650$) analyses presented in Tables~\ref{table:chi2_ACT_alone} and~\ref{table:params-ACT-DR4-P18TTlmax650}, respectively, the best-fit parameters shift between $\Lambda$CDM and EDE when fit to these data, although not as significantly as in those analyses, due to the influence of the CMB lensing and BAO data here.  The latter serve in particular to constrain $S_8$ and $\Omega_m$ independently of the other cosmological parameters, which breaks degeneracies in the CMB data and thereby prevents shifts in $H_0$ as large as those seen in Tables~\ref{table:chi2_ACT_alone} and~\ref{table:params-ACT-DR4-P18TTlmax650}.  However, non-negligible $f_{\rm EDE}$ is allowed by the data considered here, and the statistical constraining power is sufficiently high to yield moderately significant evidence for non-zero $f_{\rm EDE}$: the 95\% CL constraint on the EDE fraction is $f_{\rm EDE} = 0.091^{+0.069}_{-0.056}$, while the 99.7\% CL constraint is $f_{\rm EDE} = 0.091^{+0.11}_{-0.063}$.

The inclusion of the BAO and CMB lensing data thus reduces error bars, but does not prevent the EDE model from still providing an improved fit to the ACT DR4 data over $\Lambda$CDM, as seen in the previous subsections.  The best-fit value for the maximal EDE fraction is $f_{\rm EDE} = 0.091$, somewhat lower than seen in the ACT + large-scale \emph{Planck} TT analysis in the previous subsection, with $\log_{10}(z_c) = 3.16$, yielding $H_0 = 70.9$ km/s/Mpc.  The posteriors are generally more Gaussian than those in the ACT-only and ACT + large-scale \emph{Planck} TT analyses, as seen in the agreement of the best-fit values with the posterior means (e.g., the marginalized constraint on the Hubble constant is $H_0 = 70.9^{+1.0}_{-2.0}$ km/s/Mpc).  There are a few exceptions: $\theta_i$ is essentially unconstrained and $\Omega_c h^2$ is still noticeably skewed, with a tail extending to high values.  Perhaps intriguingly from a model-building perspective, this data set combination (which is the most constraining we consider without including the full \emph{Planck} data set) prefers values of $z_c < z_{\rm eq}$, close to the recombination epoch.\footnote{Hints that $z_c \sim z_{\rm eq}$ may suggest that the ``coincidence problem'' in the EDE scenario could be resolved by linking the dynamics to relevant physics at matter-radiation equality.  The fit here suggests that this model-building clue may have been spurious.}  Formally, the best-fit $z_c = 1450$, very close to recombination at $z_* = 1100$.

While non-zero values of $f_{\rm EDE}$ remain moderately preferred as in the previous two subsections, the downward shift in $H_0$ after folding in the BAO and CMB lensing data is notable.  This shift occurs because the BAO + CMB lensing combination favors $\Omega_m \approx 0.3$,\footnote{Formally, $\Omega_m = 0.303^{+0.016}_{-0.018}$ from \emph{Planck} 2018 CMB lensing and BAO data, within $\Lambda$CDM~\cite{Planck2018lensing}.} thereby forbidding the decrease in $\Omega_m$ seen in Tables~\ref{table:params-ACT-DR4} and~\ref{table:params-ACT-DR4-P18TTlmax650}.  The upward shift in $\Omega_c h^2$ required to counteract the EDE-induced early ISW effect thus cannot simultaneously accommodate a very high $H_0$ value, as it would drive $\Omega_m$ too low to match the BAO and CMB lensing data.  This result is in accord with the general argument presented by Ref.~\cite{Jedamzik:2020zmd}, who showed that models that attempt to increase the CMB-inferred $H_0$ solely by reducing the sound horizon cannot simultaneously accommodate BAO data and an $H_0$ value larger than roughly 70 km/s/Mpc.

The impact of the BAO and CMB lensing data is also seen in $S_8$.  In the ACT-only and ACT + large-scale \emph{Planck} TT analyses, $S_8$ was able to take quite low values in the EDE fits.  The shift in $\Omega_m$ discussed above now forbids this, and $S_8$ takes on a noticeably higher value, $S_8 = 0.828^{+0.015}_{-0.018}$.  Even more strikingly, the error bar on $S_8$ in Table~\ref{table:params-ACT-DR4-P18TTlmax650-Lens-BAO} decreases by a factor of nearly three relative to that in Table~\ref{table:chi2_ACT_P18TTlmax650}.  This reflects the extremely important role played by LSS data in constraining models that aim to increase the value of $H_0$ inferred from indirect, cosmological probes~\cite{Hill:2020osr,Ivanov:2019hqk,DAmico:2020ods}.  With the inclusion of the BAO and CMB lensing data, the apparent resolution of the ``$S_8$ tension'' seen in the previous two subsections is now no longer present.  However, even with these additional data sets included, the error bars remain sufficiently large in the EDE fit that strong conclusions cannot be drawn.

As in the previous subsections, we assess the preference for EDE over $\Lambda$CDM in this analysis via the difference in the best-fit $\chi^2$ values.  Table~\ref{table:chi2_ACT_P18TTlmax650_Lens_BAO} presents $\chi^2$ values for the best-fit $\Lambda$CDM and EDE models to the ACT DR4, \emph{Planck} TT ($\ell_{\rm max} = 650$), \emph{Planck} 2018 CMB lensing, and BAO data.  The ACT $\chi^2$ are further broken down into contributions from the TT, TE, and EE power spectra.

As in Sec.~\ref{subsec:ACT_alone} and~\ref{subsec:ACT_P18TTlmax650}, both the $\Lambda$CDM and EDE models remain a good fit to the ACT power spectra within the joint ACT + \emph{Planck} TT ($\ell_{\rm max} = 650$) + CMB lensing + BAO analysis.  The best-fit $\Lambda$CDM model has $\chi^2_{\rm ACT} = 292.1$ for 254 degrees of freedom, corresponding to PTE $= 0.050$.  The best-fit EDE model has $\chi^2_{\rm ACT} = 276.1$ for 251 degrees of freedom, corresponding to PTE $= 0.133$.  Thus, both models provide acceptable fits to the ACT data.

The best-fit EDE model yields an improvement of $\Delta \chi^2 = -12.7$ over the best-fit $\Lambda$CDM model.  As in Table~\ref{table:chi2_ACT_P18TTlmax650}, the improvement in $\chi^2$ is driven entirely by the ACT data here; the fit to every other data set is worsened (albeit most only by a negligible amount).  Again as in the analysis in Sec.~\ref{subsec:ACT_P18TTlmax650}, the improvement here does not come solely from a single observable channel, with the ACT TT, TE, and EE data all contributing.  However, again the largest contribution to $\Delta \chi^2$ comes from the ACT TE power spectrum with $\Delta \chi^2_{\rm TE} = -9.8$.  The EDE preference is not as highly localized as in the ACT-only case (where the lowest several multipole bins in EE dominated), but it does appear to be driven by the same features in the TE data that partially drove the results in the ACT + \emph{Planck} TT ($\ell_{\rm max} = 650$) analysis.

However, the EE residuals play less of a role here than in Sec.~\ref{subsec:ACT_alone} and~\ref{subsec:ACT_P18TTlmax650}.  In particular, computing a diagonal-only $\Delta \chi^2$ as in the previous subsections, we find for the EE data that $\Delta \chi^2_{\rm EE, wide, diag} = -1.6$ and $\Delta \chi^2_{\rm EE, deep, diag} = -1.6$, i.e., the prominent role played by the lowest seven EE-wide multipole bins is no longer seen.  This is likely due to the restrictions placed on the model by the matter density constraints from the BAO and CMB lensing data, which do not allow the EE residuals to be accommodated.  For comparison, for the TE data we find $\Delta \chi^2_{\rm TE, wide, diag} = -3.4$ and $\Delta \chi^2_{\rm TE, deep, diag} = -5.6$, indicating that the TE-deep data again have the most significant impact in the overall EDE preference, although the TE-wide data also exhibit more noticeable improvement than in Sec.~\ref{subsec:ACT_P18TTlmax650}.  In general, the residuals to the best-fit model here are sufficiently similar to those in Fig.~\ref{fig:ACT_P18TTlmax650_residuals} that we refrain from plotting them.

Finally, we use the value of $\Delta \chi^2$ (assumed to be $\chi^2$-distributed with three degrees of freedom) and the AIC to assess the robustness of the overall improvement in fit seen here, given that the EDE model has three additional free parameters beyond $\Lambda$CDM.  We find that $\Delta \chi^2 = -12.7$ corresponds to a preference for EDE over $\Lambda$CDM at the 99.5\% CL, or $2.8\sigma$, which is not significant, albeit not completely negligible.  We find $\Delta {\rm AIC} = -6.7$ for the improvement of the EDE fit over $\Lambda$CDM, which is not significant.

\begin{table*}[htb!]
Constraints on $\Lambda$CDM and EDE ($n=3$) from ACT DR4 TT+TE+EE + \emph{Planck} 2018 TT+TE+EE (full $\ell$ range, but no low-$\ell$ EE) + $\tau$ prior \vspace{2pt} \\
  \centering
  \begin{tabular}{|c|c|c|c|c|}
    \hline\hline Parameter &$\Lambda$CDM Best-Fit~~&$\Lambda$CDM Marg.~~&~~~EDE ($n=3$) Best-Fit~~~&~~~EDE ($n=3$) Marg.\\ \hline \hline

{\boldmath$\log(10^{10} A_\mathrm{s})$} & $3.089                    $ & $3.082\pm 0.024            $ & $3.086                    $ & $3.091\pm 0.025            $\\

{\boldmath$n_\mathrm{s}   $} & $0.9727                  $ & $0.9697\pm 0.0039          $ & $0.9838                    $ & $0.9771^{+0.0069}_{-0.0099}$\\

{\boldmath$100\theta_\mathrm{s}$} & $1.04218                  $ & $1.04205\pm 0.00026        $ & $1.04164                  $ & $1.04182^{+0.00037}_{-0.00031}$ \\

{\boldmath$\Omega_\mathrm{b} h^2$} & $0.02242                  $ & $0.02238\pm 0.00013        $ & $0.02242                  $ & $0.02247\pm 0.00018        $\\

{\boldmath$\Omega_\mathrm{c} h^2$} & $0.1188                   $ & $0.1195\pm 0.0012          $ & $0.1283                   $ & $0.1244^{+0.0025}_{-0.0051}$ \\

{\boldmath$\tau_\mathrm{reio}$} & $0.076                    $ & $0.070\pm 0.012            $ & $0.065                   $ & $0.070\pm 0.012            $\\

{\boldmath$y_p            $} & $1.0003                   $ & $1.0017\pm 0.0046          $ & $1.0000                   $ & $1.0015\pm 0.0047          $\\

{\boldmath$f_\mathrm{EDE} $} & $-$ & $-$ & $0.0904                    $ & $< 0.124                  $\\

{\boldmath$\mathrm{log}_{10}(z_c)$} & $-$ & $-$ & $3.52                     $ & $3.54^{+0.28}_{-0.20}      $ \\

{\boldmath$\theta_i$} & $-$ & $-$ & $2.83                     $ & $> 0.51                    $        \\

    \hline

$H_0                       $ & $67.94                     $ & $67.64\pm 0.54             $ & $70.63                     $ & $69.17^{+0.83}_{-1.70}      $\\

$\Omega_\mathrm{m}         $ & $0.3074                    $ & $0.3115\pm 0.0074          $ & $0.3034                    $ & $0.3084\pm 0.0084          $\\

$\sigma_8                  $ & $0.8278                    $ & $0.8261\pm 0.0097          $ & $0.843                    $ & $0.838^{+0.013}_{-0.015}   $\\

$S_8$ & $0.838                    $ & $0.842\pm 0.015          $ & $0.848                    $ & $0.850\pm 0.017          $\\

$\mathrm{log}_{10}(f/{\mathrm{eV}})$ & $-$ & $-$ & $26.53                    $ & $26.51^{+0.22}_{-0.36}     $\\

$\mathrm{log}_{10}(m/{\mathrm{eV}})$ & $-$ & $-$ & $-27.37                   $ & $-27.22^{+0.44}_{-0.39}    $\\

    \hline
  \end{tabular} 
  \caption{Best-fit and marginalized 68\% CL constraints on cosmological parameters in the $\Lambda$CDM and EDE ($n=3$) models, inferred from ACT DR4 TT+TE+EE and \emph{Planck} 2018 TT+TE+EE (full $\ell$ range, but no low-$\ell$ EE data to be consistent with the other analyses in this paper), in combination with a Gaussian prior on the optical depth $\tau$.  Upper and lower bounds are quoted at 95\% CL.  The associated posteriors are shown in Fig.~\ref{fig:EDE_ACT_Planckfull} (EDE parameters) and in Fig.~\ref{fig:ACT_Planckfull} in Appendix~\ref{app:plots} (standard $\Lambda$CDM parameters).}
  \label{table:params-ACT-DR4-P18full}
\end{table*}

\begin{table}[t!]
\centering
  \begin{tabular}{|c|c|c|}
    \hline
    Data set &~~$\Lambda$CDM~~&~~~EDE ($n=3$) ~~~\\ \hline \hline
    ACT DR4 TT &  53.8  &  52.3 \\
    ACT DR4 TE &  90.3   & 85.6 \\
    ACT DR4 EE &  97.5  &  95.7 \\
    \hline
    ACT DR4 TT+TE+EE &  244.6  &  238.5 \\
    \emph{Planck} 2018 low-$\ell$ TT & 22.6  & 21.7 \\
    \emph{Planck} 2018 TT+TE+EE & 2343.1 & 2343.3 \\
    \hline
    Total $\chi^2 $   & 2610.3 & 2603.5\\
     $\Delta \chi^2 $ &  & $-6.8$ \\ 
    \hline
  \end{tabular}
  \caption{$\chi^2$ values for the best-fit $\Lambda$CDM and EDE models to the ACT DR4 TT+TE+EE and \emph{Planck} 2018 TT+TE+EE (full $\ell$ range) data.  The low-$\ell$ EE data from \emph{Planck} are excluded for consistency with the other analyses presented in this paper (in lieu of the $\tau$ prior used throughout), although we verify that this choice has negligible impact on these results.  Note that the joint $\chi^2$ for ACT DR4 TT+TE+EE is not equal to the sum of the individual $\chi^2$ due to non-negligible off-diagonal blocks in the covariance matrix.  Also, the ACT $\chi^2$ values in this table cannot be directly compared to those in Tables~\ref{table:chi2_ACT_alone},~\ref{table:chi2_ACT_P18TTlmax650}, or~\ref{table:chi2_ACT_P18TTlmax650_Lens_BAO}, as the $\ell$ range used here for the ACT TT data is reduced in order to avoid double-counting information in combination with \emph{Planck}~\cite{Choi2020,Aiola2020}.  The decrease in $\chi^2$ is 6.8 for the three-parameter EDE extension of $\Lambda$CDM, driven primarily in this case by the ACT TE data.}
  \label{table:chi2_ACT_P18full}
\end{table}


\subsection{Constraints from ACT DR4 + \emph{Planck} 2018}
\label{subsec:ACT_P18full}

As a final analysis, we consider the combination of the ACT DR4 data with the full \emph{Planck} 2018 TT, TE, and EE data.  The parameter error bars are sufficiently large in the ACT-only analysis (see Sec.~\ref{subsec:ACT_alone}) that it is statistically acceptable to combine the ACT data with the full \emph{Planck} data set, despite their somewhat different preferred regions in the EDE parameter space.  The statistical weight of the full \emph{Planck} data set is sufficiently large compared to ACT DR4 that it dominates the combined posteriors, as seen below.  To avoid double-counting information contained in both data sets, we restrict the $\ell$ range of the ACT TT likelihood here such that $\ell_{\rm min} = 1800$~\cite{Aiola2020}.

We fit the $\Lambda$CDM and EDE ($n=3$) models to the ACT and \emph{Planck} data in combination with the $\tau$ prior discussed in Sec.~\ref{sec:data}.  The marginalized parameter constraints are presented in Table~\ref{table:params-ACT-DR4-P18full}, while the posteriors are shown in Fig.~\ref{fig:EDE_ACT_Planckfull} (EDE parameters) and in Fig.~\ref{fig:ACT_Planckfull} in Appendix~\ref{app:plots} (standard $\Lambda$CDM parameters).  Note that Fig.~\ref{fig:EDE_ACT_Planckfull} also shows posteriors from the \emph{Planck}-only analysis presented in Ref.~\cite{Hill:2020osr}.  For consistency with the analyses presented elsewhere in this paper, we exclude the \emph{Planck} low-$\ell$ EE likelihood ($\ell < 30$) and instead use the Gaussian prior on $\tau$ discussed in Sec.~\ref{sec:data}.  As a cross-check, we perform the analyses described in this subsection using the low-$\ell$ EE likelihood instead of the $\tau$ prior, and find negligible changes in all results, apart from a $1\sigma$ shift in $A_s$ due to the slightly lower central value of $\tau$ preferred by the \emph{Planck} low-$\ell$ EE likelihood~\cite{Planck2018parameters} as compared to our $\tau$ prior.

Compared to the analyses presented in the previous subsections (see Tables~\ref{table:params-ACT-DR4}, \ref{table:params-ACT-DR4-P18TTlmax650}, and~\ref{table:params-ACT-DR4-P18TTlmax650-Lens-BAO}), the best-fit parameters shift much less between $\Lambda$CDM and EDE when fit to these data, due to the strong statistical weight carried by the full \emph{Planck} data set, which does not prefer the existence of EDE~\cite{Hill:2020osr}.  However, as ACT alone weakly prefers non-zero $f_{\rm EDE}$ (see Table~\ref{table:params-ACT-DR4}), the upper limit on this parameter weakens in the joint ACT + \emph{Planck} analysis here ($f_{\rm EDE} < 0.124$ at 95\% CL), as compared to the limit obtained from \emph{Planck} alone ($f_{\rm EDE} < 0.087$ at 95\% CL~\cite{Hill:2020osr}).  The impact of the inclusion of ACT DR4 data on other parameters of interest is relatively minimal, e.g., we find $H_0 = 69.17^{+0.83}_{-1.70}$ km/s/Mpc (ACT+\emph{Planck}) vs.~$H_0 = 68.29^{+0.73}_{-1.20}$ km/s/Mpc (\emph{Planck} alone)~\cite{Hill:2020osr}.  Similarly, we find $S_8 = 0.850 \pm 0.017$ (ACT+\emph{Planck}) vs.~$S_8 = 0.839 \pm 0.017$ (\emph{Planck} alone)~\cite{Hill:2020osr}.  The inclusion of ACT data does lead to somewhat more skewness in the posteriors, however, as reflected in differences between the best-fit and posterior mean values.  For example, the best-fit $H_0 = 70.63$ km/s/Mpc and the best-fit $f_{\rm EDE} = 0.090$, which is relatively close to the 95\% CL bound quoted above.

The small change in parameter posteriors compared to those found from \emph{Planck} data alone is also reflected in a much smaller $\chi^2$ improvement in the EDE fit over $\Lambda$CDM for the ACT + \emph{Planck} analysis here, as compared to those seen in the previous subsections.  Table~\ref{table:chi2_ACT_P18full} presents $\chi^2$ values for the best-fit $\Lambda$CDM and EDE models to the ACT DR4 and \emph{Planck} primary CMB data.  The ACT $\chi^2$ are further broken down into contributions from the TT, TE, and EE power spectra.  The best-fit EDE model yields an improvement of $\Delta \chi^2 = -6.8$ over the best-fit $\Lambda$CDM model.  As in Tables~\ref{table:chi2_ACT_P18TTlmax650} and~\ref{table:chi2_ACT_P18TTlmax650_Lens_BAO}, the improvement in $\chi^2$ -- albeit moderate here -- is driven primarily by the ACT data; the fit to \emph{Planck} is essentially unchanged ($\Delta \chi^2_{Planck} = -0.7$).  Also as seen in Sec.~\ref{subsec:ACT_P18TTlmax650} and~\ref{subsec:ACT_P18TTlmax650_Lens_BAO}, the largest contribution to $\Delta \chi^2$ comes from the ACT TE power spectrum, with $\Delta \chi^2_{\rm TE} = -4.7$.  This suggests that the EDE model attempts to fit similar residuals to those that drive the fits in the previous two subsections, but is prevented from fully doing so by the highly constraining \emph{Planck} TT data.  In fact, even the \emph{Planck} EE error bars are smaller than those from ACT at $\ell \lesssim 800$, and this acts to prevent the ACT-wide EE residual in the lowest several multipole bins from driving the fit toward EDE.  Differences between the best-fit EDE models to ACT, ACT + large-scale \emph{Planck} TT, and \emph{Planck} alone are further investigated in Sec.~\ref{sec:discussion} below.

As a brief additional investigation, we compute diagonal-only $\Delta \chi^2$ for the wide and deep patches separately, as in the previous subsections.  We find  $\Delta \chi^2_{\rm TE, wide, diag} = -2.6$ and $\Delta \chi^2_{\rm TE, deep, diag} = -2.0$, indicating that both regions contribute roughly equally; in particular, the TE-deep data do not dominate, as had been seen earlier in Sec.~\ref{subsec:ACT_P18TTlmax650}.  We find $\Delta \chi^2_{\rm EE, wide, diag} = -2.5$ and $\Delta \chi^2_{\rm EE, deep, diag} = -0.6$, indicating that the EE-wide data still play a larger role in the EDE preference than the EE-deep data, but far less overall than seen in the ACT-only analysis in Sec.~\ref{subsec:ACT_alone}.

Finally, we use the value of $\Delta \chi^2$ (assumed to be $\chi^2$-distributed with three degrees of freedom) and the AIC to assess the robustness of the overall improvement in fit seen here, given that the EDE model has three additional free parameters beyond $\Lambda$CDM.  We find that $\Delta \chi^2 = -6.8$ corresponds to a preference for EDE over $\Lambda$CDM at the 92.2\% CL, or $1.8\sigma$, which is not significant.  We find $\Delta {\rm AIC} = -0.8$ for the improvement of the EDE fit over $\Lambda$CDM, corresponding to no preference.  This result is very similar to that found in the \emph{Planck}-only case, with $\Delta \chi^2 = -4.1$ and $\Delta {\rm AIC} = +1.9$~\cite{Hill:2020osr}.


\section{Discussion and Outlook}
\label{sec:discussion}

A summary of the main results of this work is provided in Sec.~\ref{subsec:summary}.  The main qualitative takeaway is that current conclusions about the EDE scenario do not appear to be fully robust to the choice of CMB data set.  The full \emph{Planck} data set and the ACT data in combination with large-scale TT data from \emph{Planck} (or from \emph{WMAP}~\cite{Schoneberg:2021qvd}) yield different conclusions as to the viability of EDE for significantly increasing the value of $H_0$ inferred from CMB data, with the latter preferring EDE over $\Lambda$CDM at roughly $3\sigma$ significance (as assessed via $\Delta\chi^2$).  However, we emphasize that the full \emph{Planck} data set carries significantly more statistical weight than ACT DR4 (or ACT + large-scale \emph{Planck} TT).

\begin{figure}[ht!]
\includegraphics[width=\columnwidth]{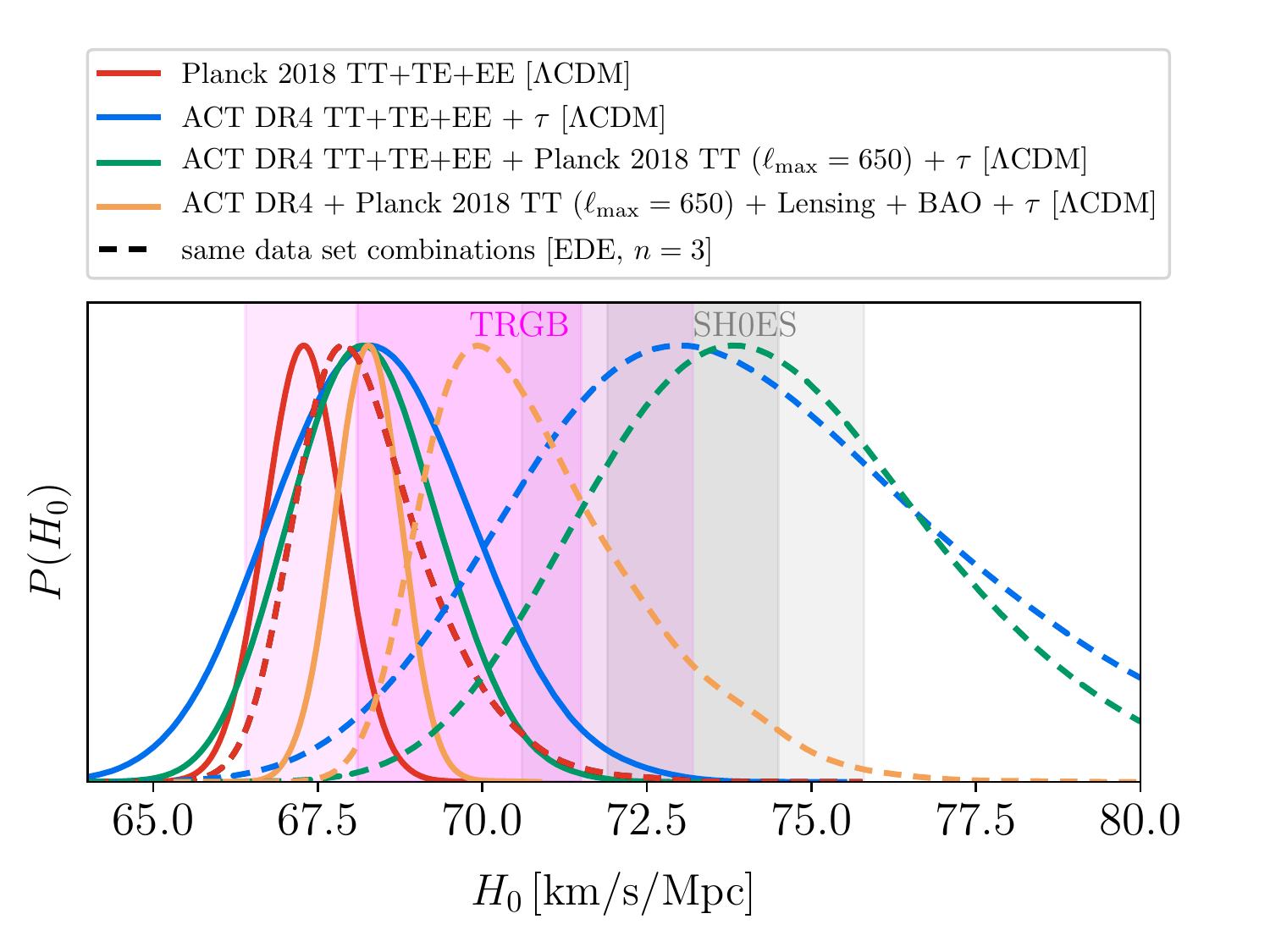}
\caption{Summary of marginalized $H_0$ posteriors in the $\Lambda$CDM (solid) and EDE (dashed) models, as inferred from the data set combinations listed in the legend.  Results for \emph{Planck} alone (red) are from Ref.~\cite{Hill:2020osr}.  Results for the other data set combinations are from Sec.~\ref{subsec:ACT_alone}-\ref{subsec:ACT_P18TTlmax650_Lens_BAO}.  For visual comparison, the latest direct $H_0$ measurements from TRGB~\cite{Freedman:2021ahq} (magenta) and SH0ES~\cite{Riess:2020fzl} (grey) are also shown, with the shaded bands denoting the 68\% and 95\% CL constraints.
\label{fig:H0_summary}}
\end{figure}

Fig.~\ref{fig:H0_summary} presents a compact summary of the implications for the inference of $H_0$ from the CMB, BAO, and CMB lensing data analyzed in this work.  We show the marginalized posterior for $H_0$ inferred within the $\Lambda$CDM and EDE models, inferred from the \emph{Planck} primary CMB data alone~\cite{Hill:2020osr}, and inferred from the ACT DR4 primary CMB data alone or in combination with other probes (Sec.~\ref{subsec:ACT_alone}-\ref{subsec:ACT_P18TTlmax650_Lens_BAO}).  We do not plot the posteriors derived from the combination of ACT with the full \emph{Planck} data (Sec.~\ref{subsec:ACT_P18full}), as they are similar to the \emph{Planck}-only results.  The ACT-only and ACT + large-scale \emph{Planck} TT analyses exhibit significant increases in the inferred value of $H_0$ between $\Lambda$CDM and EDE, whereas \emph{Planck} does not.  In both cases, the $H_0$ error bar increases as well, although more substantially in the ACT-based analyses.  The inclusion of BAO and CMB lensing data with ACT + large-scale \emph{Planck} TT somewhat moderates the increase in $H_0$, but a long tail out to high $H_0$ values is nevertheless still seen in the posterior.  For visual comparison, the plot also displays the latest direct $H_0$ measurements from TRGB~\cite{Freedman:2021ahq} and SH0ES~\cite{Riess:2020fzl}.  It is clear that the differences amongst the indirect (CMB) probes when analyzed in the EDE context must be understood before a careful, quantitative comparison to local $H_0$ measurements can be performed.

An interesting feature of the ACT-based EDE fits is the lack of a clear preference for large values of $\theta_i$, the initial EDE field value.  As discussed in detail in Ref.~\cite{Smith:2019ihp}, EDE fits to \emph{Planck} data (and \emph{Planck} data in combination with other data sets) generally prefer values of $\theta_i \approx \pi$.\footnote{The preference arises from the need to maximize the number of modes within the horizon at $z_c$ that have a low effective EDE perturbation sound speed; see Ref.~\cite{Smith:2019ihp} for details.}  An important consequence is that power-law potentials do not appear to serve as good candidate EDE potentials when \emph{Planck} data are considered~\cite{Agrawal:2019lmo,Smith:2019ihp}.  In contrast, our results suggest that ACT data can accommodate low values of $\theta_i$ while obtaining high $f_{\rm EDE}$ and high $H_0$.  The results of Sec.~\ref{subsec:ACT_P18TTlmax650_Lens_BAO}, in which we analyze ACT + large-scale \emph{Planck} TT + CMB lensing + BAO data, are particularly striking in this regard -- see the orange curve in the $\theta_i$ panel in Fig.~\ref{fig:EDE_ACT_PlanckTTlmax650_BAO_Lens}.

\begin{figure}[ht!]
\includegraphics[width=\columnwidth]{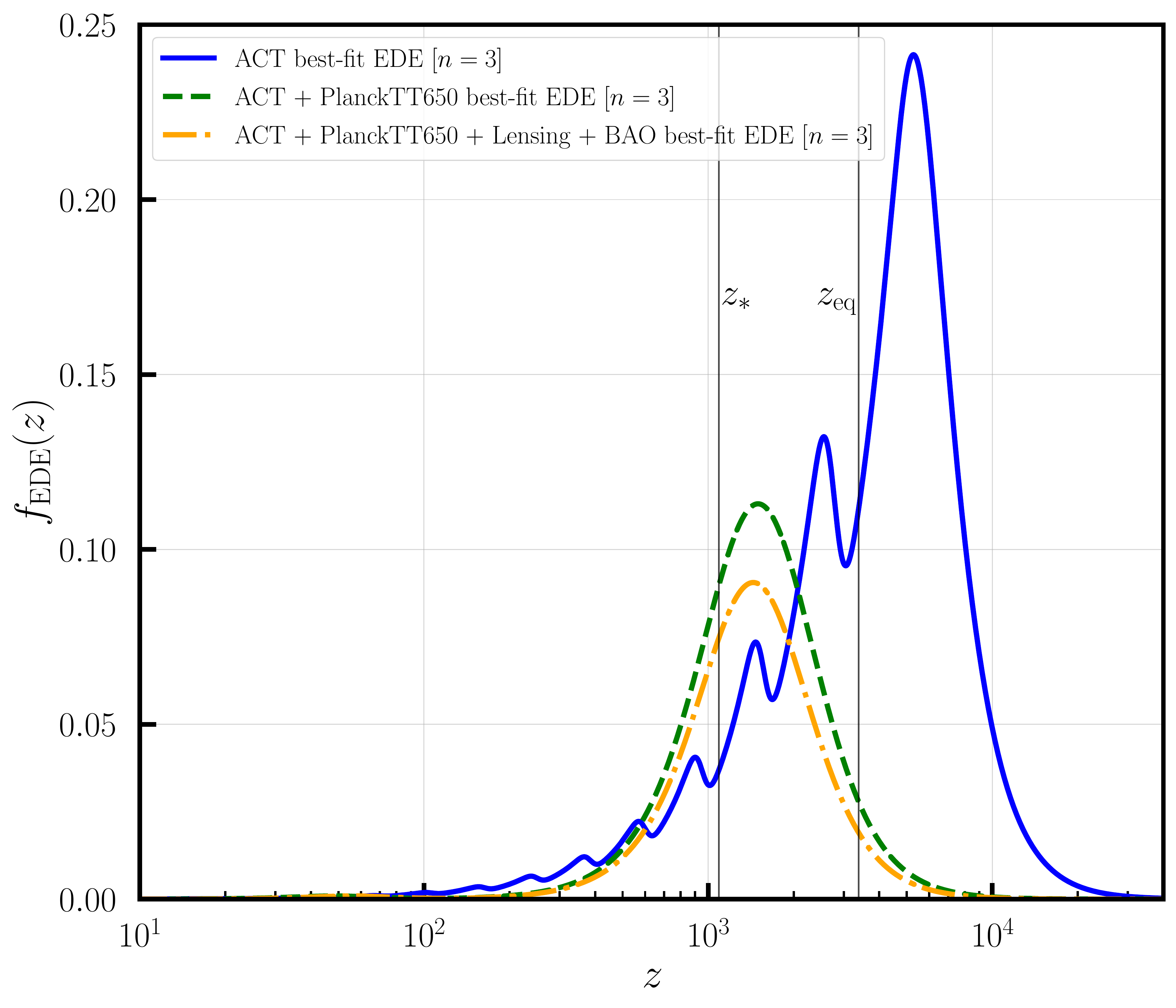}
\caption{Evolution of the cosmic energy density fraction stored in EDE as a function of redshift, $f_{\rm EDE}(z)$, for the best-fit EDE models to ACT DR4 (solid blue), ACT DR4 + \emph{Planck} TT ($\ell_{\rm max} = 650$) (dashed green), and ACT DR4 + \emph{Planck} TT ($\ell_{\rm max} = 650$) + CMB lensing + BAO (dot-dashed orange).  The redshifts of recombination ($z_*$) and matter-radiation equality ($z_{\rm eq}$) are indicated by the solid vertical lines.  Results for data set combinations including the full \emph{Planck} primary CMB data are not shown, as these data do not prefer non-zero $f_{\rm EDE}$.
\label{fig:fEDE_pedagogical}}
\end{figure}

The physical implications of these results are illustrated in Fig.~\ref{fig:fEDE_pedagogical}, which shows $f_{\rm EDE}(z)$ for the best-fit EDE models to ACT DR4, ACT DR4 + \emph{Planck} TT ($\ell_{\rm max} = 650$), and ACT DR4 + \emph{Planck} TT ($\ell_{\rm max} = 650$) + CMB lensing + BAO.  The redshifts of recombination and matter-radiation equality are also illustrated as solid vertical lines.  The ACT-only results, with high $f_{\rm EDE}(z_c)$, should be taken with a grain of salt, as the parameters of the EDE model are only weakly constrained by ACT data alone.  In the other cases shown, which correspond to more constraining data set combinations, it is evident that the EDE field starts at small $\theta_i$, as no oscillations in the EDE energy density can be seen by eye in the green or orange curves.  In addition, the relatively low best-fit values of $z_c$ in these fits are apparent: we find $z_c \approx z_*$, rather than near $z_{\rm eq}$ as suggested in previous analyses~\cite{Poulin:2018cxd,Smith:2019ihp,Hill:2020osr}.  If similar results are found at higher significance in upcoming data, there are numerous consequences for theoretical model-building.  An immediate implication is that power-law potentials may serve as reasonable candidate EDE models in the context of the ACT-driven analyses considered here, since the potential in Eq.~\eqref{eq.V} is well-approximated by a power-law near $\phi=0$.  From a theoretical standpoint, this would bring significant advantages, as a successful model could be constructed with less fine-tuning in the potential.  We also note that the analysis in this work could be extended in a more model-independent approach: although we have focused here on a specific EDE model, more generally the shape of the CMB power spectrum can be used in an analogous way to detect deviations from a fiducial expansion history out to high redshift due to the effect of expansion on the growth of perturbations.  We leave such an investigation to future work.

Another interesting theoretical implication of our results is the preference for models in which the decay constant $f$ in the EDE potential is near or even above the Planck scale.  For example, the best-fit EDE model to ACT DR4 + \emph{Planck} 2018 TT ($\ell_{\rm max} = 650$) in Table~\ref{table:params-ACT-DR4-P18TTlmax650} has $f = 1.23 \times 10^{28}$ eV, well above the Planck scale, $M_{\rm Pl} = 2.435 \times 10^{27}$ eV.  Similarly, the best-fit EDE model in the analysis of these data sets supplemented with \emph{Planck} CMB lensing and BAO data in Table~\ref{table:params-ACT-DR4-P18TTlmax650-Lens-BAO} has $f = 2.97 \times 10^{27}$ eV.  Taken at face value, these super-Planckian decay constants represent a theoretical challenge to the EDE scenario~\cite{Banks:2003sx,Rudelius:2014wla}.  Alternatively, one may choose to impose a prior $f < M_{\rm Pl}$ in the data analysis, an approach that was pursued in Ref.~\cite{Hill:2020osr}.  In that work, this prior cut did not have a significant impact, as the use of the full \emph{Planck} data set pulled $f_{\rm EDE}$ toward zero and hence $f$ toward sub-Planckian values.  Here, the preference for non-negligible $f_{\rm EDE}$ seen in ACT leads to significant posterior support for super-Planckian $f$ values, and thus it is likely that imposing a cut $f < M_{\rm Pl}$ would have a more noticeable impact.  We leave such theoretical interpretation to future work, focusing instead on the data-driven differences seen between ACT and \emph{Planck} in the context of the EDE scenario.

\begin{figure*}[!tp]
\includegraphics[width=0.87\textwidth]{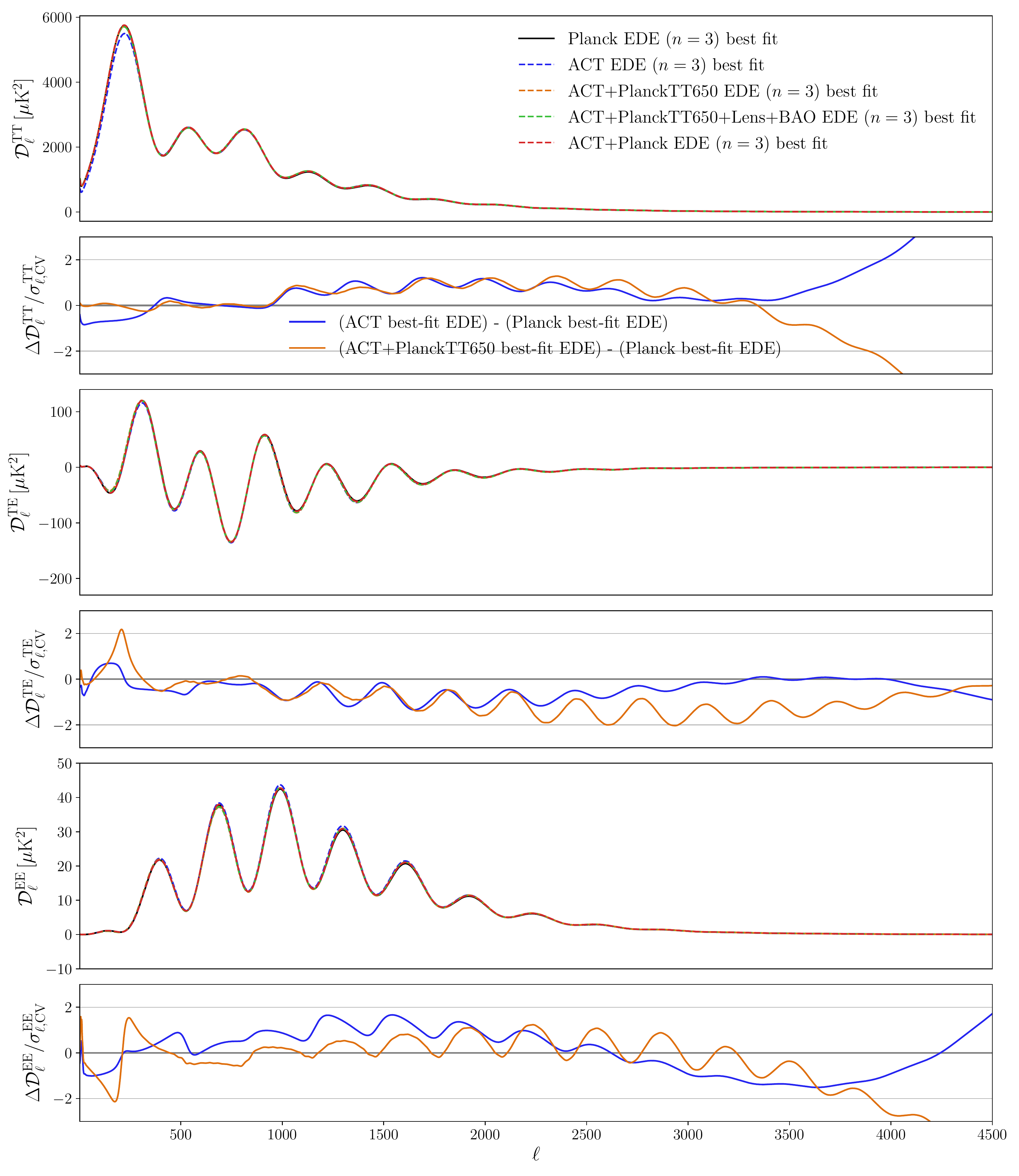}
\caption{Comparison of TT (top), TE (middle), and EE power spectra in the best-fit EDE models to the data set combinations considered in Sec.~\ref{sec:analysis} and the best-fit EDE model to \emph{Planck} data alone (the latter from Ref.~\cite{Hill:2020osr}).  The smaller panels show differences with respect to the \emph{Planck} best-fit EDE model in units of the CV-limited error bar at each $\ell$, with the $\pm 2 \sigma$ range demarcated by the thin grey lines.  The blue curve shows residuals for the best-fit EDE model to ACT DR4 alone (Table~\ref{table:params-ACT-DR4}) and the orange curve shows residuals for the best-fit EDE model to ACT + large-scale \emph{Planck} TT data (Table~\ref{table:params-ACT-DR4-P18TTlmax650}).  Fig.~\ref{fig:EDE_LCDM_bf_comp_CV} in Appendix~\ref{app:plots} shows analogous residuals to the best-fit $\Lambda$CDM model to \emph{Planck} alone, which are similar to those shown here.}
\label{fig:EDE_bf_comp_CV}
\end{figure*}

To further understand the origin of these differences between ACT and \emph{Planck}, in Fig.~\ref{fig:EDE_bf_comp_CV} we plot the best-fit EDE models to all of the data set combinations considered in this paper, as well as to the full \emph{Planck} 2018 data set on its own (from~\cite{Hill:2020osr}).  We also show the residuals of the best-fit EDE model to ACT DR4 alone with respect to the best-fit EDE model to \emph{Planck} alone, and analogous residuals for the best-fit EDE model to ACT DR4 + \emph{Planck} 2018 TT ($\ell_{\rm max} = 650$).  These residuals are shown in units of the CV-limited error bar at each multipole:
\begin{eqnarray}
\label{eq.CVerrs}
\sigma_{\ell, {\rm CV}}^{TT} & = & \sqrt{\frac{2}{2\ell+1}} C_{\ell}^{\rm TT} \\
\sigma_{\ell, {\rm CV}}^{TE} & = & \sqrt{\frac{1}{2\ell+1}} \sqrt{C_{\ell}^{\rm TT} C_{\ell}^{\rm EE} + \left(C_{\ell}^{\rm TE} \right)^2} \\
\sigma_{\ell, {\rm CV}}^{EE} & = & \sqrt{\frac{2}{2\ell+1}} C_{\ell}^{\rm EE} 
\end{eqnarray}

The residuals in Fig.~\ref{fig:EDE_bf_comp_CV} clearly illustrate the data trends that drive the fits in Sec.~\ref{subsec:ACT_alone}-~\ref{subsec:ACT_P18TTlmax650_Lens_BAO}.  The best-fit EDE model to ACT DR4 alone exhibits noticeable residuals in the EE power spectrum, particularly over $700 \lesssim \ell \lesssim 2500$.  This matches the behavior noted in Fig.~\ref{fig:ACT_alone_residuals}, which is driven by the wide-patch EE data shown in Fig.~\ref{fig:ACT_alone_wide_residuals} in Appendix~\ref{app:plots}.  The best-fit EDE model to ACT and \emph{Planck} TT ($\ell_{\rm max} = 650$) exhibits clear residuals in the TE power spectrum, deviating at the $-2\sigma_{\rm CV}$ level from the best-fit \emph{Planck} EDE model in several places, and generally lying below the \emph{Planck} model over $1000 \lesssim \ell \lesssim 4500$.  The EE residuals are less pronounced in this case, apart from a noticeable feature below the multipole range of the ACT EE data, but there is a trend in TT that is non-negligible (and also seen in the ACT-only case).

To quantify the discrepancy between the best-fit ACT EDE model and the \emph{Planck} data, we compute $\chi^2$ values using the {\tt Plik\_lite} high-$\ell$ likelihood for \emph{Planck} TT+TE+EE, in which the effects of foregrounds have already been marginalized over, thus allowing comparison of CMB-only theory spectra~\cite{Planck2018likelihood}.  This likelihood has only one free parameter, the overall calibration, which we hold fixed to unity ($A_{\rm Planck} = 1$) for these calculations.  We emphasize that the full \emph{Planck} likelihood (not the ``lite'' likelihood) is used in our primary analysis, and thus the $\chi^2$ values here should not be compared to those in Sec.~\ref{sec:analysis}.  We use the ``lite'' likelihood here to compare theory power spectra directly to the \emph{Planck} data, without needing to re-fit foreground parameters.  We obtain {\tt Plik\_lite} $\chi^2$ values for the following best-fit theory models:\footnote{Note that we do not consider the ACT-only best-fit models from Sec.~\ref{subsec:ACT_alone} here because of their anomalously low prediction for the first TT acoustic peak (see the top panel of Fig.~\ref{fig:EDE_bf_comp_CV}), which is not measured by ACT, and thus leads to misleading $\chi^2$ results when compared to \emph{Planck}.}
\begin{itemize}
    \item Best-fit $\Lambda$CDM model to \emph{Planck} TT+TE+EE + $\tau$ prior (from Ref.~\cite{Aiola2020}): $\chi^2_{\tt plik\_lite} = 581$ 
    \item Best-fit $\Lambda$CDM model to ACT DR4 TT+TE+EE + \emph{Planck} TT ($\ell_{\rm max} = 650$) + $\tau$ prior (see Table~\ref{table:params-ACT-DR4-P18TTlmax650}): $\chi^2_{\tt plik\_lite} = 662$ 
    \item Best-fit EDE model to ACT DR4 TT+TE+EE + \emph{Planck} TT ($\ell_{\rm max} = 650$) + $\tau$ prior (see Table~\ref{table:params-ACT-DR4-P18TTlmax650}): $\chi^2_{\tt plik\_lite} = 807$ 
\end{itemize}
Thus, the best-fit EDE model from Sec.~\ref{subsec:ACT_P18TTlmax650} is a significantly worse fit to \emph{Planck} than the best-fit $\Lambda$CDM model (note that there are 613 multipole bins in the {\tt Plik\_lite} likelihood, although these are partially correlated due to the effects of foreground marginalization).  However, we emphasize that this exercise is a frequentist point estimate, which does not propagate uncertainties on the parameters, as is required for a careful assessment of consistency (as done for $\Lambda$CDM in Appendix~\ref{app:ha}).  Nevertheless, it provides an intuitive assessment of the level of differences in the best-fit models to these data.

At present, we are not aware of a known systematic effect in ACT (or in \emph{Planck}) that could easily explain the differences seen here.  However, it was shown in Ref.~\cite{Aiola2020} that dividing the ACT TE power spectrum data by a factor of 1.05 yielded $\Lambda$CDM parameters in closer agreement with those from \emph{WMAP} and \emph{Planck}.  We emphasize that the factor of 1.05 has no obvious origin; see Ref.~\cite{Aiola2020} for further discussion.  Here, we find that this rescaling of the TE data leads to a broadened posterior on $f_{\rm EDE}$ that is more consistent with zero (see Sec.~\ref{subsec:ACT_alone} and~\ref{subsec:ACT_P18TTlmax650}).  However, the relatively high value of $H_0$ inferred in the analysis persists.  Understanding whether this rescaling is associated with a real effect is a critical question.

Ref.~\cite{Choi2020} also found that multiple null tests between ACT TE power spectra and ACT $\times$ \emph{Planck} cross-TE power spectra failed, in both the D56 and BOSS-N regions (see Table 17 in that work).\footnote{Note that the failed TE null tests from Ref.~\cite{Choi2020} were preliminary, in the sense that all relevant effects for the ACT vs.~\emph{Planck} comparison were not included (e.g., \emph{Planck} temperature-to-polarization leakage beams).  Robust comparisons between the data sets are the subject of ongoing work.}  It was suggested that these null test failures could be related to the mild differences in $\Lambda$CDM parameters preferred between ACT and \emph{Planck}.  It seems natural to speculate that this issue could be connected to the parameter differences seen in the EDE model here as well.  Elucidating the origin of these differences -- and whether they may be related to systematic effects in ACT -- is of utmost importance and is the focus of ongoing work.

An additional test that will be useful in future analyses will be restricting the ACT power spectra to $\ell > 1000$ (or considering other multipole cuts).  With the ACT DR4 data, this would non-negligibly weaken the overall constraining power, but for future data releases the statistical weight will be sufficiently large to allow precise tests from such data subsets.

Looking ahead in an optimistic light, the differences in Fig.~\ref{fig:EDE_bf_comp_CV} provide clear targets for upcoming CMB measurements, including those from ACT~\cite{Henderson:2015nzj}, SPT-3G~\cite{Benson2014}, Simons Observatory~\cite{Ade:2018sbj}, and CMB-S4~\cite{2019arXiv190704473A}.  To quantify this outlook, we compute the $\Delta \chi^2$ at which the models shown in Fig.~\ref{fig:EDE_bf_comp_CV} could be distinguished in upcoming analyses.  We assume an observed sky fraction $f_{\rm sky} = 0.3$ (which rescales the CV errors as $\sqrt{f_{\rm sky}}$) and CV-limited data on the following multipole ranges: $2 \leq \ell \leq 2500$ (TT); $350 \leq \ell \leq 1800$ (TE); $350 \lesssim \ell \leq 1000$ (EE).\footnote{In TT, we assume that \emph{WMAP} or large-scale \emph{Planck} data are used to reach $\ell=2$.}  We find that the best-fit ACT EDE model in this work can be distinguished from the best-fit \emph{Planck} EDE model at $\Delta \chi^2 = 550$ in such an analysis, corresponding to a $23\sigma$ preference for one model or the other.  Similarly, we find that the best-fit EDE model to ACT and large-scale \emph{Planck} TT data in this work can be distinguished from the best-fit \emph{Planck} EDE model at $\Delta \chi^2 = 470$, corresponding to a $22\sigma$ preference.  Thus, near-future CMB data will clearly distinguish between $\Lambda$CDM (or a $\Lambda$CDM-like EDE model) and an EDE model capable of yielding a significant increase in the value of $H_0$ inferred from the CMB.

As a cross-check, Fig.~\ref{fig:EDE_LCDM_bf_comp_CV} in Appendix~\ref{app:plots} shows analogous residual difference plots when comparing the ACT or ACT + large-scale \emph{Planck} TT best-fit EDE models to the best-fit $\Lambda$CDM model to \emph{Planck} 2018 data (rather than best-fit EDE model as in Fig.~\ref{fig:EDE_bf_comp_CV}).  As expected, the residuals are similar in the two plots, as the best-fit $\Lambda$CDM and EDE models to \emph{Planck} yield nearly identical power spectra ($\Delta \chi^2 = -4.1$~\cite{Hill:2020osr}).  However, minor differences can be seen: the TE residuals in Fig.~\ref{fig:EDE_LCDM_bf_comp_CV} are somewhat smaller than those in Fig.~\ref{fig:EDE_bf_comp_CV}, while the EE residuals at high multipoles are larger.  Nevertheless, it is clear that the differences between the best-fit EDE model to ACT (or ACT + large-scale \emph{Planck} TT) and either the \emph{Planck} best-fit EDE or $\Lambda$CDM models are sufficiently large to be distinguishable in upcoming CMB data.  This sets a clear target for near-future measurements.

As a final illustration of the moderate parameter differences uncovered in the EDE analyses in this work, Fig.~\ref{fig:EDE_ACT_PlanckTTlmax650_Planckfull} in Appendix~\ref{app:plots} shows the marginalized posteriors for the standard $\Lambda$CDM parameters within the EDE model fit to ACT, ACT + \emph{Planck} TT ($\ell_{\rm max} = 650$), and the full \emph{Planck} data on their own.  While the agreement is reasonable in most of the 2D posterior plots, there are nevertheless multiple 2D posteriors for which the \emph{Planck} and ACT contours do not overlap, even at 95\% CL.  Uncovering the origin of the preferences for these different regions of EDE parameter space is necessary to re-establish the robustness of cosmological constraints on this scenario.

We conclude by noting that in most CMB analyses to date, it has sufficed to establish consistency across data sets by comparing $\Lambda$CDM parameter constraints.  With this work, we have demonstrated that current CMB data are now sufficiently powerful that this is no longer the case: ACT and \emph{Planck} are in agreement within $\Lambda$CDM, but in some disagreement within EDE.  Understanding this disagreement will be a crucial focus of upcoming work.


\acknowledgments
We are grateful to Evan McDonough and Michael Toomey for their contributions to the development of {\tt CLASS\_EDE} and for useful conversations, and we thank the Scientific Computing Core staff at the Flatiron Institute for computational support.  The Flatiron Institute is supported by the Simons Foundation.  EC acknowledges support from the STFC Ernest Rutherford Fellowship ST/M004856/2 and STFC Consolidated Grant ST/S00033X/1.  EC and UN acknowledge support from the European Research Council (ERC) under the European Union’s Horizon 2020 research and innovation programme (Grant agreement No. 849169).  JD and ES acknowledge support from NSF grant AST-1814971.  NS acknowledges support from NSF grant number AST-1907657.  MHi and KM acknowledge support from the National Research Foundation of South Africa.  VG is supported by the National Science Foundation under Grant No.~PHY-2013951.  ZX is supported by the Gordon and Betty Moore Foundation.  Research at Perimeter Institute is supported in part by the Government of Canada through the Department of Innovation, Science and Industry Canada and by the Province of Ontario through the Ministry of Colleges and Universities.  ADH acknowledges support from the Sutton Family Chair in Science, Christianity and Cultures and from the Faculty of Arts and Science, University of Toronto.  SKC acknowledges support from NSF award AST-2001866.  EV acknowledges support from the NSF GRFP via Grant No.~DGE-1650441.  CS acknowledges support from the Agencia Nacional de Investigaci\'on y Desarrollo (ANID) under FONDECYT grant no.~11191125.  This work was completed at the Aspen Center for Physics, which is supported by National Science Foundation grant PHY-1607611.

Support for ACT was through the U.S.~National Science Foundation through awards AST-0408698, AST-0965625, and AST-1440226 for the ACT project, as well as awards PHY-0355328, PHY-0855887 and PHY-1214379. Funding was also provided by Princeton University, the University of Pennsylvania, and a Canada Foundation for Innovation (CFI) award to UBC.  ACT operates in the Parque Astron\'omico Atacama in northern Chile under the auspices of the Agencia Nacional de Investigaci\'on y Desarrollo (ANID).  The development of multichroic detectors and lenses was supported by NASA grants NNX13AE56G and NNX14AB58G.  Detector research at NIST was supported by the NIST Innovations in Measurement Science program. 

We acknowledge use of the {\tt matplotlib}~\cite{Hunter2007}, {\tt numpy}~\cite{2020Natur.585..357H}, {\tt GetDist}~\cite{Lewis:2019xzd}, {\tt Cobaya}~\cite{Torrado:2020dgo}, and {\tt CosmoMC}~\cite{LewisBridle2002} packages and use of the Boltzmann codes {\tt CAMB}~\cite{Lewis:1999bs} and {\tt CLASS}~\cite{Blas2011}.

\appendix

\section{High-accuracy CMB lensing calculations and ACT DR4 cosmology}
\label{app:ha}

In this appendix, we investigate the accuracy settings used in standard Einstein-Boltzmann codes for theoretical calculations of the primary CMB power spectra.  We show that the accuracy settings in  calculations used for the analysis of the \textit{Planck} CMB data are not high enough for analysis of current and future high-resolution, low-noise CMB observations. We show that this already has a small, but notable, effect on the $\Lambda$CDM cosmological parameter constraints from the ACT DR4 release.  For future experiments (e.g., SO~\cite{Ade:2018sbj} and CMB-S4~\cite{2019arXiv190704473A}), biases in parameter inference due to the use of \emph{Planck}-level accuracy settings in Einstein-Boltzmann codes can be many times larger than the statistical error bars~\cite{McCarthy2021}.

\subsection{Introduction}
Ref.~\cite{Aiola2020}, hereafter A20, presented cosmological parameter constraints inferred from ACT DR4. Cosmological parameters were derived from the ACT TT, TE, and EE CMB-only (foreground-marginalized) power spectra~\citep{Choi2020} alone or in combination with WMAP~\citep{Bennett2013} or \textit{Planck}~\citep{Planck2018likelihood}. 

The parameter estimation pipeline in A20 made use of the Fortran-90 version of the \texttt{CAMB/CosmoMC} package~\citep{Lewis:1999bs,LewisBridle2002}\footnote{In A20 we used the January 2017 versions of these codes, but later releases lead to minimal, negligible changes.} and theory predictions were computed to a maximum multipole $\ell_{\rm max}=6000$, well beyond the maximum CMB multipole retained in the ACT DR4 cosmological analyses, $\ell_{\rm max, data} = 4500$. Other settings in the theory calculations were left to the standard \texttt{CAMB} high-accuracy defaults, which are set to give a target accuracy of 0.1\% in the lensing power spectrum for multipoles $500<L<2000$ (optimal for \textit{Planck} cosmology). In particular, in A20, the parameter \texttt{k\_eta\_max\_scalar}, the maximum of the product of the wavenumber of scalar perturbations and the curvature perturbation variable,\footnote{See the CAMB documentation at \url{https://cosmologist.info/notes/CAMB.pdf} .} $k\eta$, was internally computed from the $\ell_{\rm max}$ value as $2\ell_{\rm max}$. Subsequent investigations and in particular, comparisons between \texttt{CAMB} and \texttt{CLASS}~\citep{Blas2011}\footnote{All \texttt{CLASS} calculations in this appendix (and elsewhere in the paper) use \texttt{CLASS-2.8}.}, determined that the use of higher accuracy settings changes the theoretical predictions for the lensed CMB power spectra at a level that is marginally non-negligible for the ACT DR4 data.  For upcoming CMB data sets, the use of higher accuracy settings is evidently required.

In this appendix, we show how the use of high-accuracy settings impacts the baseline ACT DR4 cosmology results and make recommendations for theory predictions used for analyzing these data and future low-noise, high-resolution CMB observations.

\subsection{High-accuracy lensing}
At CMB multipoles $\ell > 3000$, the CMB damping-tail predictions  depend on the CMB lensing potential power spectrum at scales $L>2000$. Achieving high accuracy in the theory calculations on those lensing scales requires setting a number of parameters in the \texttt{CAMB} and \texttt{CLASS} initialization (``ini'') files. To identify settings providing accuracy high enough for the analysis of high-precision small-scale CMB data, we fine-tune a number of parameters that enter the lensing calculations, similarly to what was done for the \textit{Planck} analyses.\footnote{The default accuracy settings of \texttt{CLASS} are specified such that the theory predictions are sufficiently accurate for \emph{Planck}-like data --- see~\cite{Lesgourgues_2011} for details.}  This includes exploration of the parameters \texttt{lens\_potential\_accuracy}, \texttt{lens\_margin}, \texttt{AccuracyBoost}, \texttt{lSampleBoost}, \texttt{lAccuracyBoost}, and \texttt{DoLateRadTruncation} in the Python-CAMB package, \texttt{pycamb}, as well as using increased values of $\ell_{\rm max}$. 

As a way to identify suitable settings, we explore convergence in the ACT DR4 $\chi^2$ for a $\Lambda$CDM cosmology, i.e., monitoring changes in the ACT likelihood in response to changes in the theoretical predictions while holding the cosmological parameters fixed to the ACT+WMAP solutions in A20 but increasing the accuracy parameters.  We verify that the following accuracy parameters suffice for ACT DR4:
\begin{itemize}
    \item \texttt{lens\_potential\_accuracy = 8}
    \item \texttt{lens\_margin = 1050}
    \item \texttt{AccuracyBoost = 2.0} 
    \item \texttt{lSampleBoost = 2.0} 
    \item \texttt{lAccuracyBoost = 2.0}
    \item \texttt{DoLateRadTruncation = False} 
\end{itemize}
in \texttt{pycamb}, or 
\begin{itemize}
    \item \texttt{neglect\_CMB\_sources\_below\_visibility = 1.e-30}
    \item \texttt{transfer\_neglect\_late\_source = 3000.}
    \item \texttt{halofit\_k\_per\_decade = 3000.}
    \item \texttt{accurate\_lensing = 1}
    \item \texttt{num\_mu\_minus\_lmax = 1000.}
    \item \texttt{delta\_l\_max = 1000.}
    \item \texttt{k\_min\_tau0 = 0.002}
    \item \texttt{k\_max\_tau0\_over\_l\_max = 3.}
    \item \texttt{k\_step\_sub = 0.015}
    \item \texttt{k\_step\_super = 0.0001}
    \item \texttt{k\_step\_super\_reduction = 0.1}
\end{itemize}
in \texttt{CLASS}.  In both codes, the requested maximum multipole for the lensed CMB power spectra, $\ell_{\rm max}$, should be set well above $\ell_{\rm max, data}$, e.g., $\ell_{\rm max} = 10000$ is (more than) sufficient for ACT DR4.

However, note that we do not explore how \emph{low} the accuracy settings can be set such that one still obtains converged $\chi^2$ for the ACT DR4 data. Lowering the parameters to the point that they are just above the necessary tolerance level would be needed to optimize the overall computation efficiency in an MCMC analysis, but we find that this is not required for the ACT cosmological parameter runs, as the theory predictions are still sufficiently fast.

\begin{figure}[!t]
\includegraphics[width=\columnwidth]{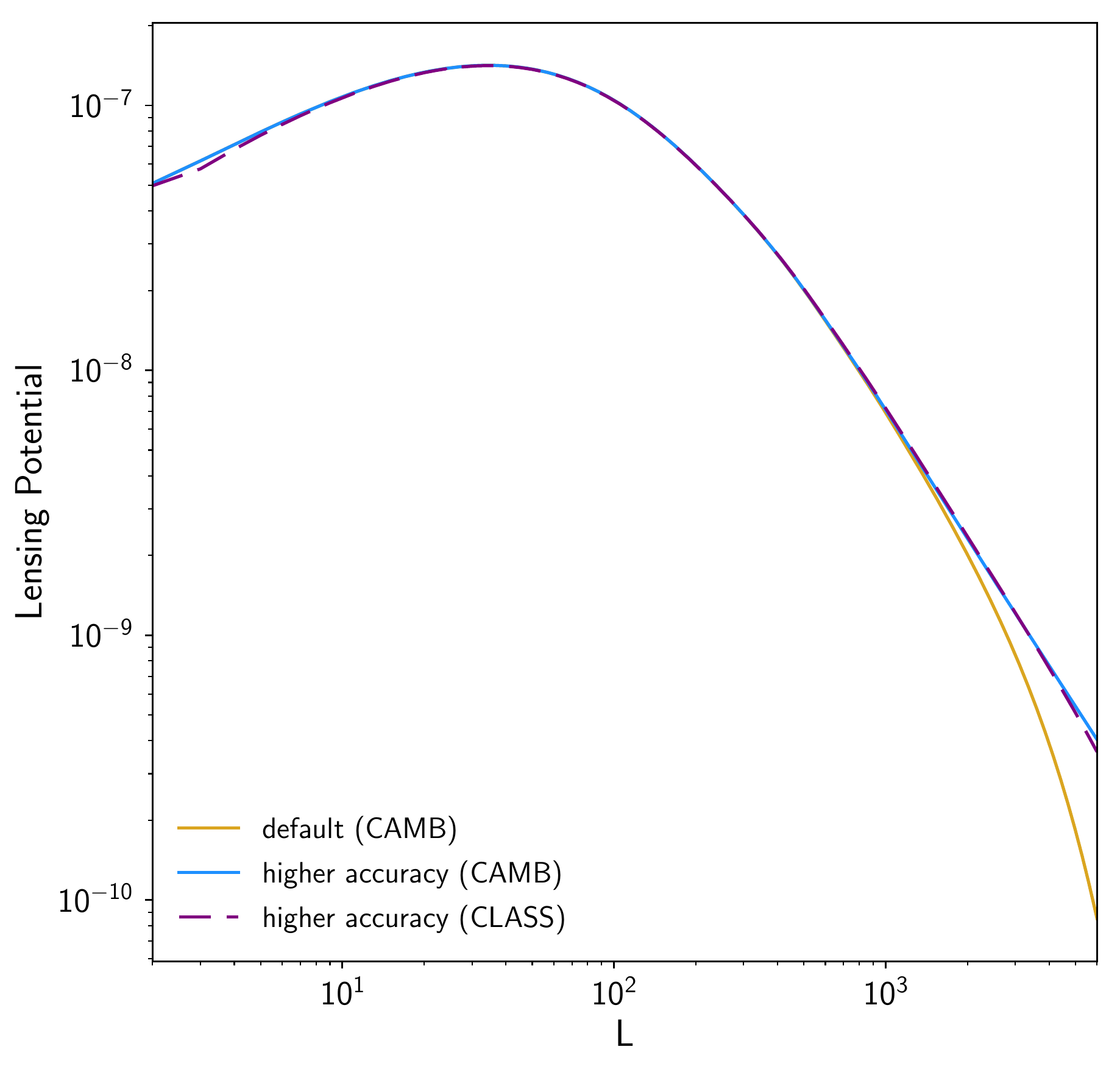}
\caption{CMB lensing potential power spectrum, $[L(L+1)]^2 C_L^{\phi\phi}/(2\pi)$, obtained with the \texttt{CAMB} default high-accuracy settings (solid orange), and with the higher-accuracy settings presented here from \texttt{CAMB} (solid blue) or \texttt{CLASS} (dashed purple). All curves are computed with identical cosmological parameters (column 4 of Table 4 in A20).}
\label{fig:ppha}
\end{figure}

\begin{figure}[!tp]
\includegraphics[width=\columnwidth]{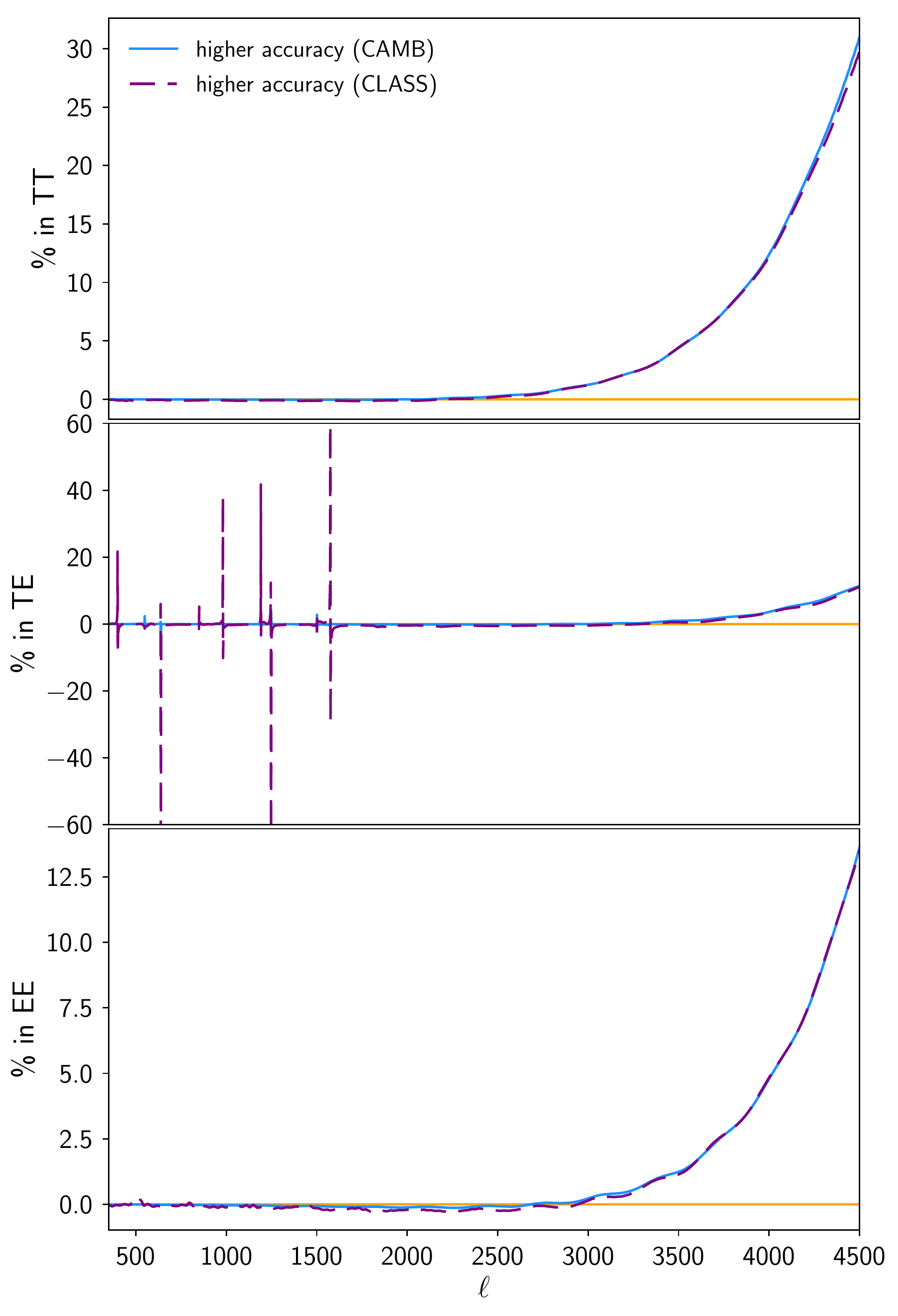}
\caption{Fractional differences in the lensed CMB TT, TE, and EE power spectra, (higher-accuracy -- default)/default, showing the deviation of the two sets of accuracy settings.  The models and settings shown are identical to those used in Fig.~\ref{fig:ppha}.}
\label{fig:hattteee}
\end{figure}

Fig.~\ref{fig:ppha} shows the impact of the default high-accuracy settings and the higher-accuracy ones presented above for the CMB lensing power spectrum prediction.  The higher-accuracy and default-accuracy calculations differ by 15\% at $L=2000$, 43\% at $L=3000$, and 97\% at $L=4000$.  We obtain excellent agreement between \texttt{CAMB} and \texttt{CLASS} for the higher-accuracy runs.

This change in the lensing power spectrum corresponds to a change in the tails of the TT, TE, and EE power spectra, as seen in Fig.~\ref{fig:hattteee}, which shows the fractional change in these quantities when computed with the higher-accuracy versus default-accuracy settings.  In particular, fractional differences greater than $1\%$ are seen in the TT tail at $\ell>3000$.  At $\ell=3000$, $3500$, and $4000$, the fractional differences in TT are 1.2\%, 4.4\%, and 12\%, respectively.  At a given $\ell$ value in the damping tail, the fractional contribution from lensing to the total power is larger in TT than in EE because of the larger gradient in the primary CMB temperature field, which is why the effects seen here are larger in TT. This coherent increase in amplitude of the signal predicted in the TT tail will be evident at the $\chi^2$ level (since the ACT DR4 error bars are larger in EE than TT, the same trend seen in EE will have a much smaller impact).

To match these accuracy settings with the parameters in the \texttt{CAMB/CosmoMC} runs one needs to explicitly define the maximum $k\eta$ and set it to \texttt{k\_eta\_max\_scalar=144000}. This specific value converts the \texttt{pycamb} higher-accuracy settings (in particular, \texttt{lens\_potential\_accuracy}) to the Fortran-CAMB settings.\footnote{Antony Lewis, priv.\ comm.}

It is important to note that on these small scales, the lensing calculation itself is subject to uncertainty due to nonlinear evolution and baryonic effects.  Here we adopt a fixed nonlinear model,  the Halofit fitting function~\citep{Takahashi2012}, and simply explore the effect of increasing the numerical accuracy settings in \texttt{CAMB} and \texttt{CLASS}.  For a detailed assessment of the impact of nonlinear and baryonic uncertainties on high-$\ell$ CMB theory predictions for upcoming experiments, see Ref.~\cite{McCarthy2021}.

\subsection{Updated ACT DR4 cosmology}

\begin{figure}[tp]
\includegraphics[width=\columnwidth]{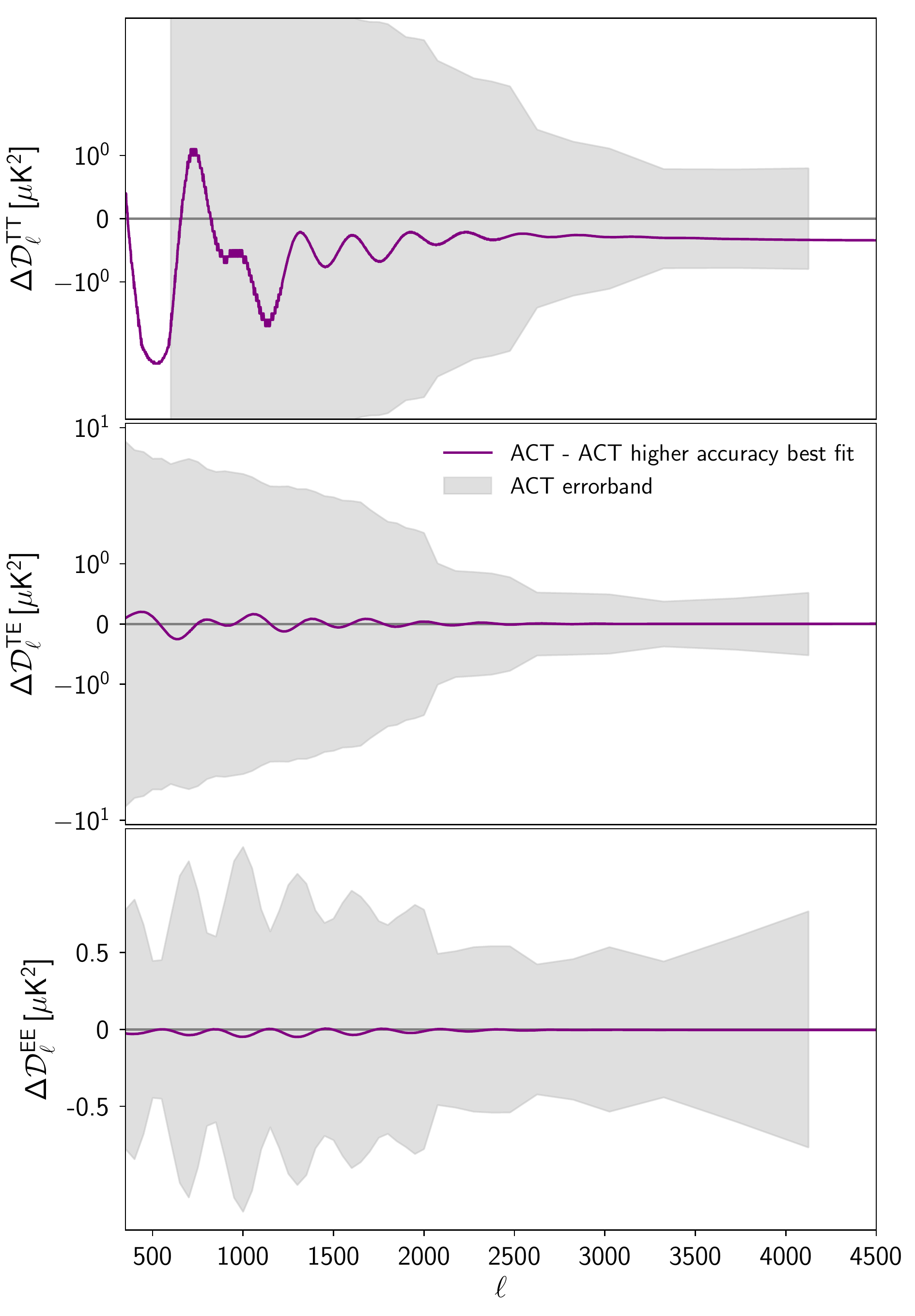}
\caption{Differences in the $\Lambda$CDM best-fit CMB power spectra to the ACT DR4 data using default or higher-accuracy settings (purple lines). The grey bands show the $1\sigma$ error-band on the ACT DR4 spectra. We note that, although small, the differences are coherent over a large range of multipoles in the TT tail.}
\label{fig:hamodel}
\end{figure}

\begin{figure*}[ht!]
\includegraphics[width=\textwidth]{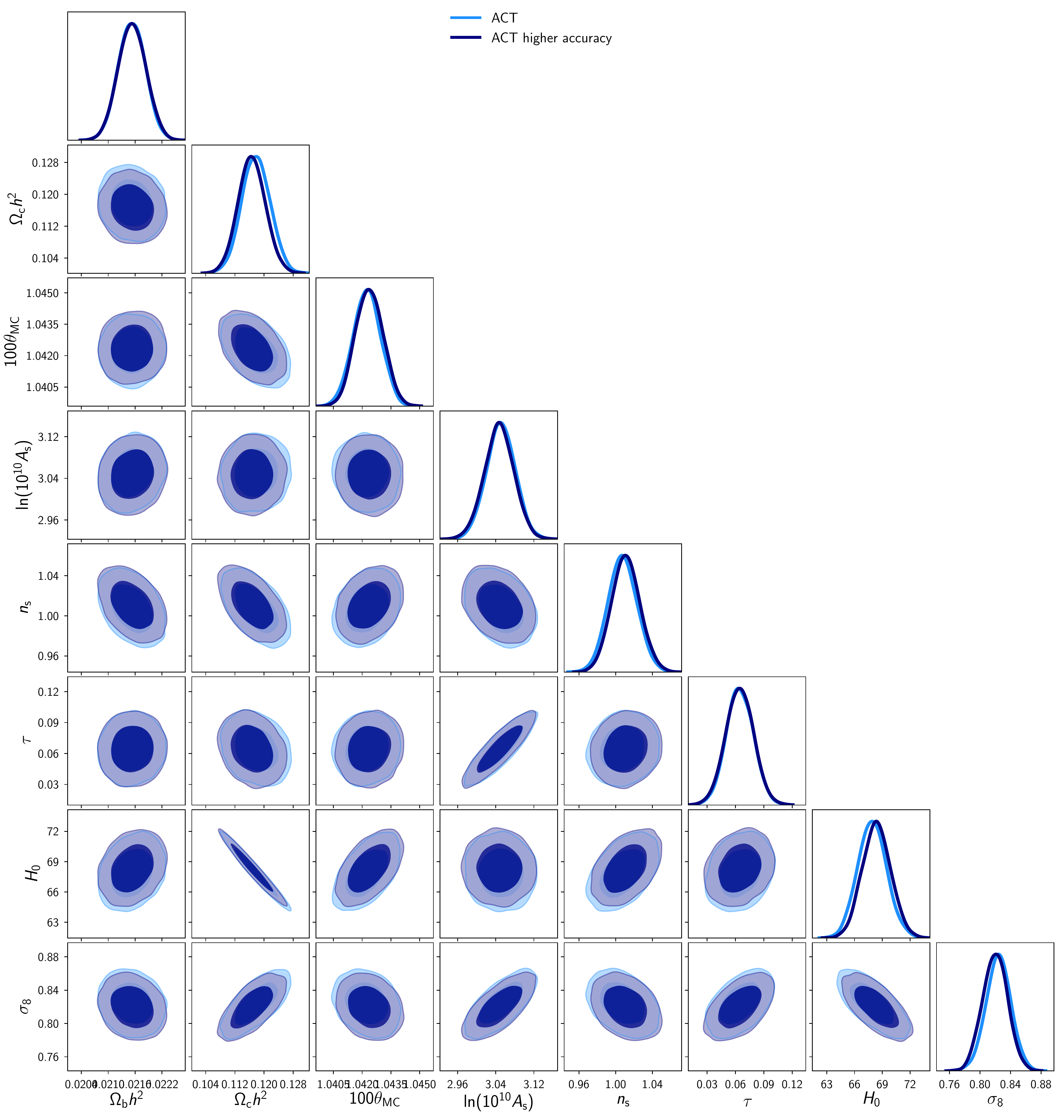}
\caption{$\Lambda$CDM and derived parameters from A20 ACT-alone (light blue) compared with results from higher-accuracy settings (dark blue).}
\label{fig:hatri}
\end{figure*}
\begin{figure*}[ht!]
\includegraphics[width=\textwidth]{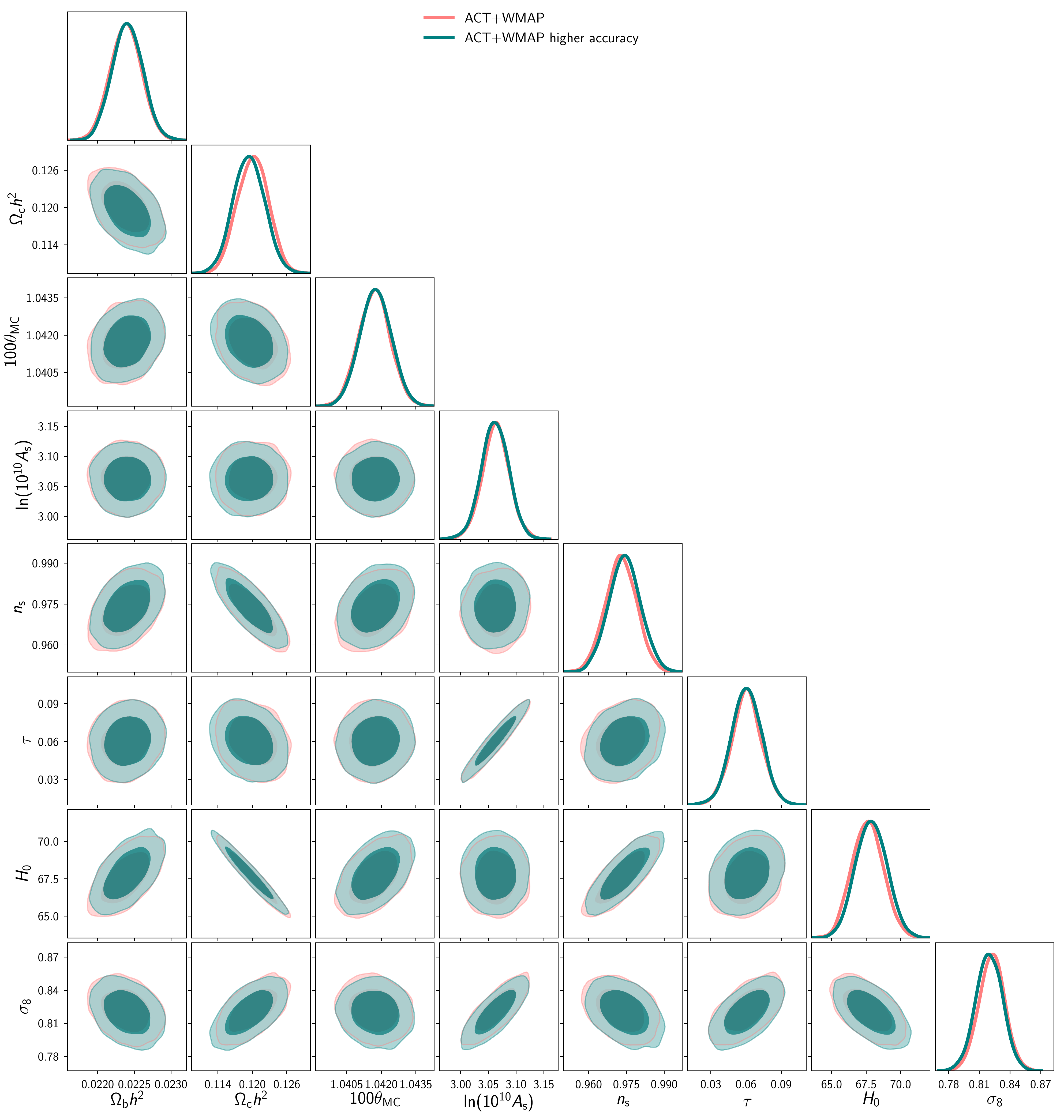}
\caption{$\Lambda$CDM and derived parameters from A20 ACT+WMAP (orange) compared with results from higher-accuracy settings (teal).}
\label{fig:haawtri}
\end{figure*}

\begin{table*}[tp!]
\begin{center}
\begin{tabular}{lcccc}
\hline
\hline
Parameter & ACT (A20) & ACT HA & ACT+WMAP (A20) & ACT+WMAP HA\\
\hline
{$100\Omega_b h^2$} & $2.153 \pm 0.030$ & $2.153\pm0.030$ & $2.239\pm0.021$ & $2.241\pm0.020$\\
{$100\Omega_c h^2$} & $11.78\pm0.38$ & $11.66\pm0.37$ & $12.00\pm0.26$ & $11.94\pm0.27$\\
{$10^4\theta_{MC}$} & $104.225\pm0.071$ & $104.238\pm0.071$ & $104.170\pm0.067$ & $104.175\pm0.067$\\
{$\tau$}& $0.065\pm0.014$ & $0.064\pm0.015$ & $0.061\pm0.012$ & $0.060\pm0.013$\\
{$n_s$} & $1.008\pm0.015 $ & $1.011\pm0.015$ & $0.9729\pm0.0061$ & $0.9743\pm0.0062$ \\
{${\rm{ln}}(10^{10} A_s)$}& $3.050\pm0.030$ & $3.047\pm0.031$ & $3.064\pm0.024$ & $3.062\pm0.024$\\
{$H_0$ [km/s/Mpc]} & $67.9\pm1.5$ & $68.4\pm1.5$ & $67.6\pm1.1$ & $67.9\pm1.1$\\
{$\sigma_8$} & $0.824\pm0.016$ & $0.820\pm0.016$ & $0.822\pm0.012$ & $0.820\pm0.013$\\
\hline
\hline
\end{tabular}
\caption{$\Lambda$CDM parameters as presented in A20 and as obtained with the higher-accuracy (HA) settings defined here. }
\label{tab:comparison}
\end{center}
\end{table*}

The amplitude of this effect compared to the sensitivity of the ACT DR4 data is shown in Fig.~\ref{fig:hamodel}.  The figure shows the difference between the $\Lambda$CDM best-fit model to the ACT data as determined with the default or with the higher-accuracy calculations. These best-fits are obtained by re-running the \texttt{CosmoMC} minimizer as in A20.  We note that in this figure an adjustment in the best-fit cosmological parameters absorbs and changes some of the differences seen in Fig.~\ref{fig:hattteee}, where the cosmology was kept fixed.  Although small, we note the presence of a coherent trend in the TT tail as well as features at other scales.

We re-run the ACT DR4 cosmological parameter constraints with these higher-accuracy settings in {\tt CosmoMC}.  We show and quantify differences compared to A20 in Figs.~\ref{fig:hatri} and \ref{fig:haawtri}, and Table~\ref{tab:comparison}.  The main differences are summarized below:

\begin{itemize}
    \item The most notable impact is a change in the best-fit $\chi^2$. For the ACT-alone best-fit, we find $\Delta \chi^2_{\rm ACT}=3.7$, i.e., the best-fit $\Lambda$CDM model to the ACT data has a worse $\chi^2$ here than found in A20.  This corresponds to a decrease in the PTE from 0.13 (A20) to 0.10 (here).  For the ACT+WMAP best-fit, we find $\Delta \chi^2_{\rm ACT}=4.0$, i.e., again the best-fit $\Lambda$CDM model to the ACT+WMAP data has a worse $\chi_{\rm ACT}^2$ here than found in A20.
    
    \item Small differences appear in the $\Omega_c h^2$ and $n_s$ distributions, with small changes in the mean and width ($0.2-0.3\sigma$ for the ACT-only analysis and $0.2\sigma$ for ACT+WMAP). Changes in the derived parameters $\sigma_8$ and $H_0$ are also at the level of $0.2-0.3\sigma$.
    
    \item Within $\Lambda$CDM, the ACT DR4 parameters are consistent with WMAP at $2.4\sigma$ (unchanged from A20), while the ACT DR4 consistency with \textit{Planck} improves to $2.5\sigma$ (here) from $2.7\sigma$ (A20).
    
    \item No impact is seen in the marginalized constraints on beyond-$\Lambda$CDM parameters, e.g., $N_{\rm eff}$ and $A_L$ runs yield the same results for these parameters as in A20.  The stability of $A_L$, which may seem surprising, is explained by the fact that this parameter is defined as the ratio of the measured amount of lensing compared to that predicted by a given set of $\Lambda$CDM parameters; thus the change in the $\Lambda$CDM parameters described above absorbs the effect, and $A_L$ is unchanged.
\end{itemize}

\subsection{Outlook}
We recommend the use of the higher-accuracy settings defined here for ACT DR4 and future releases, and for other high-resolution CMB data.

\section{Additional Posterior and Residual Plots}
\label{app:plots}

This appendix contains additional parameter posterior plots and power spectrum residual plots associated with the analyses presented in Sec.~\ref{sec:analysis} and~\ref{sec:discussion}.  Relevant details are provided in the figure captions.

\begin{figure*}[!tp]
\includegraphics[width=\textwidth]{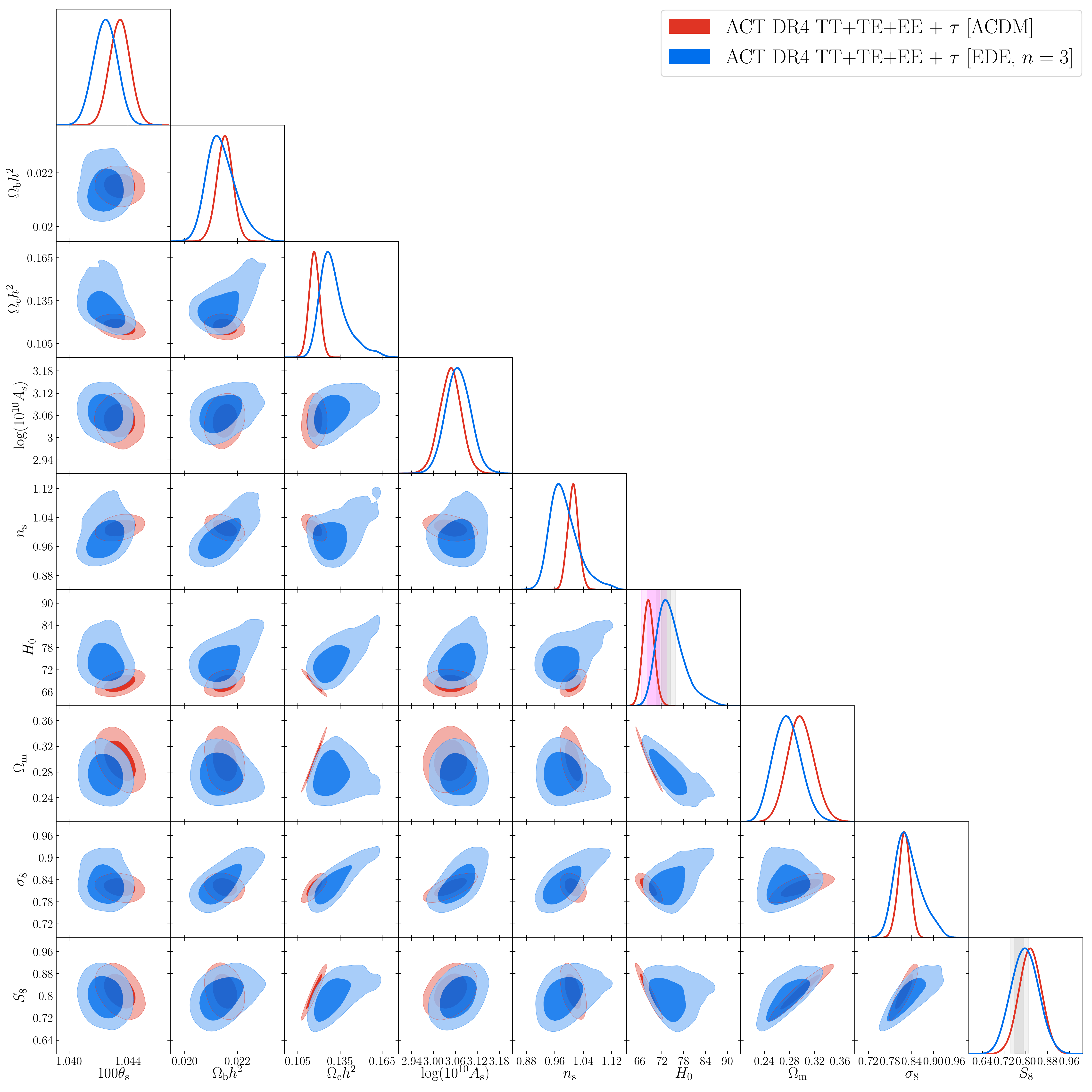}
\caption{Marginalized posteriors for the standard cosmological parameters in the $\Lambda$CDM (red) and EDE (blue) models fit to ACT DR4 TT+TE+EE data in combination with a Gaussian prior on $\tau$, as presented in Sec.~\ref{subsec:ACT_alone}.  The vertical grey and magenta bands in the $H_0$ panel show the latest SH0ES~\cite{Riess:2020fzl} and TRGB~\cite{Freedman:2021ahq} constraints, respectively.  The vertical grey band in the $S_8$ panel shows the DES-Y3 constraint~\cite{DES:2021wwk}.}
\label{fig:ACT_alone}
\end{figure*}

\begin{figure*}[!tp]
\includegraphics[width=\textwidth]{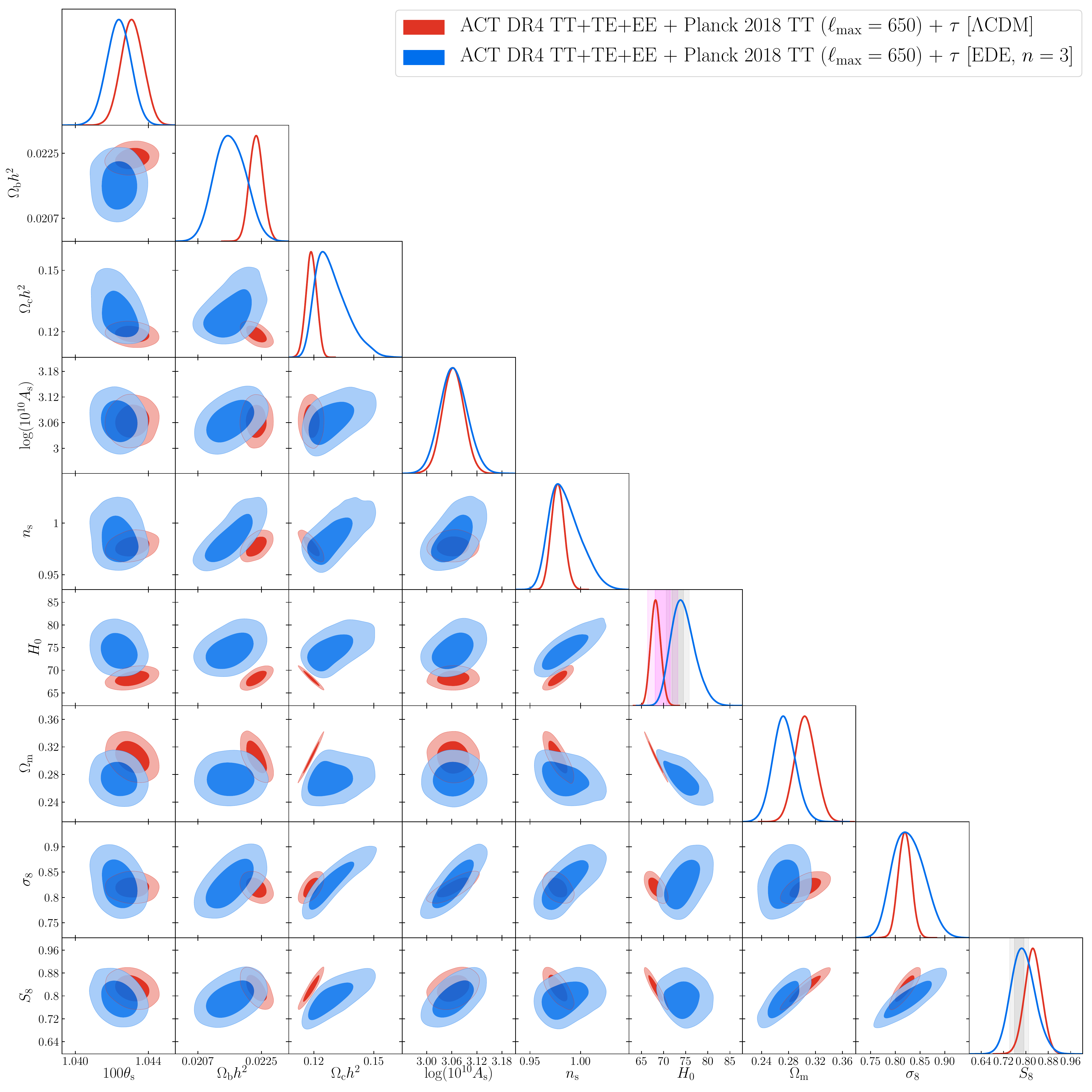}
\caption{Marginalized posteriors for the standard cosmological parameters in the $\Lambda$CDM (red) and EDE (blue) models fit to ACT DR4 TT+TE+EE and \emph{Planck} 2018 TT ($\ell_{\rm max} = 650$) data in combination with a Gaussian prior on $\tau$, as presented in Sec.~\ref{subsec:ACT_P18TTlmax650}.  The vertical grey and magenta bands in the $H_0$ panel show the latest SH0ES~\cite{Riess:2020fzl} and TRGB~\cite{Freedman:2021ahq} constraints, respectively.  The vertical grey band in the $S_8$ panel shows the DES-Y3 constraint~\cite{DES:2021wwk}.}
\label{fig:ACT_PlanckTTlmax650}
\end{figure*}

\begin{figure*}[!tp]
\includegraphics[width=\textwidth]{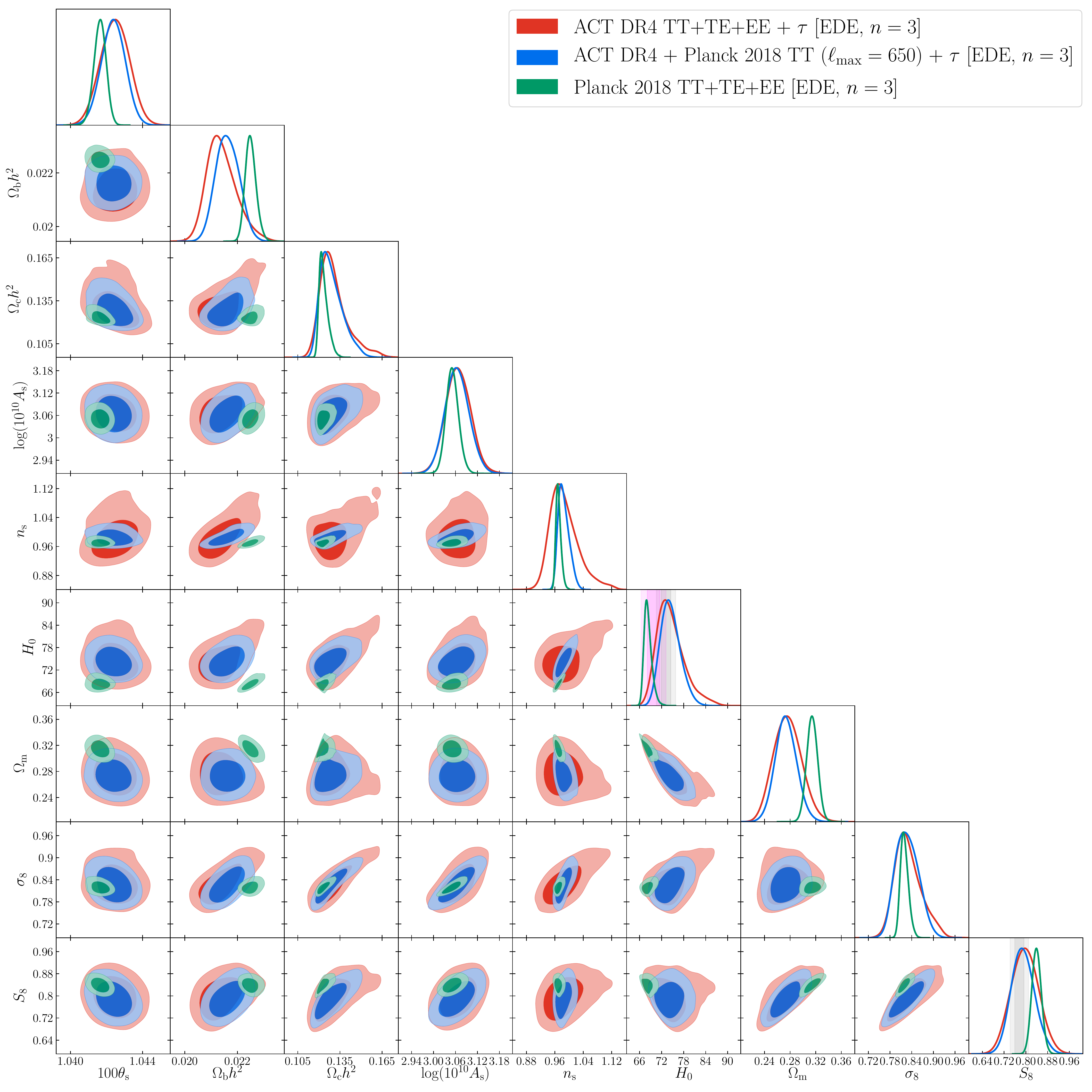}
\caption{Marginalized posteriors for the standard cosmological parameters in the EDE model fit to ACT DR4 TT+TE+EE data (red), ACT DR4 and \emph{Planck} 2018 TT ($\ell_{\rm max} = 650$) data (blue), and \emph{Planck} 2018 TT+TE+EE data (green, results from Ref.~\cite{Hill:2020osr}).  The vertical grey and magenta bands in the $H_0$ panel show the latest SH0ES~\cite{Riess:2020fzl} and TRGB~\cite{Freedman:2021ahq} constraints, respectively.  The vertical grey band in the $S_8$ panel shows the DES-Y3 constraint~\cite{DES:2021wwk}.  Within the EDE model, both ACT and ACT + large-scale \emph{Planck} TT data appear to prefer somewhat different regions of parameter space than that preferred by the full \emph{Planck} data set on its own, but the uncertainties in the ACT-based analyses are too large to draw strong conclusions at present.}
\label{fig:EDE_ACT_PlanckTTlmax650_Planckfull}
\end{figure*}

\begin{figure*}[!tp]
\includegraphics[width=\textwidth]{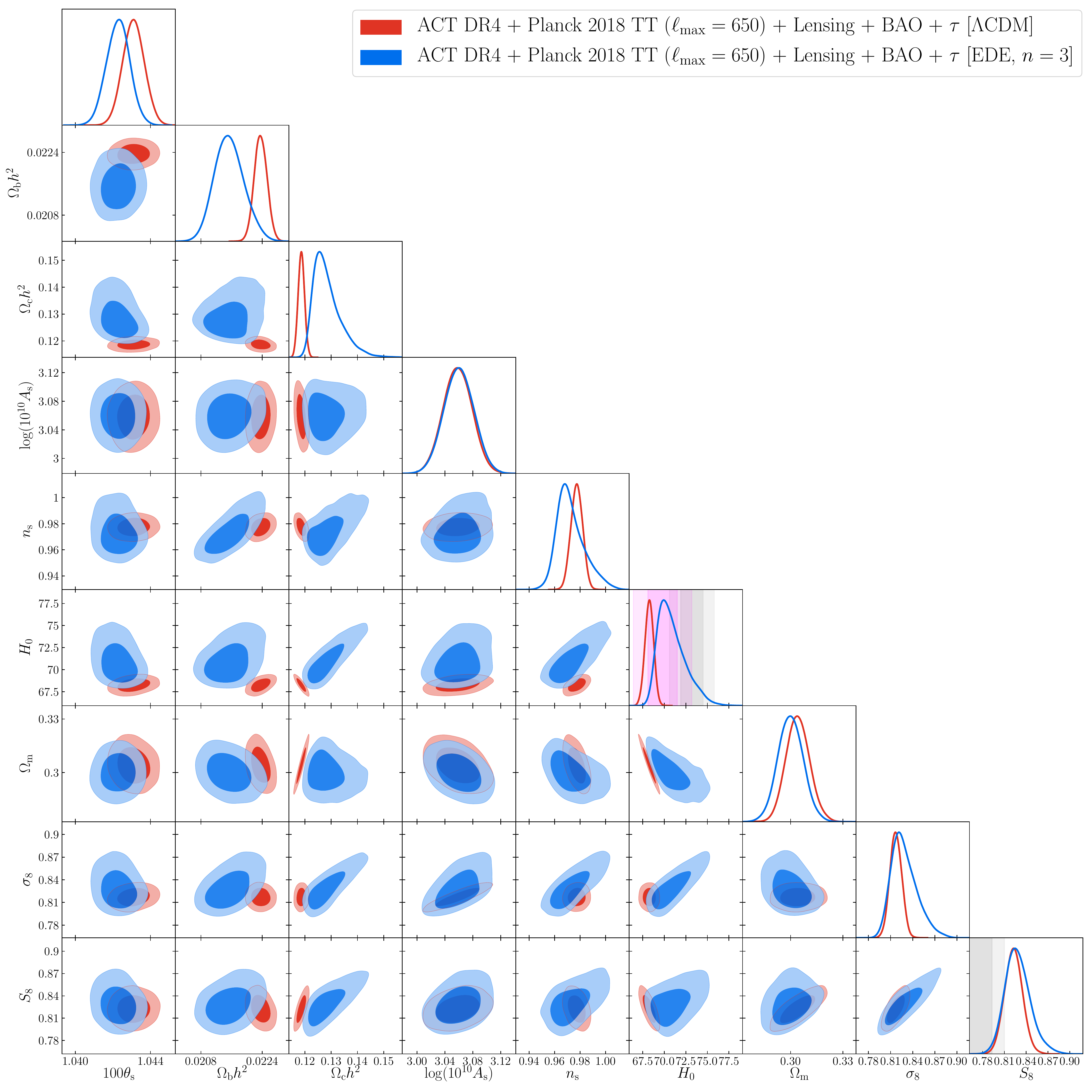}
\caption{Marginalized posteriors for the standard cosmological parameters in the $\Lambda$CDM (red) and EDE (blue) models fit to ACT DR4 TT+TE+EE, \emph{Planck} 2018 TT ($\ell_{\rm max} = 650$), \emph{Planck} 2018 CMB lensing, and BAO data in combination with a Gaussian prior on $\tau$, as presented in Sec.~\ref{subsec:ACT_P18TTlmax650_Lens_BAO}.  The vertical grey and magenta bands in the $H_0$ panel show the latest SH0ES~\cite{Riess:2020fzl} and TRGB~\cite{Freedman:2021ahq} constraints, respectively.  The vertical grey band in the $S_8$ panel shows the DES-Y3 constraint~\cite{DES:2021wwk}.}
\label{fig:ACT_PlanckTTlmax650_BAO_Lens}
\end{figure*}

\begin{figure*}[!tp]
\includegraphics[width=\textwidth]{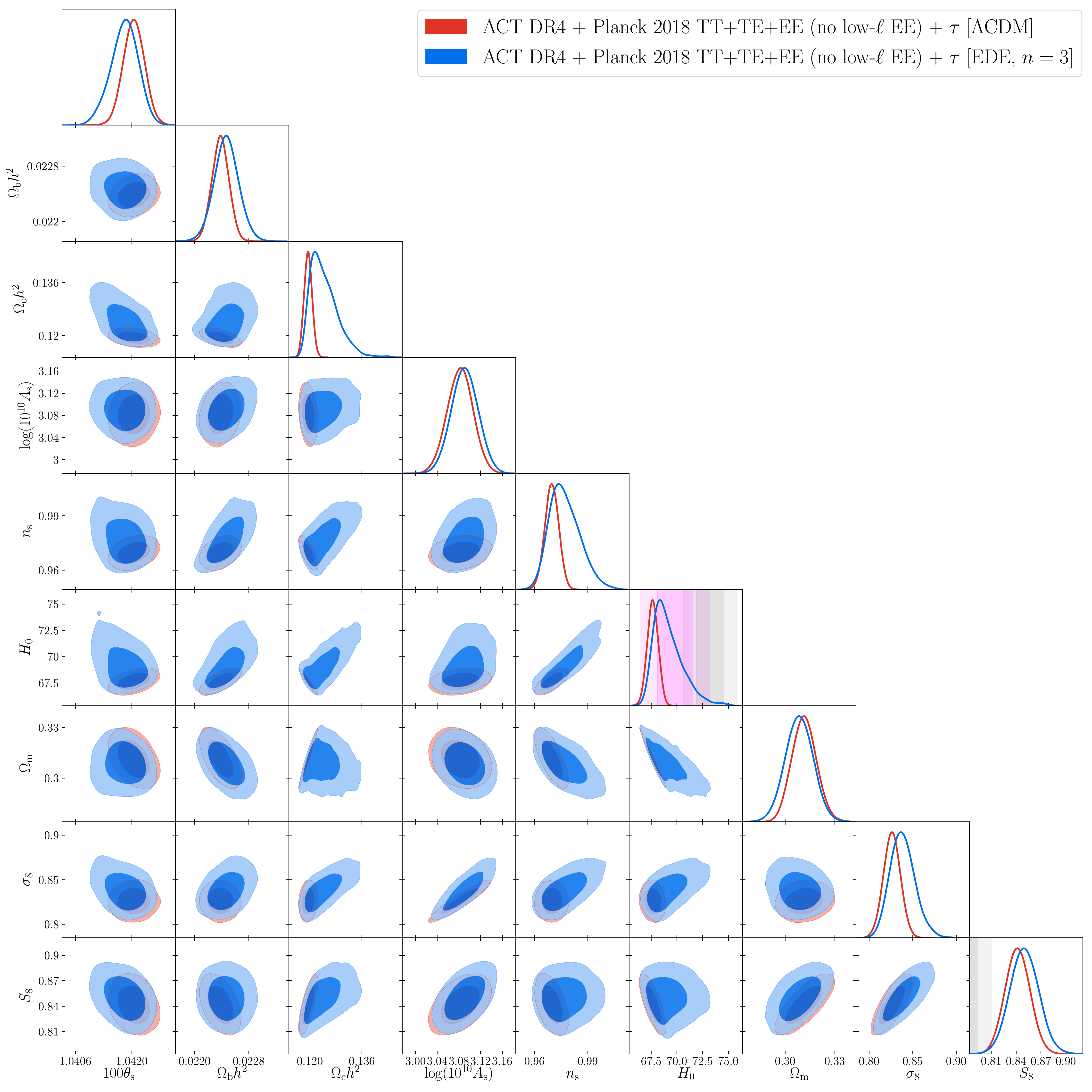}
\caption{Marginalized posteriors for the standard cosmological parameters in the $\Lambda$CDM (red) and EDE (blue) models fit to ACT DR4 TT+TE+EE and \emph{Planck} 2018 TT+TE+EE (full $\ell$ range, but no low-$\ell$ EE) data in combination with a Gaussian prior on $\tau$, as presented in Sec.~\ref{subsec:ACT_P18full}.  The vertical grey and magenta bands in the $H_0$ panel show the latest SH0ES~\cite{Riess:2020fzl} and TRGB~\cite{Freedman:2021ahq} constraints, respectively.  The vertical grey band in the $S_8$ panel shows the DES-Y3 constraint~\cite{DES:2021wwk}.}
\label{fig:ACT_Planckfull}
\end{figure*}

\begin{figure*}[!tp]
\includegraphics[width=\textwidth]{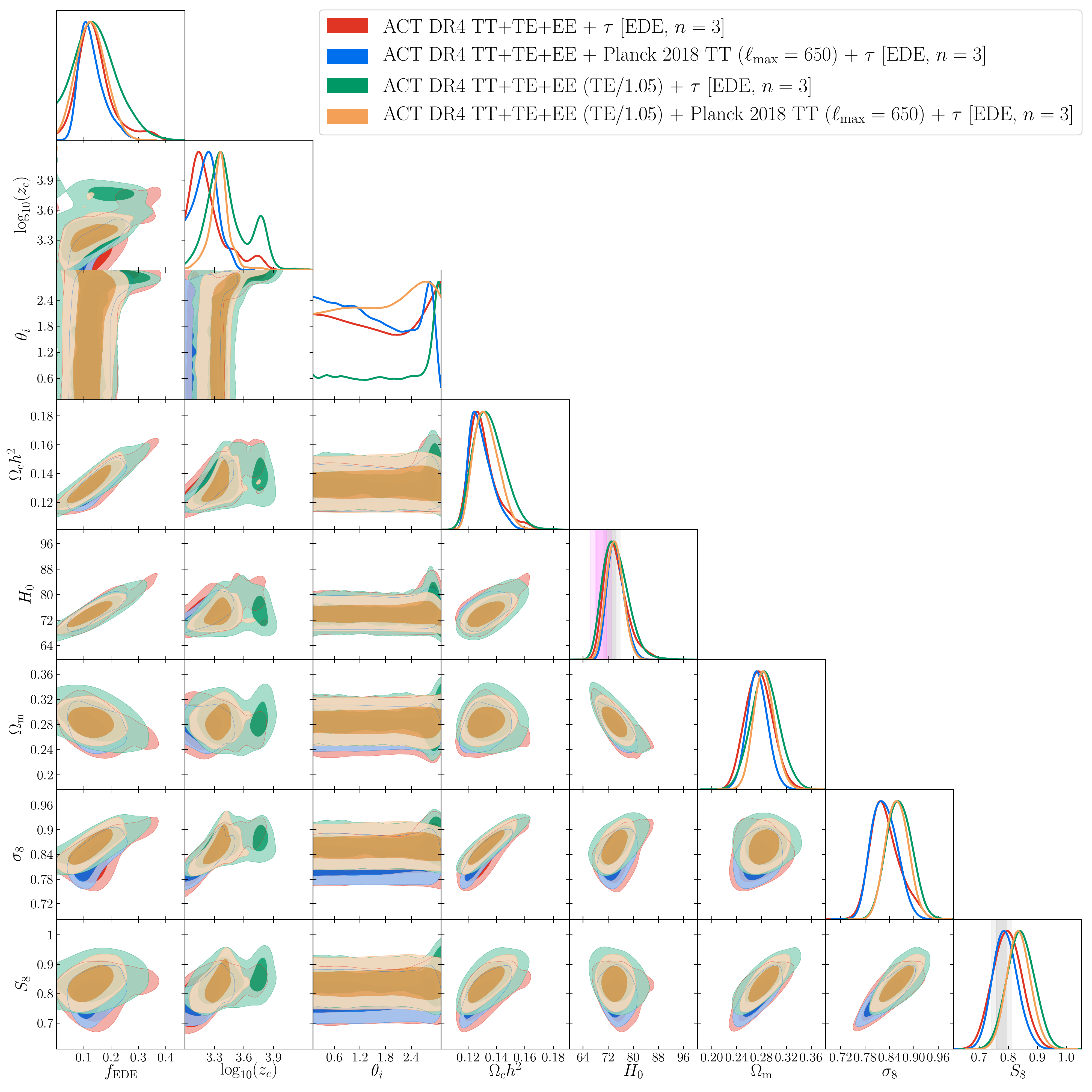}
\caption{Marginalized posteriors for the EDE parameters and a subset of other parameters of interest in fits to ACT DR4 TT+TE+EE (red) and ACT DR4 TT+TE+EE with \emph{Planck} 2018 TT ($\ell_{\rm max} = 650$) (blue).  The green and orange contours show results for the same data set combinations, but with the ACT TE data divided by 1.05.  These results are discussed in Sec.~\ref{subsec:ACT_alone} and~\ref{subsec:ACT_P18TTlmax650}.  The vertical grey and magenta bands in the $H_0$ panel show the latest SH0ES~\cite{Riess:2020fzl} and TRGB~\cite{Freedman:2021ahq} constraints, respectively.  The vertical grey band in the $S_8$ panel shows the DES-Y3 constraint~\cite{DES:2021wwk}.}
\label{fig:EDE_ACT_TEresc}
\end{figure*}

\begin{figure*}[!tp]
\includegraphics[width=\textwidth]{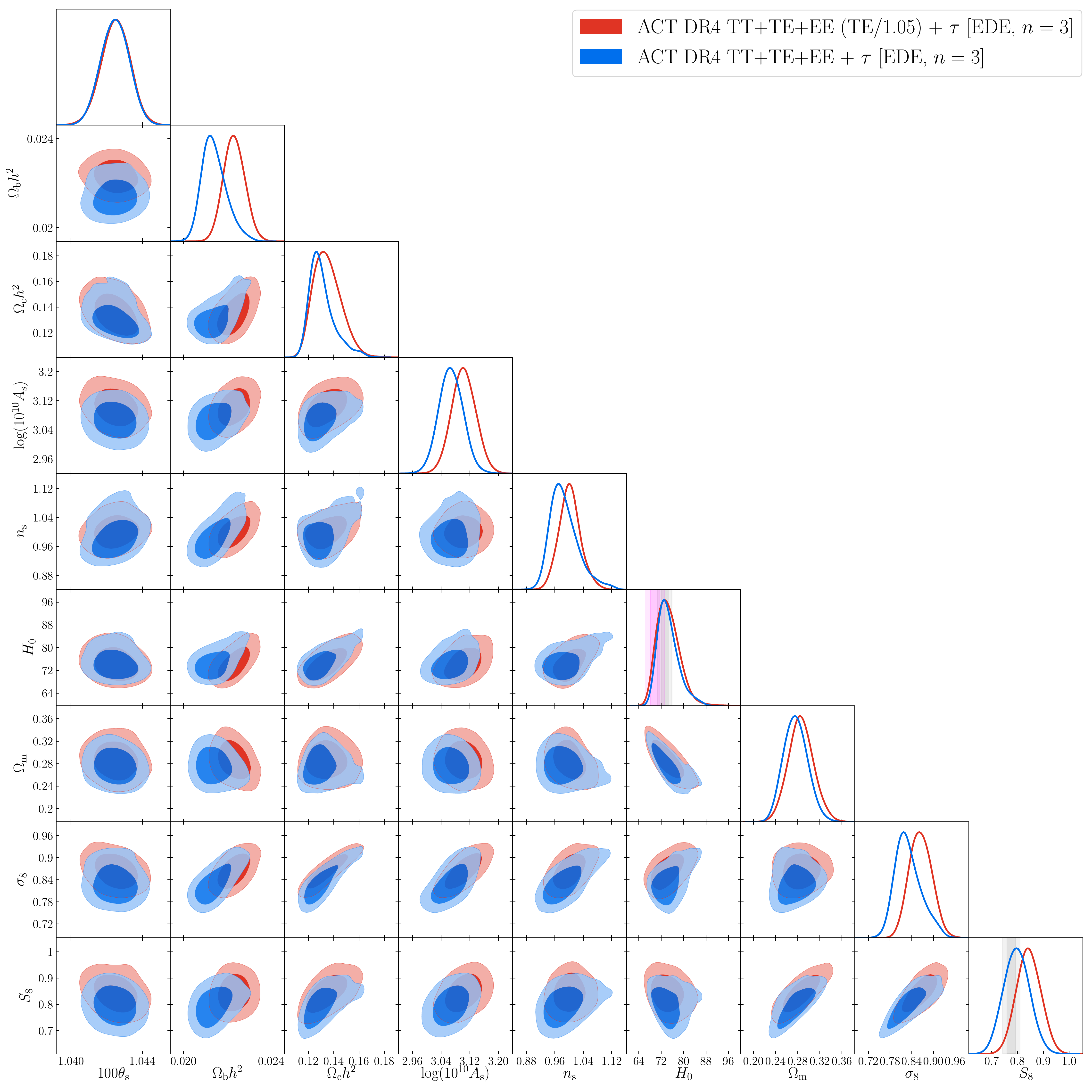}
\caption{Marginalized posteriors for the standard cosmological parameters in the EDE model fit to ACT DR4 TT+TE+EE data in combination with a Gaussian prior on $\tau$, as presented in Sec.~\ref{subsec:ACT_alone}.  The blue contours show the same ACT results presented in Fig.~\ref{fig:ACT_alone}, while the red contours show results found after dividing the ACT TE data by 1.05.  The vertical grey and magenta bands in the $H_0$ panel show the latest SH0ES~\cite{Riess:2020fzl} and TRGB~\cite{Freedman:2021ahq} constraints, respectively.  The vertical grey band in the $S_8$ panel shows the DES-Y3 constraint~\cite{DES:2021wwk}.}
\label{fig:ACT_alone_TEresc}
\end{figure*}

\begin{figure*}[!tp]
\includegraphics[width=\textwidth]{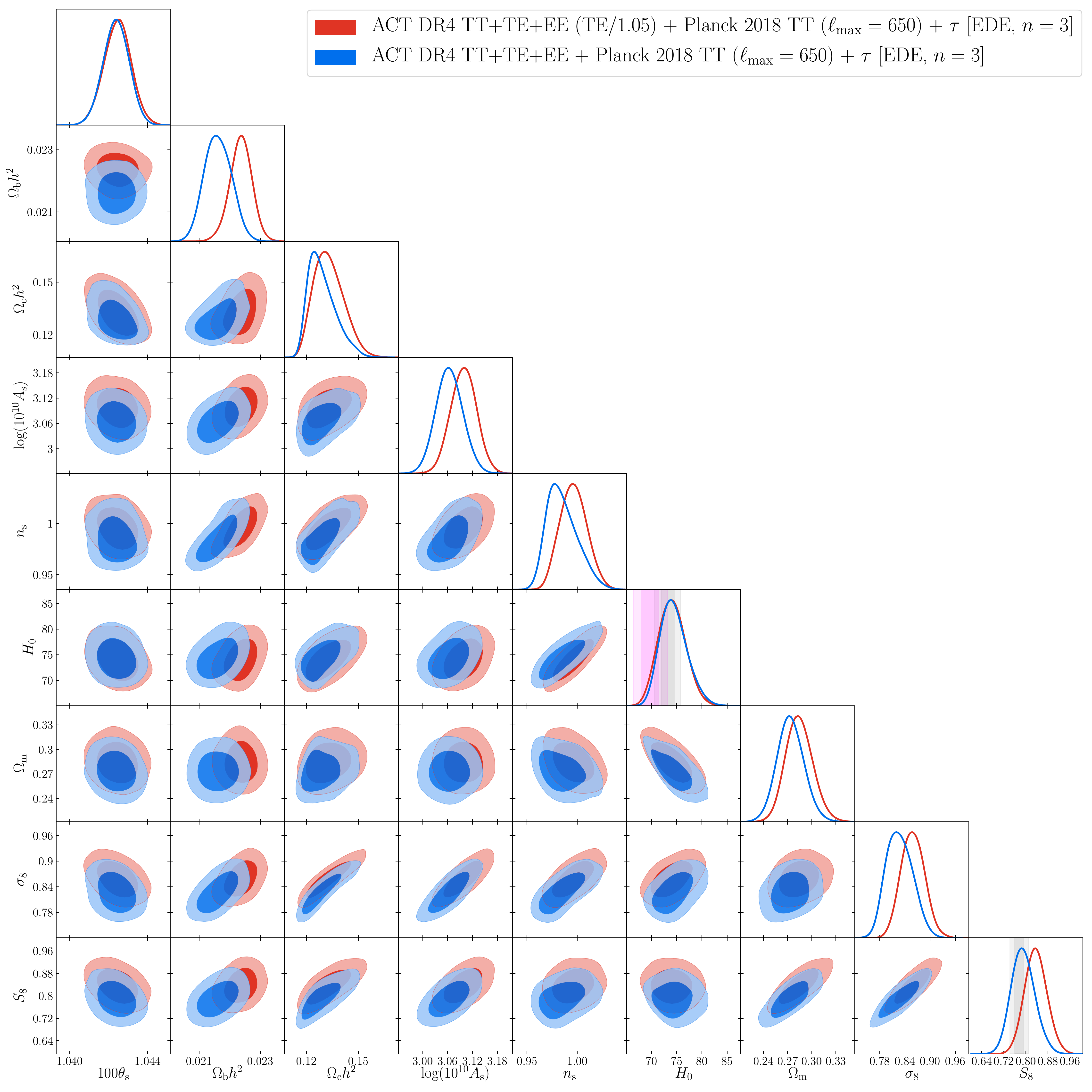}
\caption{Marginalized posteriors for the standard cosmological parameters in the EDE model fit to ACT DR4 TT+TE+EE and \emph{Planck} 2018 TT ($\ell_{\rm max} = 650$) data in combination with a Gaussian prior on $\tau$, as presented in Sec.~\ref{subsec:ACT_P18TTlmax650}.  The blue contours show the results presented previously in Fig.~\ref{fig:ACT_PlanckTTlmax650}, while the red contours show results found after dividing the ACT TE data by 1.05.  The vertical grey and magenta bands in the $H_0$ panel show the latest SH0ES~\cite{Riess:2020fzl} and TRGB~\cite{Freedman:2021ahq} constraints, respectively.  The vertical grey band in the $S_8$ panel shows the DES-Y3 constraint~\cite{DES:2021wwk}.}
\label{fig:ACT_PlanckTTlmax650_TEresc}
\end{figure*}

\begin{figure*}[!tp]
\includegraphics[width=0.87\textwidth]{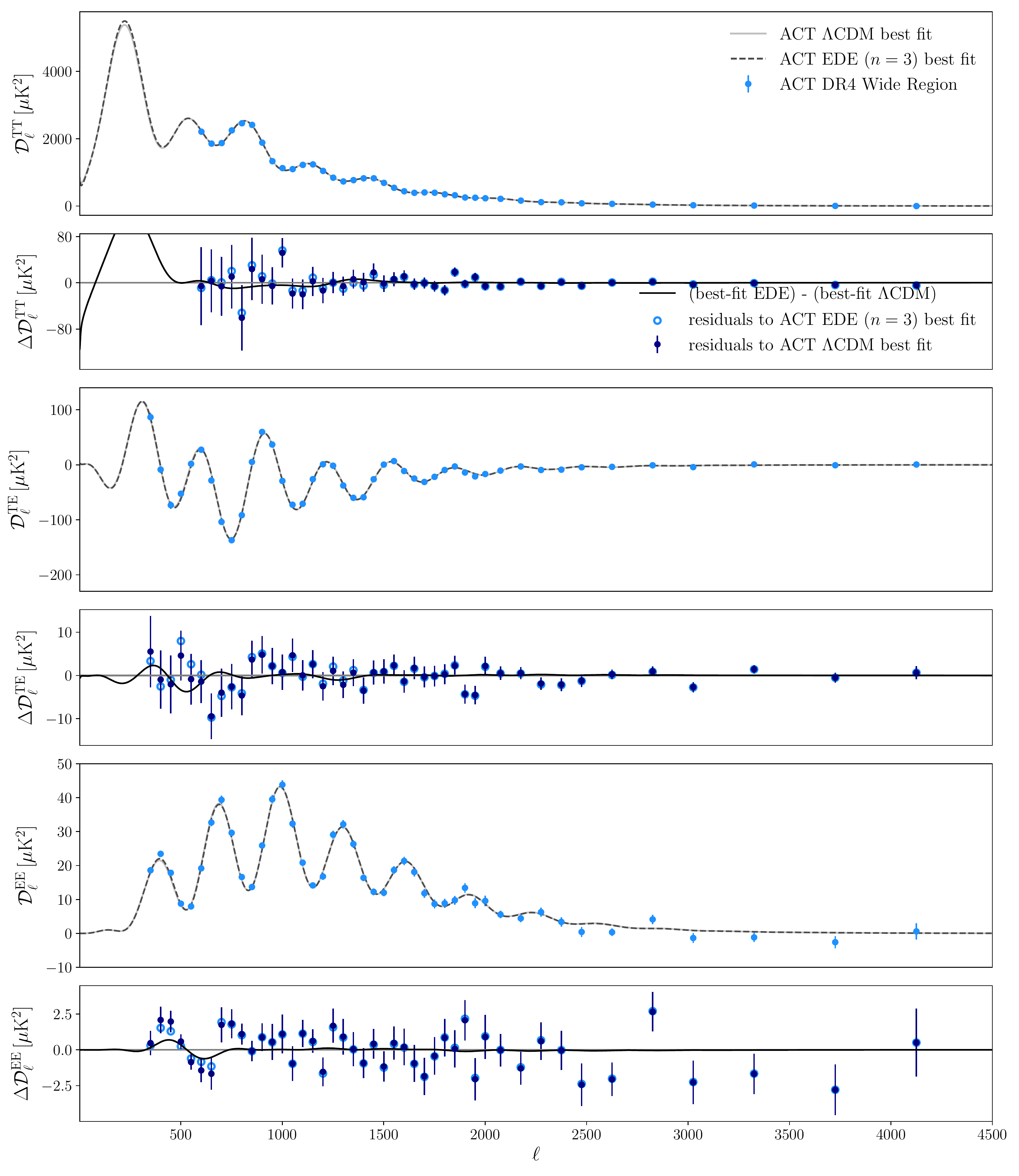}
\caption{Best-fit $\Lambda$CDM and EDE ($n=3$) models to the ACT DR4 TT (top), TE (middle), and EE (bottom) power spectra, shown here in comparison to the wide-patch data.  (The best-fit models are determined by the full data set shown in Fig.~\ref{fig:ACT_alone_residuals}.)  The smaller panels show the residuals of the best-fit models with respect to the wide-patch data, as well as the difference between the best-fit EDE and $\Lambda$CDM models.  The overall EDE $\chi^2$ improvement over $\Lambda$CDM (c.f. Table~\ref{table:chi2_ACT_alone}) is driven entirely by the seven lowest multipole bins in the wide-patch EE power spectrum shown here.}
\label{fig:ACT_alone_wide_residuals}
\end{figure*}

\begin{figure*}[!tp]
\includegraphics[width=0.87\textwidth]{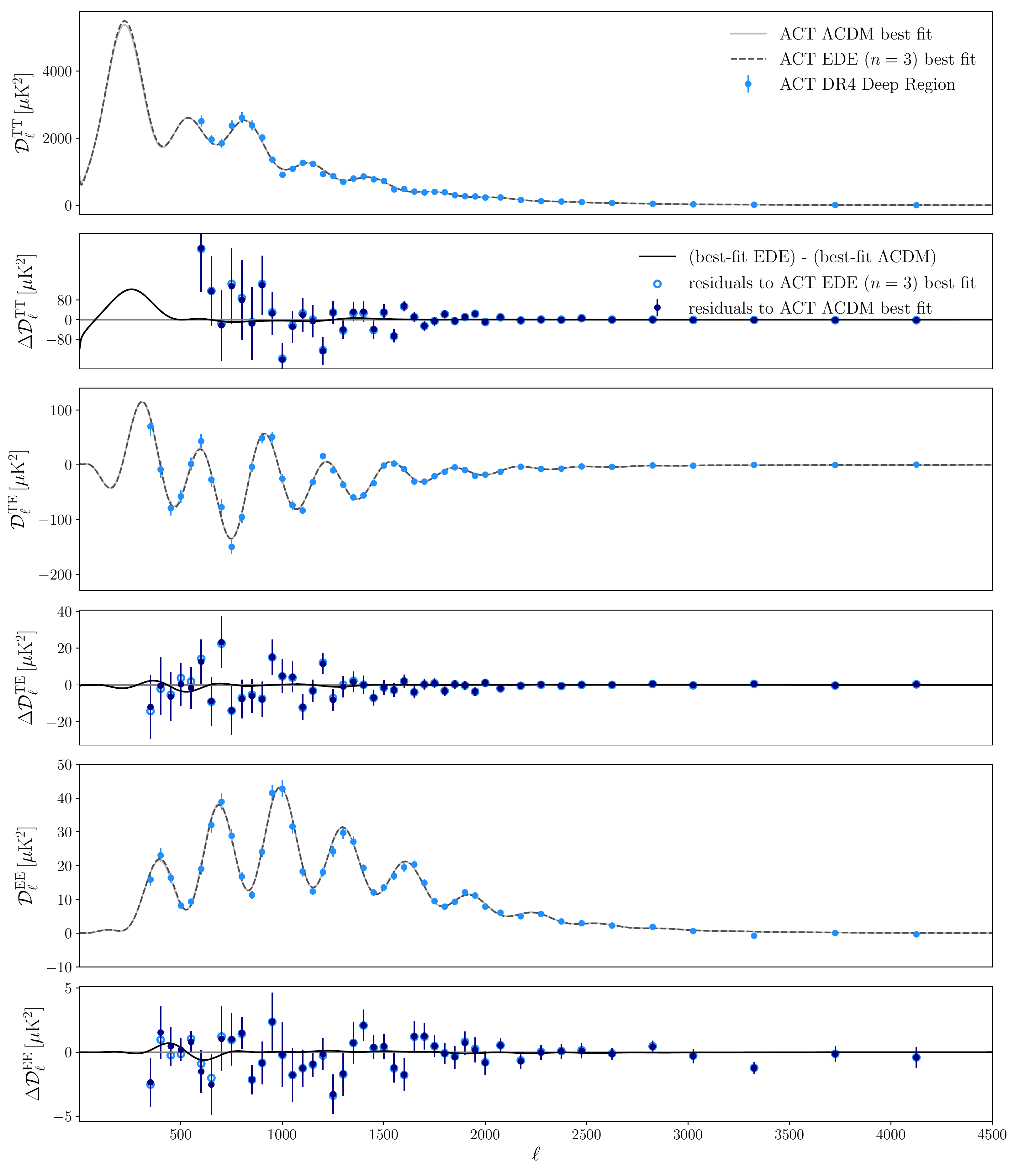}
\caption{Best-fit $\Lambda$CDM and EDE ($n=3$) models to the ACT DR4 TT (top), TE (middle), and EE (bottom) power spectra, shown here in comparison to the deep-patch data.  (The best-fit models are determined by the full data set shown in Fig.~\ref{fig:ACT_alone_residuals}.)  The smaller panels show the residuals of the best-fit models with respect to the deep-patch data, as well as the difference between the best-fit EDE and $\Lambda$CDM models.  The EDE model provides negligible improvement over $\Lambda$CDM in the fit to the deep-patch data shown here. }
\label{fig:ACT_alone_deep_residuals}
\end{figure*}

\begin{figure*}[!tp]
\includegraphics[width=0.87\textwidth]{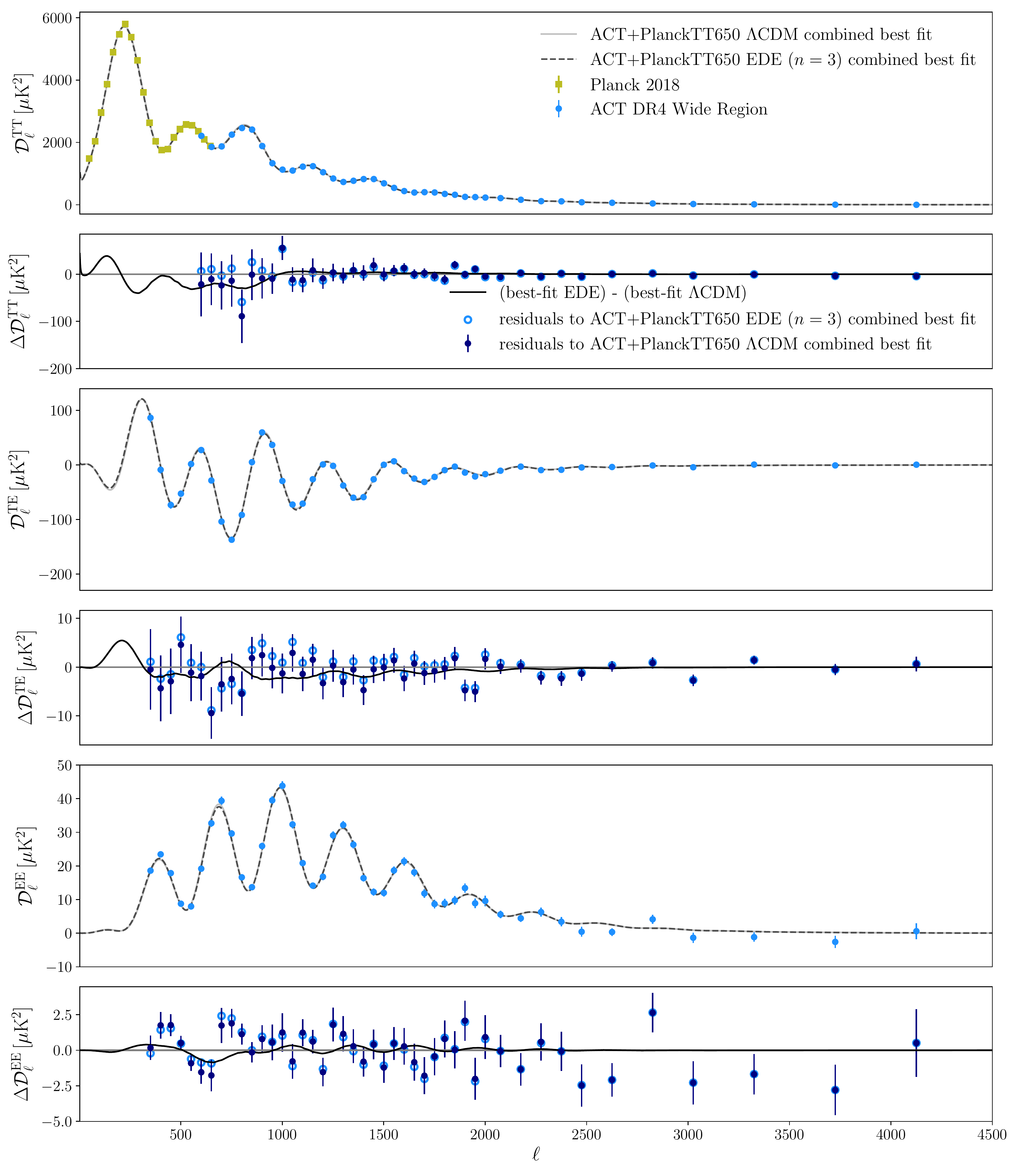}
\caption{Best-fit $\Lambda$CDM and EDE ($n=3$) models to the ACT DR4 TT (top), TE (middle), and EE (bottom) power spectrum data, fit in combination with large-scale ($\ell_{\rm max} = 650$) \emph{Planck} 2018 TT power spectrum data (yellow squares in top panel), shown here in comparison to the wide-patch ACT data.  (The best-fit models are determined by the full data set shown in Fig.~\ref{fig:ACT_P18TTlmax650_residuals}.)  The smaller panels show the residuals of the best-fit models with respect to the wide-patch data, as well as the difference between the best-fit EDE and $\Lambda$CDM models.  The most significant difference seen here is in the low-$\ell$ EE data, similar to that in the ACT-only analysis in Fig.~\ref{fig:ACT_alone_wide_residuals}.}
\label{fig:ACT_P18TTlmax650_wide_residuals}
\end{figure*}

\begin{figure*}[!tp]
\includegraphics[width=0.87\textwidth]{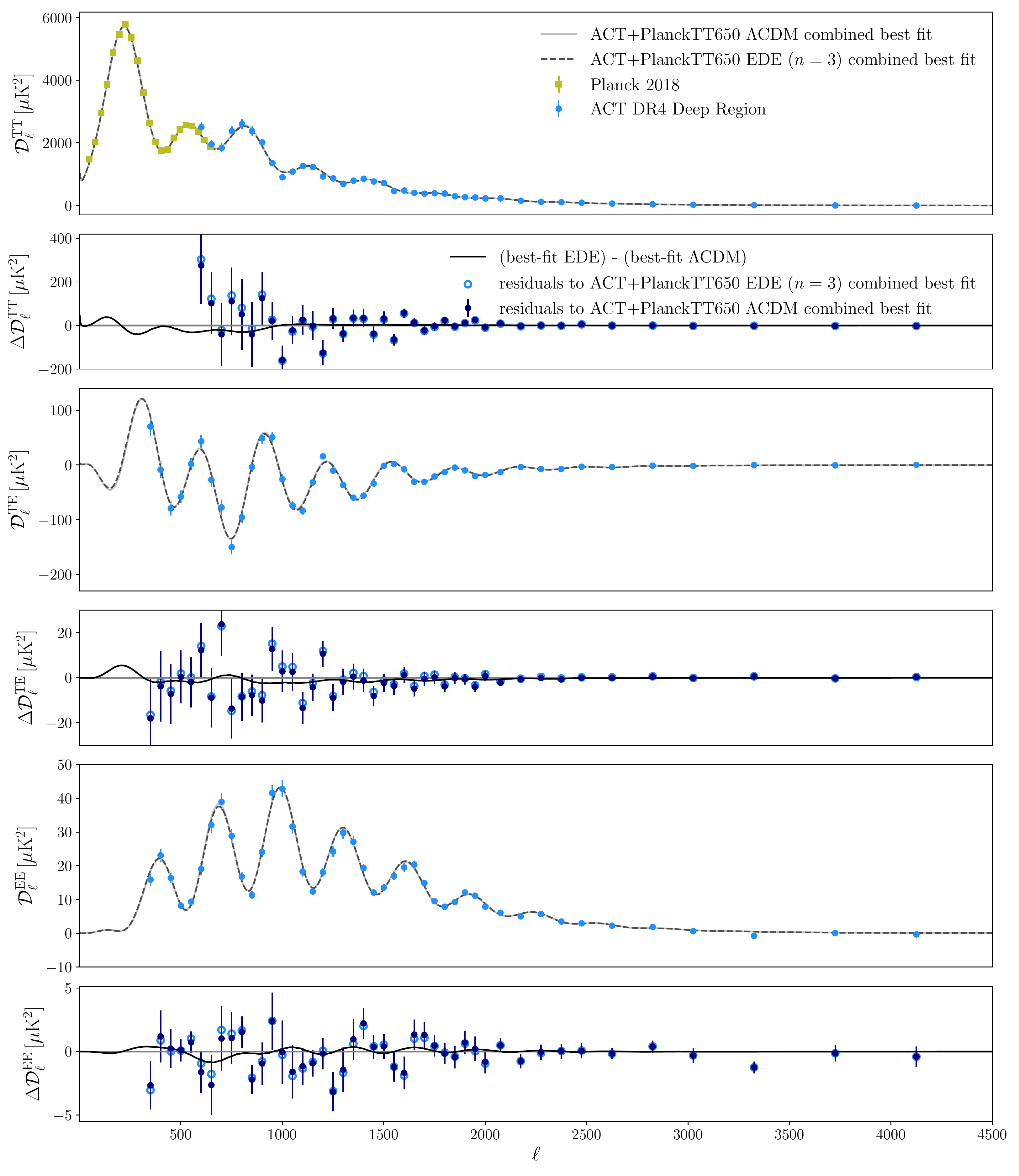}
\caption{Best-fit $\Lambda$CDM and EDE ($n=3$) models to the ACT DR4 TT (top), TE (middle), and EE (bottom) power spectrum data, fit in combination with large-scale ($\ell_{\rm max} = 650$) \emph{Planck} 2018 TT power spectrum data (yellow squares in top panel), shown here in comparison to the deep-patch ACT data.  (The best-fit models are determined by the full data set shown in Fig.~\ref{fig:ACT_P18TTlmax650_residuals}.)  The smaller panels show the residuals of the best-fit models with respect to the deep-patch data, as well as the difference between the best-fit EDE and $\Lambda$CDM models.  The EDE $\chi^2$ improvement over $\Lambda$CDM (c.f. Table~\ref{table:chi2_ACT_P18TTlmax650}) receives its largest contribution from the ACT deep-patch TE power spectrum shown here, although moderate improvements in TT and EE are also seen.  The EDE-$\Lambda$CDM residuals in TE are distributed very evenly in $\ell$, with no individual bin differing by $>0.5\sigma$.}
\label{fig:ACT_P18TTlmax650_deep_residuals}
\end{figure*}

\begin{figure*}[!tp]
\includegraphics[width=0.87\textwidth]{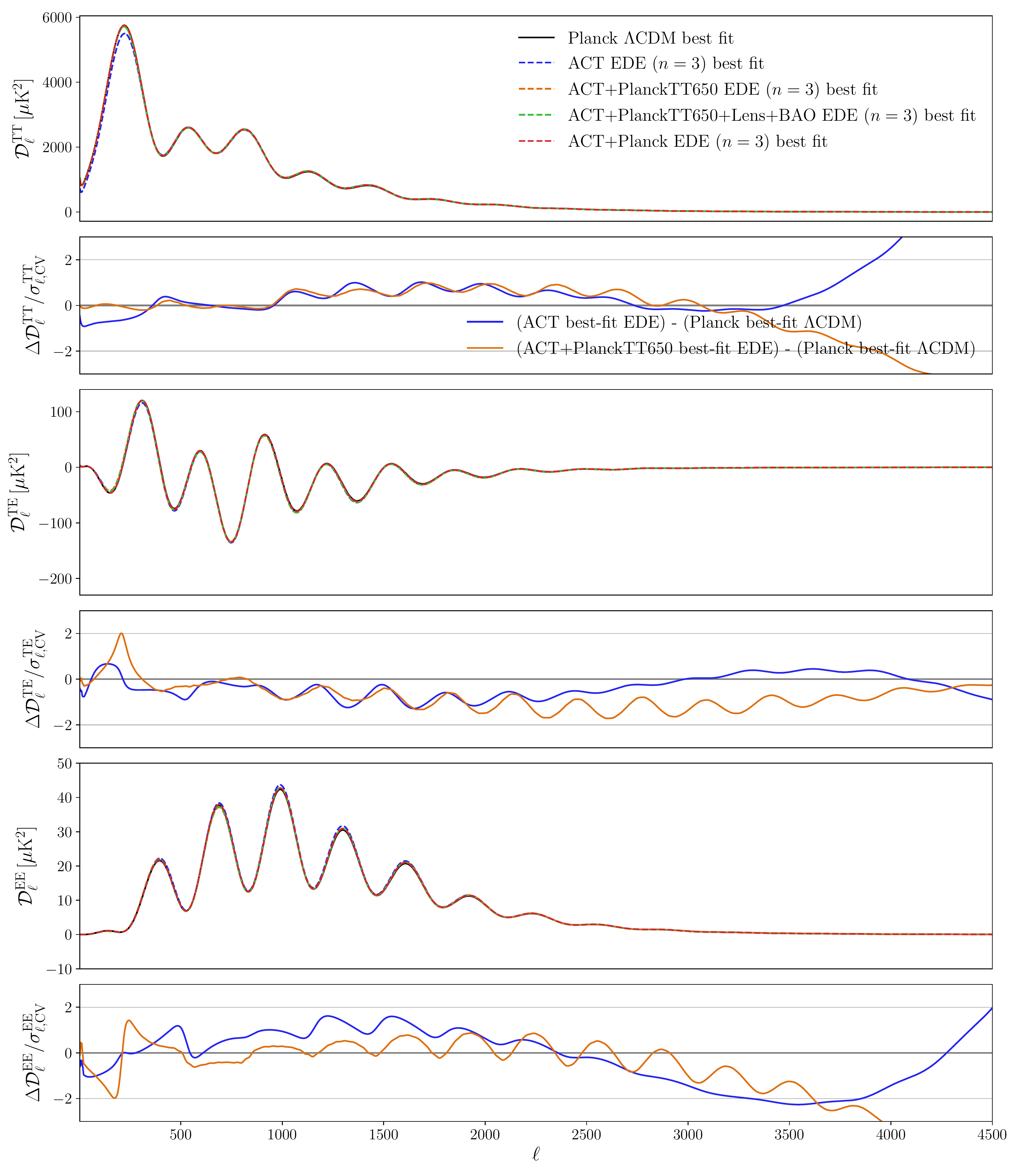}
\caption{Comparison of TT (top), TE (middle), and EE power spectra in the best-fit EDE models to the data set combinations considered in Sec.~\ref{sec:analysis} and the best-fit $\Lambda$CDM model to \emph{Planck} data alone (the latter from Ref.~\cite{Hill:2020osr}, but in excellent agreement with Ref.~\cite{Planck2018parameters}).  The smaller panels show differences with respect to the \emph{Planck} best-fit $\Lambda$CDM model in units of the CV-limited error bar at each $\ell$, with the $\pm 2 \sigma$ range demarcated by the thin grey lines.  The blue curve shows residuals for the best-fit EDE model to ACT DR4 alone (Table~\ref{table:params-ACT-DR4}) and the orange curve shows residuals for the best-fit EDE model to ACT + large-scale \emph{Planck} TT data (Table~\ref{table:params-ACT-DR4-P18TTlmax650}).  The residuals shown here are similar to those seen with respect to the best-fit EDE model to \emph{Planck} alone, shown in Fig.~\ref{fig:EDE_bf_comp_CV}.}
\label{fig:EDE_LCDM_bf_comp_CV}
\end{figure*}


\bibliographystyle{JHEP}
\bibliography{refs}

\end{document}